\documentclass{pasj01}

\usepackage{natbib}
\usepackage{ulem}
\usepackage{url}
\usepackage[switch,mathlines]{lineno}

\definecolor{nsgreen}{rgb}{0.1,0.5,0.1}
\definecolor{orange}{rgb}{0.8, 0.40, 0.0}

\newcommand{\comment}[1]{}
\newcommand{\SgrA}[0]{Sgr\,A\textsuperscript{$\ast$}}
\def\jasmine{{\it JASMINE}}
\def\uas{$\mathrm{\mu}$as}
\def\um{$\mathrm{\mu}$m}

\Received{11-Jul-2023}
\Accepted{27-Feb-2024}
\Published{$\langle$publication date$\rangle$}

\title{\jasmine{}: Near-Infrared Astrometry and Time Series Photometry Science}
\author{Daisuke Kawata$^{1,2,*}$}
\author{Hajime Kawahara$^{3,12}$}
\author{Naoteru Gouda$^{1,4}$}
\author{Nathan J.\ Secrest$^{5}$}
\author{Ryouhei Kano$^{1,3}$}
\author{Hirokazu Kataza$^{1,3}$}
\author{Naoki Isobe$^{3}$}
\author{Ryou Ohsawa$^{1}$}
\author{Fumihiko Usui$^{3}$}
\author{Yoshiyuki Yamada$^{6}$}
\author{Alister W. Graham$^{7}$}
\author{Alex R. Pettitt$^{8}$}
\author{Hideki Asada$^{9}$}
\author{Junichi Baba$^{1,10}$}
\author{Kenji Bekki$^{11}$}
\author{Bryan N.\ Dorland$^{5}$}
\author{Michiko Fujii$^{12}$}
\author{Akihiko Fukui$^{13}$}
\author{Kohei Hattori$^{1,14}$}
\author{Teruyuki Hirano$^{15}$}
\author{Takafumi Kamizuka$^{16}$}
\author{Shingo Kashima$^{1}$}
\author{Norita Kawanaka$^{17}$}
\author{Yui Kawashima$^{3,18}$}
\author{Sergei A.\ Klioner$^{19}$}
\author{Takanori Kodama$^{20}$}
\author{Naoki Koshimoto$^{21,22}$}
\author{Takayuki Kotani$^{4,15}$}
\author{Masayuki Kuzuhara$^{15}$}
\author{Stephen E.\ Levine$^{23,24}$}
\author{Steven R.\ Majewski$^{25}$}
\author{Kento Masuda$^{26}$}
\author{Noriyuki Matsunaga$^{12}$}
\author{Kohei Miyakawa$^{1}$}
\author{Makoko Miyoshi$^{1}$}
\author{Kumiko Morihana$^{27}$}
\author{Ryoichi Nishi$^{28}$}
\author{Yuta Notsu$^{29,30}$}
\author{Masashi Omiya$^{15}$}
\author{Jason Sanders$^{31}$}
\author{Ataru Tanikawa$^{32}$}
\author{Masahiro Tsujimoto$^{1}$}
\author{Taihei Yano$^{1}$}
\author{Masataka Aizawa$^{33}$}
\author{Ko Arimatsu$^{34}$}
\author{Michael Biermann$^{35}$}
\author{Celine Boehm$^{36}$}
\author{Masashi Chiba$^{37}$}
\author{Victor P. Debattista$^{38}$}
\author{Ortwin Gerhard$^{39}$}
\author{Masayuki Hirabayashi$^{1}$}
\author{David Hobbs$^{40}$}
\author{Bungo Ikenoue$^{1}$}
\author{Hideyuki Izumiura$^{41}$}
\author{Carme Jordi$^{42,43,44}$}
\author{Naoki Kohara$^{1}$}
\author{Wolfgang L\"offler$^{35}$}
\author{Xavier Luri$^{42,43,44}$}
\author{Ichiro Mase$^{1}$}
\author{Andrea Miglio$^{45,46}$}
\author{Kazuhisa Mitsuda$^{1}$}
\author{Trent Newswander$^{47}$}
\author{Shogo Nishiyama$^{48}$}
\author{Yoshiyuki Obuchi$^{1}$}
\author{Takafumi Ootsubo$^{1}$}
\author{Masami Ouchi$^{1,49,50}$}
\author{Masanobu Ozaki$^{1}$}
\author{Michael Perryman$^{51}$}
\author{Timo Prusti$^{52}$}
\author{Pau Ramos$^{1}$}
\author{Justin I. Read$^{53}$}
\author{R. Michael Rich$^{54}$}
\author{Ralph Sch\"onrich$^{2}$}
\author{Minori Shikauchi$^{55,56}$}
\author{Risa Shimizu$^{1}$}
\author{Yoshinori Suematsu$^{1}$}
\author{Shotaro Tada$^{4}$}
\author{Aoi Takahashi$^{15}$}
\author{Takayuki Tatekawa$^{57,58}$}
\author{Daisuke Tatsumi$^{1}$}
\author{Takuji Tsujimoto$^{1}$}
\author{Toshihiro Tsuzuki$^{1}$}
\author{Seitaro Urakawa$^{59}$}
\author{Fumihiro Uraguchi$^{1}$}
\author{Shin Utsunomiya$^{1}$}
\author{Vincent Van Eylen$^{2}$}
\author{Floor van Leeuwen$^{60}$}
\author{Takehiko Wada$^{1}$}
\author{Nicholas A. Walton$^{60}$}
\altaffiltext{1}{National Astronomical Observatory of Japan, 2-21-1 Osawa, Mitaka, Tokyo 181-8588, Japan }
\altaffiltext{2}{Mullard Space Science Laboratory, University College London, \\ Holmbury St. Mary, Dorking, Surrey RH5 6NT, UK}
\altaffiltext{3}{Institute of Space and Astronautical Science, Japan Aerospace Exploration Agency, 3-1-1 Yoshinodai, Chuo-ku, Sagamihara, Kanagawa, 252-5210, Japan}
\altaffiltext{4}{Astronomical Science Program, Graduate Institute for Advanced Studies, SOKENDAI, 2-21-1 Osawa, Mitaka, Tokyo, 181-1855 Japan}
\altaffiltext{5}{U.S.\ Naval Observatory, 3450 Massachusetts Ave NW, Washington, DC 20392-5420, USA}
\altaffiltext{6}{Department of Physics, Kyoto University, Kitashirakawa-oiwake-cho, Sakyo-ku, Kyoto 606-8502, Japan}
\altaffiltext{7}{Centre for Astrophysics and Supercomputing, Swinburne University of Technology, Hawthorn, VIC 3122, Australia}
\altaffiltext{8}{Department of Physics and Astronomy, California State University, Sacramento, 6000 J Street, Sacramento, CA 95819-6041, USA}
\altaffiltext{9}{Graduate School of Science and Technology, Hirosaki University, Aomori 036-8561, Japan}
\altaffiltext{10}{Graduate School of Science and Engineering, Kagoshima University, 1-21-35 Korimoto, Kagoshima 890-0065, Japan}
\altaffiltext{11}{International Centre for Radio Astronomy Research, The University of Western Australia, 7 Fairway, Crawley, WA 6009, Australia}
\altaffiltext{12}{Department of Astronomy, Graduate School of Science, The University of Tokyo, 7-3-1 Hongo, Bunkyo-ku, Tokyo 113-0033, Japan}
\altaffiltext{13}{Department of Earth and Planetary Science, Graduate School of Science, The University of Tokyo, 7-3-1 Hongo, Bunkyo-ku, Tokyo 113-0033, Japan}
\altaffiltext{14}{Institute of Statistical Mathematics, 10-3 Midoricho, Tachikawa, Tokyo, 190-8562, Japan}
\altaffiltext{15}{Astrobiology Center,  2-21-1 Osawa, Mitaka, Tokyo 181-8588, Japan }
\altaffiltext{16}{Institute of Astronomy, Graduate School of Science, The University of Tokyo, 2-21-1 Osawa, Mitaka, Tokyo 181-0015, Japan}
\altaffiltext{17}{Center for Gravitational Physics and Quantum Information, Yukawa Institute for Theoretical Physics, Kitashirakawa-oiwake-cho, Sakyo-ku, Kyoto University, Kyoto, 606-8502, Japan}
\altaffiltext{18}{Cluster for Pioneering Research, RIKEN, 2-1 Hirosawa, Wako, Saitama 351-0198, Japan}
\altaffiltext{19}{Lohrmann Observatory, Technische Universit\"at Dresden, 01062 Dresden, Germany}
\altaffiltext{20}{Earth-Life Institute (ELSI), Tokyo Institute of Technology, 2-12-1 Ookayama, Meguro, Tokyo 152-8550, Japan}
\altaffiltext{21}{Code 667, NASA Goddard Space Flight Center, Greenbelt, MD 20771, USA}
\altaffiltext{22}{Department of Astronomy, University of Maryland, College Park, MD 20742, USA}
\altaffiltext{23}{Lowell Observatory, 1400 W Mars Hill Rd, Flagstaff, AZ 86001 USA}
\altaffiltext{24}{Massachusetts Institute of Technology, Department of Earth, Atmospheric and Planetary Sciences, 77 Massachusetts Ave. Cambridge, MA 02139 USA}
\altaffiltext{25}{Department of Astronomy, University of Virginia, P.O. Box 400325, Charlottesville, VA 22904-4325, USA}
\altaffiltext{26}{Department of Earth and Space Science, Osaka University, Osaka 560-0043, Japan}
\altaffiltext{27}{Subaru Telescope, National Astronomical Observatory of Japan, Hilo, HI 96720, USA}
\altaffiltext{28}{Department of Physics, Niigata University, 8050 Ikarashi-2, Nishi-ku, Niigata, Niigata 950-2181, Japan}
\altaffiltext{29}{Laboratory for Atmospheric and Space Physics, University of Colorado Boulder,
3665 Discovery Drive, Boulder, CO 80303, USA}
\altaffiltext{30}{National Solar Observatory, University of Colorado Boulder, 3665 Discovery Drive, Boulder, CO 80303, USA}
\altaffiltext{31}{Department of Physics and Astronomy, University College London, London WC1E 6BT, UK}
\altaffiltext{32}{Department of Earth Science and Astronomy, College of Arts and Sciences, University of Tokyo, 3-8-1 Komaba, Meguro-ku, Tokyo, 153-8902, Japan}
\altaffiltext{33}{Tsung-Dao Lee Institute, Shanghai Jiao Tong University, Shengrong Road 520, 201210 Shanghai, China}
\altaffiltext{34}{The Hakubi Center/Astronomical Observatory, Graduate School of Science, Kyoto University, Kitashirakawa-oiwake-cho, Sakyo-ku, Kyoto 606-8502, Japan} 
\altaffiltext{35}{Astronomisches Rechen-Institut, Zentrum f\"ur Astronomie der Universit\"at Heidelberg, M\"onchhofstr. 12-14, 69120 Heidelberg, Germany}
\altaffiltext{36}{School of Physics, Faculty of Science, University of Sydney, NSW 2006, Australia}
\altaffiltext{37}{Astronomical Institute, Tohoku University, Aoba-ku, Sendai 980-8578, Japan}
\altaffiltext{38}{Jeremiah Horrocks Institute, University of Central Lancashire, Preston, PR1 2HE, UK}
\altaffiltext{39}{Max-Planck-Institut f\"ur extraterrestrische Physik, Giessenbachstraße, 85748 Garching, Germany}
\altaffiltext{40}{Lund Observatory, Division of Astrophysics, Department of Physics, Lund University, Box 43, SE-221, Lund, Sweden}
\altaffiltext{41}{Subaru Telescope Okayama Branch, National Astronomical Observatory of Japan, National Institutes of Natural Sciences, Honjo 3037-5, Kamogata, Asakuchi, Okayama 719-0232, Japan}
\altaffiltext{42}{Institut de Ci\`encies del Cosmos (ICCUB), Universitat de Barcelona (UB), Mart\'i i Franqu\`es 1, E-08028 Barcelona, Spain}
\altaffiltext{43}{Departament de F\'isica Qu\`antica i Astrof\'isica (FQA), Universitat de Barcelona (UB), Mart\'i i Franqu\`es 1, E-08028 Barcelona, Spain}
\altaffiltext{44}{Institut d'Estudis Espacials de Catalunya (IEEC), c. Gran Capit\`a, 2-4, 08034 Barcelona, Spain}
\altaffiltext{45}{Dipartimento di Fisica e Astronomia ``Augusto Righi'', Universit\`a di Bologna, via Gobetti 93/2, 40129 Bologna, Italy}
\altaffiltext{46}{INAF-Osservatorio di Astrofisica e Scienza dello Spazio di Bologna, via Gobetti 93/3, 40129 Bologna, Italy}
\altaffiltext{47}{Utah State University Space Dynamics Laboratory, 416 East Innovation Avenue, North Logan, UT 84341}
\altaffiltext{48}{Miyagi University of Education, Aoba-ku, Sendai, Miyagi 980-0845, Japan}
\altaffiltext{49}{Institute for Cosmic Ray Research, The University of Tokyo, 5-1-5 Kashiwanoha, Kashiwa, Chiba 277-8582, Japan}
\altaffiltext{50}{Kavli Institute for the Physics and Mathematics of the Universe (Kavli IPMU, WPI), The University of Tokyo, 5-1-5 Kashiwanoha, Kashiwa, Chiba 277-8583, Japan}
\altaffiltext{51}{School of Physics, University College Dublin, Belfield, Dublin 4, Ireland}
\altaffiltext{52}{European Space Agency (ESA), European Space Research and Technology Centre (ESTEC), Keplerlaan 1, 2201 AZ Noordwijk, The Netherlands}
\altaffiltext{53}{Department of Physics, University of Surrey, Guildford GU2 7XH, UK}
\altaffiltext{54}{Department of Physics and Astronomy, University of California, Los Angeles, LA, CA 90095-1547, USA}
\altaffiltext{55}{Department of Physics, the University of Tokyo, 7-3-1 Hongo, Bunkyo, Tokyo 113-0033, Japan}
\altaffiltext{56}{Research Center for the Early Universe (RESCEU), the University of Tokyo, 7-3-1 Hongo, Bunkyo, Tokyo 113-0033, Japan}
\altaffiltext{57}{National Institute of Technology (KOSEN), Kochi college, 200-1 Monobe, Nankoku, Kochi 783-8508 Japan}
\altaffiltext{58}{Waseda Research Institute for Science and Engineering, Waseda University, 3-4-1 Okubo, Shinjuku, Tokyo 169-8555 Japan}
\altaffiltext{59}{Japan Spaceguard Association, Bisei Spaceguard Center 1716-3 Okura, Bisei, Ibara, Okayama 714-1411, Japan}
\altaffiltext{60}{Institute of Astronomy, University of Cambridge, Madingley Road, Cambridge CB3 0HA, UK}

\begin{document}


\maketitle

\begin{abstract}

{\it Japan Astrometry Satellite Mission for INfrared Exploration} (\jasmine{}) is a planned M-class science space mission by the Institute of Space and Astronautical Science, the Japan Aerospace Exploration Agency. \jasmine{} has two main science goals. One is the Galactic archaeology with Galactic Center Survey, which aims to reveal the Milky Way's central core structure and formation history from Gaia-level ($\sim25$ \uas{}) astrometry in the Near-Infrared (NIR) $H_\mathrm{w}$-band  (1.0--1.6~\um{}). The other is the Exoplanet Survey, which aims to discover transiting Earth-like exoplanets in the habitable zone from NIR time-series photometry of M dwarfs when the Galactic center is not accessible. We introduce the mission, review many science objectives, and present the instrument concept. \jasmine{} will be the first dedicated NIR astrometry space mission and provide precise astrometric information of the stars in the Galactic center, taking advantage of the significantly lower extinction in the NIR.
The precise astrometry is obtained by taking many short-exposure images. Hence, the \jasmine{} Galactic center survey data will be valuable for studies of exoplanet transits, asteroseismology, variable stars and microlensing studies, including discovery of (intermediate mass) black holes. We highlight a swath of such potential science, and also describe synergies with other missions. 

\end{abstract}

\section{Introduction}
\label{sec:intro}

Since the discovery that the Galaxy is one among hundreds of billions in a vast cosmic web -- a gravitational framework set in place at the very first moments of the universe -- it has been understood that the Galaxy is an ancient record, carrying within it the products and imprints of the physics of the universe and the processes that eventually resulted in our Sun, Earth, and life itself. 
The near-infrared (NIR) astrometry space mission, {\it Japan Astrometry Satellite Mission for INfrared Exploration} (\jasmine{}\footnote{The \jasmine{} mission detailed in this paper was originally known as {\it Small-\jasmine{}} in previous documentation, but now is simply referred to as \jasmine{}.}$^,$\footnote{
\url{http://jasmine.nao.ac.jp/index-en.html}}) \citep[figure~\ref{fig:jasmine_pic},][]{Gouda+2011}, has been selected for an M-class space mission by the Institute of Space and Astronautical Science (ISAS), the Japan Aerospace Exploration Agency (JAXA), with a planned launch in 2028. \jasmine{} has two main science goals. One is to decipher the ancient stellar fossil record: to reveal the Milky Way's central core structure and its formation history from Gaia-level ($\sim25$~\uas{}) astrometry in the NIR $H_\mathrm{w}$-band (1.0--1.6~\um{}, $H_\mathrm{w}\approx0.9J+0.1H-0.06(J-H)^2$).
This is referred to as the Galactic Center Survey (GCS), spanning 2.52 deg$^2$. The other goal is the ExoPlanet Survey (EPS) to discover Earth-like habitable exoplanets from the NIR time-series photometry of M dwarf transits, obtained when the Galactic center is not accessible to \jasmine{}. \jasmine{} will be the first dedicated NIR astrometry space mission, and provide precise astrometric information of the stars in the Galactic center, taking advantage of the significantly lower extinction in the NIR band than in the visual band. The precise astrometry is obtained by taking many short-exposure ($\sim12.5$~sec) images. 
Hence, \jasmine{} can also provide time series NIR photometry data in addition to precise astrometry. 
The aim of this paper is to describe the novel science opened up by the unique new capabilities of the \jasmine{} mission.

\subsection{{\it Gaia} revolution and next step: NIR astrometry}
\label{sec:intro-ga}

The European Space Agency's {\it Gaia} mission \citep[launched in 2013,][]{Gaia+Prusti16} made its second data release \citep[DR2,][]{Gaia+Brown+Vallenari+18} in April 2018 and subsequent third data release \cite[EDR3 in 2020 and DR3 in 2022,][]{Gaia+Brown+21, Gaia+Vallenari+23}, which provided measurements of the positions, motions and photometric properties for more than one billion stars with unprecedented precision. {\it Gaia} DR2 and DR3, in combination with ground-based spectroscopic surveys such as the RAdial Velocity Experiment \citep[RAVE,][]{Steinmetz+06}, the Sloan Digital Sky Survey (SDSS), the Sloan Extension for Galactic Understanding and Exploration \citep[SEGUE,][]{Yanny+Rockosi+Newberg+09}, the Apache Point Observatory Galactic Evolution Experiment \citep[APOGEE,][]{2017AJ....154...94M}, the Large Sky Area Multi-Object Fiber Spectroscopic Telescope \citep[LAMOST,][]{Zhao+Zhao+Chu12}, the Gaia-ESO survey \citep{Gilmore+Randich+Asplund+12}, and the Galactic Archaeology with HERMES survey \citep[GALAH,][]{Martel+Sharma+Bunder+17}, have revolutionized our view of the Galactic disk and halo, including their various stellar streams and the impact of satellite galaxies, and the interplay between the disks and stellar and dark halos.

For the Galactic disk, {\it Gaia} enabled us to analyze the 6D phase-space distribution (position, proper motion, radial velocity, parallax) of large samples of stars within a few kpc from the Sun for the first time, and revealed that the Galactic disk is heavily perturbed \citep{Gaia+Katz+18,Antoja+Helmi+RomeroGomez+18,Kawata+Baba+Ciuca+18,Bland-Hawthorn+Sharma+Tepper-Garcia+19}. These fine kinematic structures have provided insight into the origin of the spiral structure, possibly transient \citep[e.g.,][]{Baba+Kawata+Matsunaga+18,Hunt+Hong+Bovy+18,Hunt+Bub+Bovy+19}, and the pattern speed of the bar, including the possibility of it slowing down \citep[e.g.,][]{Monari+Famaey+Siebert+Wegg+Gerhard19b,Chiba+Schoenrich21}. In addition, the discovery of phase space features correlating in and out of plane motion of the disk stars
\citep[e.g.,][]{Antoja+Helmi+RomeroGomez+18,Hunt+Price-Whelan+Johnston+Darragh-Ford22} opened up a heated discussion of the origin of these features, which have been attributed to various causes including: the phase mixing after a tidal perturbation by a dwarf galaxy passing through the Galactic plane \citep[e.g.,][]{Binney+Schoenrich18,Laporte+Minchev+Johnston+Gomez19}, the bar buckling \citep{Khoperskov+DiMatteo+Gerhard+19}, and/or a persistent dark matter wake \citep{Grand+Pakmor+Fragkoudi+22}.  

Regarding the stellar Galactic halo, {\it Gaia} and ground-based spectroscopic data have revealed that a significant fraction of halo stars are moving on very radial orbits. This has been interpreted as a remnant of a relatively large galaxy, the so-called {\it Gaia}-Sausage-Enceladus, falling into the Milky Way about 10 Gyr ago \citep{Belokurov+Erkal+Evans+18,Helmi+Babusiaux+Koppelman+18,2019A&A...632A...4D, Gallart+Bernard+Brook+19}. 
This merger is considered to have disrupted the proto-Galactic disk and created the inner metal rich halo stars, which are suggested to be of the same population as the thick disk \citep{2019A&A...632A...4D,Gallart+Bernard+Brook+19,Belokurov+Sanders+Fattahi+20}.

Future {\it Gaia} data releases will undoubtedly yield further discoveries pertaining to the Galactic disk and halo structures and provide an even deeper insight into the formation and evolution history of the Milky Way. However, due to the high dust extinction in the optical band where {\it Gaia} operates ($\sim0.6~\micron$), {\it Gaia} cannot provide reliable astrometry for stars near the Galactic nucleus. This is an unfortunate hindrance, as this is where the history of the first structure and the formation of the Super-Massive Black Hole (SMBH) of the Galaxy should be imprinted. The \jasmine{} GCS targets the Galactic center field (spanning a projected Galactocentric radius of $R_{\rm GC}\lesssim100$~pc). 
\jasmine{} is designed to achieve {\it Gaia}-level astrometric accuracy toward the Galactic center in the NIR $H_\mathrm{w}$-band. The \jasmine{} GCS will provide precise astrometric information of the stars in the Galactic center. In this survey, for objects brighter than $H_\mathrm{w}=12.5$~mag, \jasmine{} will achieve a parallax accuracy of $\sigma_{\pi}\sim25$~\uas{} and a proper motion accuracy of $\sigma_{\mu}\sim25$~\uas{}~yr$^{-1}$. \jasmine{} plans to downlink the data for all stars brighter than $H_\mathrm{w}=14.5$~mag. For stars as bright as $H_\mathrm{w}=14.5$~mag, \jasmine{} will achieve an accuracy of $\sigma_{\mu}=125$~\uas{}~yr$^{-1}$.
Section~\ref{sec:gca} describes the scientific details of the  \jasmine{} GCS.

\subsection{Habitable Zone Exoplanet search}
\label{sec:intro-exopl}

A major ambition of modern astronomy is to identify the biosignatures of terrestrial exoplanets located in Habitable Zones (HZ). Directly imaged planets and transiting planets are the two primary targets for the astrobiological exploration of exoplanets. The space-based direct imaging missions, {\it HabEx} \citep{2020arXiv200106683G} and {\it LUVOIR} \citep{2019arXiv191206219T} and its successor {\it LUVeX} \citep{NAS21}, are capable of searching for metabolic gas biosignatures in the atmosphere of planets orbiting solar-type stars. The {\it James Webb Space Telescope} ({\it JWST}) and subsequent planned missions with larger telescopes, such as {\it LUVeX}, will characterize terrestrial planet atmospheres using stable low-dispersion spectroscopy of transiting exoplanets \citep[e.g.,][]{2019AJ....158...27L}. Extremely large ground-based telescopes such as the {\it Thirty Meter Telescope (TMT)}, the {\it Giant Magellan Telescope (GMT)}, and the {\it European Extremely Large Telescope (E-ELT)}, will search for biosignatures using transmission spectroscopy \citep[e.g.,][]{2013ApJ...764..182S,2014ApJ...781...54R} and direct imaging spectroscopy \citep[e.g.,][]{2012ApJ...758...13K,2013A&A...551A..99C} of terrestrial planets around late-type stars. There are also plans to use the UV-spectrograph on the {\it World Space Observatory-Ultraviolet (WSO-UV)} and {Life-environmentology, Astronomy, and PlanetarY Ultraviolet Telescope Assembly (\it LAPYUTA)} to search for the atomic oxygen (OI) line (130~nm), a potential biosignature, in the upper atmosphere of terrestrial transiting exoplanets orbiting late-type stars \citep[e.g.,][]{Tavrov+2018}.

The future exploration of terrestrial planet atmospheres relies crucially on identifying suitable targets prior to characterization. Among transiting planets, those in the HZ around late-type stars are best suited for two reasons. Firstly, the HZ planets around low-luminosity, late-type stars have much shorter orbital periods than the Earth. This makes it more plausible to find them via the transit method in the first place, and it also allows multiple monitorings of their transits. Secondly, the small radius of late-type stars deepens the transit signal compared to Sun-like stars. Various efforts have been made to detect terrestrial planets transiting late-type dwarfs, from both the ground and space. 
Several planets in the HZ of an ultracool dwarf, TRAPPIST-1, are good examples of transiting terrestrial planets that are ideal for further atmospheric characterization \citep{2017Natur.542..456G}. 
NASA's {\it Spitzer} telescope determined that the Earth-like planets of the TRAPPIST-1 system lie in the HZ. Unfortunately, the {\it Spitzer} mission terminated in January 2020, removing a key instrument from the exoplanet community's toolbox.
The {\it Transiting Exoplanet Survey Satellite} \citep[{\it TESS};][]{2014SPIE.9143E..20R} is an all-sky survey that discovered similar planet around an early M dwarf, TOI-700d and e \citep{2020AJ....160..116G, 2023ApJ...944L..35G}. With its 10.5~cm diameter camera apertures, TESS is mainly suited for targeting bright early- to mid-M dwarfs. In contrast, ground-based surveys such as MEarth \citep{2009IAUS..253...37I}, TRAPPIST, and SPECULOOS \citep{2018SPIE10700E..1ID}, take advantage of their larger telescope aperture to find terrestrial planets around ultracool dwarfs with even later spectral types. 

In summer and winter\footnote{\jasmine{} can observe the galactic center field only in spring and fall, because the Sun is aligned in the same direction as the galactic center in Northern Hemisphere winter. In Northern Hemisphere summer, it is prohibitively difficult for the satellite to meet the required thermal conditions for precision astrometry.}, the NIR and high-cadence time-series photometry capability of \jasmine{} will be used to conduct the exoplanet transit survey, which aims to fill a gap in {\it TESS} and ground-based surveys. \jasmine{} has a NIR photometry capability similar to that of {\it Spitzer} and offers long-term  
(a few weeks) follow-up transit observations of exoplanet systems discovered by {\it TESS} and ground-based surveys. This allows for the detection of outer orbiting planets that could be in the HZ. The scientific goals and strategies of the exoplanet transit survey are described in section~\ref{sec: exoplanets}, including other potential ways of finding exoplanets using microlensing and astrometry in the GCS field.

\subsection{The \jasmine{} mission}

\begin{figure}
\includegraphics[width=0.95\hsize]{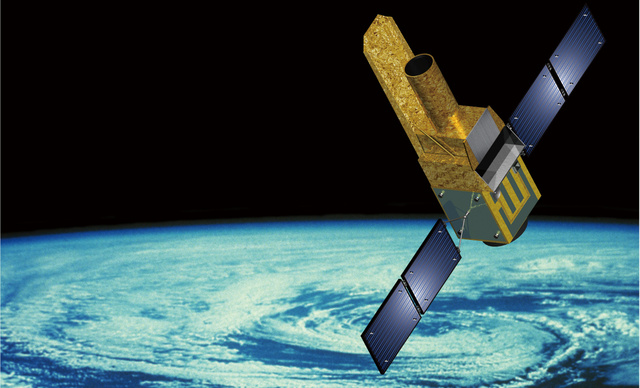}
\caption{Artist's impression of \jasmine{}.}
    \label{fig:jasmine_pic}
\end{figure}

To achieve these main science objectives of the Galactic Center Archaeology survey and the Exoplanet survey, 
\jasmine{} will have a scientific payload with a 36 cm aperture telescope with a single focal plane. \jasmine{} will provide photometric observations within a 1.0--1.6 \um{} passband over a field of view (FoV) of $0.55^\circ \times 0.55^\circ$. \jasmine{} will orbit the Earth in a temperature-stable Sun-synchronous orbit. Figure \ref{fig:jasmine_pic} shows an artist's impression of the spacecraft. The telescope structure is made of Super-super invar alloy \citep{Ona+Sakaguchi+Ohno+Utsunomiya20}. A large Sun shield and telescope hood are installed to maintain thermal stability. The satellite is expected to be launched aboard a JAXA Epsilon S Launch Vehicle from the Uchinoura Space Center in Japan in 2028. More detailed instrumental specifications and survey strategies of the mission are provided in section \ref{sec:mission}.

The unique NIR astrometric and time series photometric capabilities of \jasmine{} would be applicable for the other science targets and the unique data from the \jasmine{} GCS would be valuable for a wide range of scientific topics. In section~\ref{sec:other-science-cases}, we summarize some examples of science cases to utilise the data from \jasmine{}.
Section~\ref{sec:other_proj} summarizes the potential synergies with the other relevant projects operating when \jasmine{} launches. Finally, section~\ref{sec:sum} provides a brief summary of the paper. 

\section{Galactic Center Survey (GCS): Galactic Center Archaeology}
\label{sec:gca}

At the heart of the Galaxy, the Galactic center, lies a SMBH with a mass of
$(4.297\pm0.013)\times10^6$~$M_{\odot}$ \citep{Gravity+21AA647p59},
the origin of which remains an area of intense study in modern astrophysics. What is known is that the masses, $M_\mathrm{BH}$, of SMBHs correlate with the stellar mass, concentration, and velocity dispersion, $\sigma_\star$, of their host bulge \citep{Magorrian+98,Ferrarese+Merritt00,Gebhardt+Bender+Bower+00,2001ApJ...563L..11G}, and there is an emerging dependence on the host galaxy morphology \citep{2023MNRAS.518.2177G, 2023MNRAS.520.1975G}.  
The initial relationships were unexpected because the gravitational sphere of influence of the SMBH has a radius on the order of $\sim1$~parsec, thousands of times smaller than the scale of the Galactic bulge \citep[for reviews, see][and references therein]{2013ARA&A..51..511K, 2016ASSL..418..263G}. 

From the turn of the millennium when the initial correlations were discovered, it has arguably been the singular unifying focus of research on SMBHs and galaxies. It is slowly coming to be understood as resulting from an interplay between periods of baryonic mass accretion onto the SMBH and neighboring stellar populations, and the ``feedback'' from the resultant active galactic nucleus (AGN) activity \citep{2015ARA&A..53..115K}, a process that may have started at the earliest epochs \citep[e.g.,][]{2015ApJ...805...96S}. Indeed, the existence of SMBHs at redshifts exceeding 7.5 \citep{2018Natur.553..473B}, when the universe was barely 500~million years old, implies that much cosmic history may be imprinted upon the environment of the Galactic center, motivating Galactic archaeology \citep{2002ARA&A..40..487F} at the Galactic center.
While it is now apparent that major `dry' mergers have established the bulge-black hole (BH) scaling relations in elliptical galaxies \citep{2023MNRAS.518.2177G, 2023MNRAS.518.6293G}, with AGN feedback relegated to a maintenance role \citep{2003ApJ...599...38B}, the new morphology-aware scaling relations have recently revealed how accretion and minor mergers have likely built spiral galaxies from lenticular galaxies \citep{Graham-S0}.  Indeed, it is not just the Milky Way that shows evidence of disrupted satellites \citep[e.g.,][]{2016A&A...588A..89J,2021ApJ...907...85M}. Moreover, some of these satellites may be delivering massive BHs into the spiral galaxies, possibly evidenced by the X-ray point source in the disrupted Nikhuli system captured by NGC~4424 \citep{2021ApJ...923..146G}. 

Galactic center archaeology has been a distinctly multi-wavelength field. The highest energy observations have appropriately revealed the most conspicuous evidence of past AGN activity from the SMBH of the Milky Way, Sgittarius\,A\textsuperscript{$\ast$} (\SgrA\,), in the form of the Fermi bubble. The Fermi bubble is seen in $\mathrm{\gamma}$-rays, which extend to $\sim10$~kpc and are considered to be due to AGN activity several Myrs ago \citep{2012ApJ...756..181G,2015ApJ...799L...7F}, although the origin of the bubble is still in debate and a star formation driven bubble \citep[e.g.,][]{Crocker+Aharonian11,Sarkar+Nath+Sharma15,Nogueras-Lara+2020} is also a possible origin. The X-rays likewise implicate past AGN activity, such as the presence of Fe~K$\mathrm{\alpha}$ ($h\nu=6.4$~keV) fluorescence at the interfaces of cold molecular regions in the vicinity of \SgrA\ \citep[e.g.,][]{2010ApJ...714..732P,2013ASSP...34..331P,2013A&A...558A..32C}, or the presence of a hot-gas cavitation, symmetric about \SgrA, extending out to several kpc \citep{2016ApJ...828L..12N,Predehl+Sunyaev+Becker+20}. These ``light echos'' are also seen at lower energies \citep{1984Natur.310..568S, 1985PASJ...37..359T, 2003ApJ...582..246B}, with H$\mathrm{\alpha}$ emission seen in the Magellanic Stream requiring an illumination by UV ionizing photons generally only possible from AGNs \citep{2013ApJ...778...58B,Bland-Hawthorn+Maloney+Sutherland+19}. 

From visual to NIR wavelengths, Galactic center archaeology studies have been focused predominantly on the history and makeup of the Galaxy's stellar populations. Several key parameters, such as the stellar chemical abundances, e.g., [Fe/H] and [$\mathrm{\alpha}$/Fe], surface gravities, effective temperatures, and kinematics, are of interest to these studies. Although there are many optical photometric and spectroscopic surveys of stars in the central region of the Galaxy, because of the extremely high extinction at optical wavelengths, the optical surveys tend to avoid the Galactic center and focus instead on the global properties of the bulge/bar. To study the Galactic center, NIR observations are crucial. For example, photometric data from the {\it IRSF}/Simultaneous InfraRed Imager for Unbiased Surveys \citep[SIRIUS,][]{Nagayama+Nagashima+Nakajima+03,Nishiyama+Nagata+Kusakabe+06}, the Two Micron All Sky Survey \citep[2MASS,][]{Skrutskie+06}, {\it VISTA} Variables in the Via Lactea \citep[VVV][]{Minniti+Lucas+Emerson+10}, and GALACTICNUCLEUS \citep{GALACTICNUCLEUS+Nogueras-Lara+18} 
revealed the detailed stellar structure of the Galactic center, including the Nuclear Star Cluster (NSC, section~\ref{subsec: nsc}) and Nuclear Stellar Disk (NSD: section~\ref{subsec:nsd}). Meanwhile, the APOGEE spectroscopic survey, with its NIR (1.51--1.70~\micron) high spectral resolution ($R\sim22500$), has been critical for understanding the chemical makeup of stellar populations in the Galaxy, as well as providing line-of-sight velocities. Results from APOGEE include kinematical and structural differentiation between metal poor and metal rich stars in the inner Galaxy \citep{2016ApJ...832..132Z,2018ApJ...852...91G,Wylie+Gerhard+Ness+21}, detailed chemical abundances \citep{2019ApJ...870..138Z}, signatures of the central bar
\citep{2012ApJ...755L..25N,2021A&A...656A.156Q},
and measuring the rotation of both the bulge \citep{2016ApJ...819....2N}
and the NSD \citep{2015ApJ...812L..21S}, which may extend to kpc-scales in size \citep{Debattista+2015,2018MNRAS.473.5275D}. Recently, using the $K$-band Multi-Object Spectrograph (KMOS) on the {\it Very Large Telescope} ({\it VLT}), \citet{Fritz+Patrick+Feldmeier-Krause+21} surveyed the stars in the NSD, and analysed their line-of-sight velocities and metallicities. \citet{Schultheis+Fritz+Nandakumar+21} found that the NSD is kinematically cold, and that more metal rich stars are kinematically colder.  
The presence of the NSD in other galaxies has been used to infer the longevity of galaxy bars and their role in transporting gas to the innermost regions \citep[][see section~\ref{subsec:nsd-bar}]{2015A&A...584A..90G,Gadotti+2019}. 

However, while there has been an explosion of archaeological studies of the Galactic center as a result of these and other new facilities and methods to exploit line-of-sight velocities and chemical abundances to access relic stellar signatures that tell the history of the Galaxy formation, high-precision proper motions and parallaxes --- critical to providing the remaining three dimensions of phase space needed for mapping the gravitational potential of the Galactic center --- have thus far been unavailable. This is because the Galactic center can exhibit upwards of 30 to 60 magnitudes of extinction at visual wavelengths \citep[e.g.,][]{Fritz+11}, precluding the use of {\it Gaia} data. \jasmine{} will close this gap by mapping the positions, motions, and parallaxes of stars in the NIR ${H_\mathrm{w}}$-band, where the extinction is much lower, i.e., of order $A_{H_\mathrm{w}}\sim5$--6~mag \citep[e.g.,][]{Fritz+11,Gonzalez+Rejkuba+Zoccali+13,Nogueras-Lara+Schoedel+Neumayer21}.


\begin{figure}
 \includegraphics[width=\hsize]{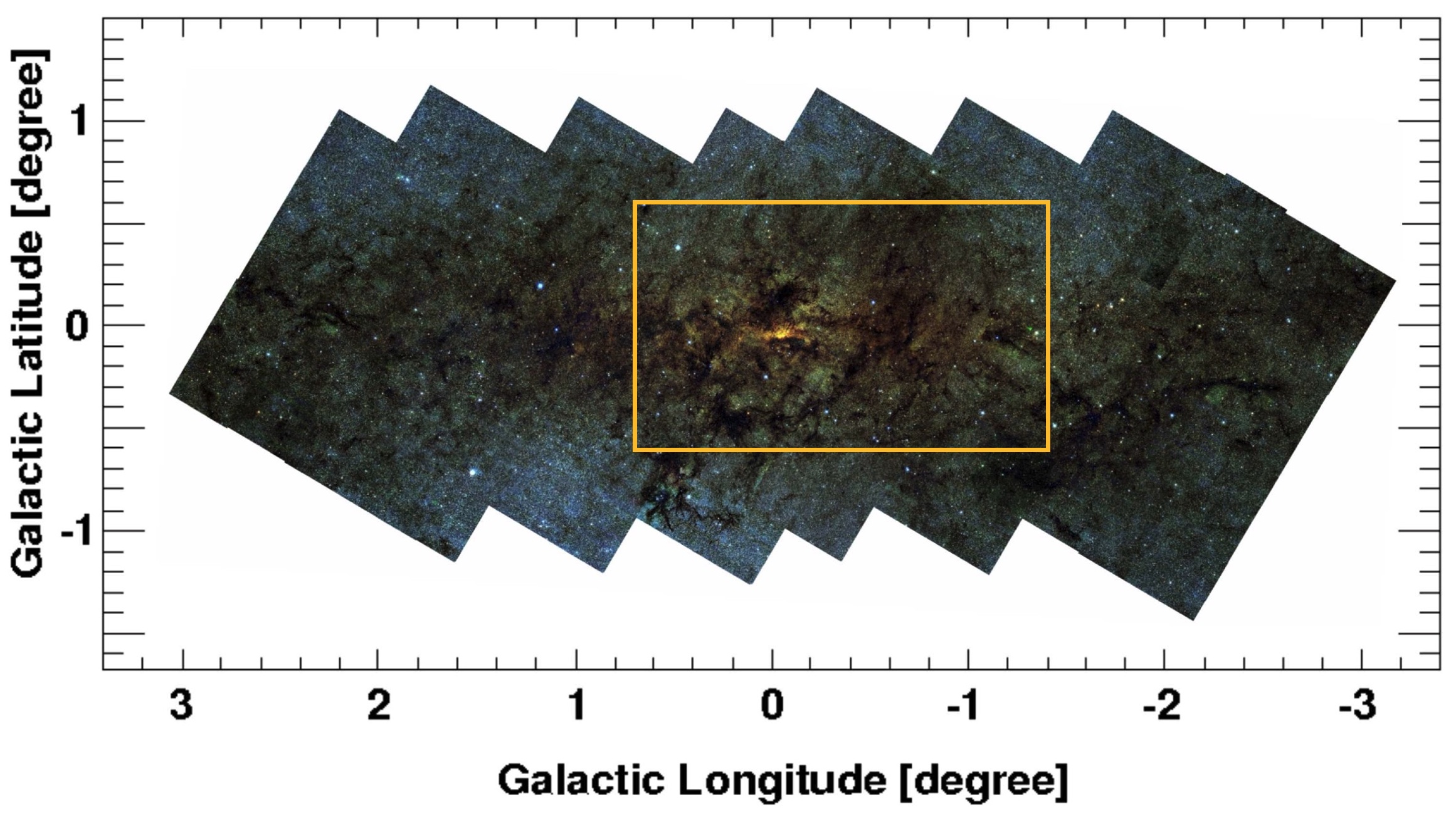}
    \caption{The planned survey area of the \jasmine{} GCS program (highlighted with a yellow open square) overlaid on an the image of the SIRIUS survey \citep{Nishiyama+Nagata+Kusakabe+06}. 
    \label{fig:jgcs_sirius}}
\end{figure}

The GCS of \jasmine{} will provide the parallax and proper motion of stars in a Galactic center field defined by a rectangular region of
$-1.4^{\circ}<l<0.7^{\circ}$ and $-0.6^{\circ}<b<0.6^{\circ}$. Note that $0.7^{\circ}$ corresponds to about 100~pc at the distance of the Galactic center. This survey region is still subject to change, and may become $-0.7^{\circ}<l<1.4^{\circ}$ and $-0.6^{\circ}<b<0.6^{\circ}$, i.e., shifting toward the positive longitude without changing the size of the survey area, depending on the scientific values and operation and/or data analysis requirements. In this paper, unless otherwise stated, we consider the region of  $-1.4^{\circ}<l<0.7^{\circ}$ and $-0.6^{\circ}<b<0.6^{\circ}$ as highlited in figure~\ref{fig:jgcs_sirius}. 
Within this main survey area, \jasmine{} is expected to produce accurate positions and proper motions for about 120000 stars up to $H_\mathrm{w}<14.5$~mag, including about 68000 Galactic center stars.\footnote{We estimate the number of stars in the Galactic center by considering that all the stars with $J-H>2$~mag are in the Galactic center, although a redder cut is likely required at lower Galactic latitudes. These numbers are obtained from the combined data of the 2MASS point source catalog and the SIRIUS catalog.}

The proper motion accuracy achieved by \jasmine{} for the bright stars with $H_\mathrm{w}<12.5$~mag is expected to be about 25~\uas{}~yr$^{-1}$, and the proper motion accuracy for the faintest stars with $H_\mathrm{w}=14.5$~mag is expected to be about 125~\uas{}~yr$^{-1}$. These accuracies respectively correspond to transverse velocity accuracies of  $\sim0.98$~km~s$^{-1}$ and $\sim4.9$~km~s$^{-1}$ at the distance of the Galactic center. Recently, careful analysis of long-timespan, NIR, ground-based imaging data has yielded sub-mas~yr$^{-1}$ proper motion accuracy. For example, the VVV InfraRed Astrometry Catalogue \citep[VIRAC,][]{Smith+Lucas+Kurtev+18} reaches a mode in the proper motion uncertainty distribution of  0.3~mas~yr$^{-1}$ in version 2 of VIRAC \citep{Sormani+Sanders+Fritz+22}. However, the systematic uncertainties of the ground-based data are difficult to quantify due to atmospheric effects and the distortion of telescope optics and deformation of detector, which are not well controlled. \jasmine{} will provide independent and improved measurements of proper motion from space, as well as a critical NIR reference frame for 
the Galactic center region, which will be a valuable asset in improving the proper motion accuracy  these legacy ground-based data. In addition, for the bright stars with $H_\mathrm{w}<12.5$~mag \jasmine{} will provide precise parallax measurements with an accuracy of $\sim25$~\uas{}. This corresponds to $\sim$20\% accuracy of the parallax distance for the stars in the Galactic center, which enables us to identify the Galactic center stars less ambiguously. These unique distance confirmations using trigonometric parallaxes will provide a critical verification of the color selection methods which are currently most predominantly used to create samples of Galactic center stars.

This section summarizes the science objectives of the GCS. The section is organized by increasing physical scale, from parsec-scale dynamics of the nuclear star cluster in section~\ref{subsec: nsc} to kiloparsec-scale dynamics of the Galactic bar, bulge and inner disk in section~\ref{subsec:inner_disk}. Note that at a Galactic center distance of 8.275~kpc \citep{Gravity+21AA647p59},
the angular scale is 0.04~pc~arcsec$^{-1}$.


\subsection{Kinematics and History of the Nuclear Star Cluster} 
\label{subsec: nsc}

On scales of only a few parsecs, the stellar structure closest to and, therefore, most affected by \SgrA\ is the nuclear star cluster (NSC), which has a stellar mass of about $\sim10^7$~$M_\odot$ \citep{2002A&A...384..112L,2014A&A...570A...2F,Chatzopoulos+15,Feldmeier-Krause+Zhu+Neumayer+17}.
The first separation of the NSC from the other components of the Galaxy (NSD/bulge/bar/disc) yielded a NSC light profile that appeared well matched by a S\'ersic model with an index of $\sim$3 and an effective half-light radius of 3.2~pc \citep{2009MNRAS.397.2148G}.  A recent estimate reports an index of 2.2$\pm$0.7 and $R_{\rm e}=5.1\pm1.0$~pc
\citep[][]{Gallego-Cano+Schoedel+Nogueras-Lara-20}. However, perhaps the core-S\'ersic model \citep{2003AJ....125.2951G}, fit separately to the NSC's different stellar populations, may prove more apt at quantifying it and, in turn, yielding insight into its origins and evolution. This could be relevant in the case of binary collisions preferentially removing red giant stars \citep{1991ApJ...381..449D}.

When present, NSCs scale with the mass of their host bulge, possibly in a manner dependent on the host galaxy morphology \citep{2003ApJ...582L..79B,2003AJ....125.2936G}. Furthermore, a positive correlation between NSC mass and the central BH mass has also been observed \citep{2016IAUS..312..269G, 2020MNRAS.492.3263G}.  
NSCs in late-type galaxies show extended periods of star formation \citep[for a review, see][]{Neumayer+Seth+Boeker20}, resulting in a younger stellar population \citep{2006ApJ...649..692W}. The NSC of the Milky Way also shows extended star formation \citep{Neumayer+Seth+Boeker20}.  Most stars formed more than five Gyr ago, followed by a low level of star formation, until the burst of star formation in the last 100 Myr \citep[e.g.,][]{Blum+Ramirez+Sellgren+03,Pfuhl+Fritz+Zilka+11,Nogueras-Lara+2020}.  Wolf-Rayet and O- and B-type stars (a few Myr age) are found in the central 0.5~pc of NSCs \citep{2006ApJ...643.1011P,2009ApJ...697.1741B,Neumayer+Seth+Boeker20}, which indicates the in-situ formation of the NSC of the Milky Way, at least for the youngest population \citep{2015ApJ...799..185A}.

On the other hand, using NIR drift-scan spectroscopic observations, \citet{2014A&A...570A...2F} found a kinematic misalignment between the NSC with respect to the Galactic plane and a rotating substructure that suggests past accretion of at least one other star cluster.   Dynamical friction may lead to the capture and central deposition of globular clusters \citep{Tremaine+Ostriker+Spitzer75}.  
This scenario has some theoretical support, as $N$-body simulations have shown that an NSC like the one in the Milky Way can be produced by the infall of a few globular clusters that are shredded by the SMBH \citep{2012ApJ...750..111A,2014ApJ...784L..44P}. Interestingly, high resolution spectroscopy of the NSC \citep{Rich+Ryde+Thorsbro+17} finds a wide abundance range spanning [M/H]$=-1.25$ to supersolar metallicities, [M/H]$>0.3$.  This is confirmed in larger low-resolution surveys, e.g., using the integral field spectrograph KMOS on {\it VLT} by \citet{Feldmeier-Krause+Kerzendorf+Nogueras-Lara+20}. They further find that the subsolar population with [M/H]$<0$ shows an asymmetric spatial distribution, which might be the signature of a recent globular cluster merger. 

Hence, the nature and origin of the NSC of the Milky Way are still in debate, and could link to the in-situ star formation at the Galactic center and/or mergers of globular clusters or ancient mergers with other galaxies. These are distinct dynamical processes, where spectroscopic and numerical studies have made great headway so far. What is missing is the other two dimensions of kinematical information in the proper motion that an astrometric mission, such as \jasmine{}, can provide. We expect \jasmine{} to observe about 100 NSC stars with $H_\mathrm{w}<14.5$~mag. 
Although this is a small number of stars, the novel $\sim$100 $\mu$as~yr$^{-1}$ level of proper motion for the NSC stars, combined with high-precision radial velocities, will likely yield accelerations and orbits.
\jasmine{} astrometry data will also be complementary to astrometry of the fainter infrared sources \citep[e.g.,][]{Schoedel+Mettitt+Eckart09} and the millimeter-submillimeter sources in the inner region of the NSC, i.e., close to \SgrA. Recently, \citet{Tsuboi+Tsutsumi+Miyazaki+22} demonstrated that the proper motions of millimeter-submillimeter bright young stars within 0.5 pc of \SgrA can be obtained with the \textit{Atacama Large
Millimeter/submillimeter Array (ALMA)}, measuring the relative astrometry with respect to 
\SgrA. Combined with this information, \jasmine{} will enhance our understanding of the nature and origin of the NSC. 


\subsection{SMBH Formation History}
\label{sec:smbh}

As alluded to in the beginning of this section, remnants of the earliest epochs can be imprinted on the centers of galaxies, and indeed the formation of the SMBHs themselves may leave a mark on their environment. The presence of $10^9$~$M_\odot$ SMBHs that powered quasars when the universe was only a few hundred Myr old \citep[e.g.,][]{2018Natur.553..473B,Yang+Wang+Fan+20,Wang+Yang+Fan+21} is a major conundrum for the mechanism of formation of SMBHs \citep[e.g.,][]{Rees+84ARA&A,2010A&ARv..18..279V,Woods+Agarwal+Bromm+19PASA}. This is especially challenging for the ``light" seed scenario, where the stellar mass BHs ( $\lesssim100$~$M_{\odot}$) formed from massive first generation stars (Pop~III stars) in mini-halos. Even if accreting at the maximum stable rate set by the balance of gravity and radiation pressure (the Eddington limit), these light BHs did not have sufficient time to grow into the monster SMBHs powering the earliest quasars. 
On the other hand, super-Eddington accretion from disk-fed BHs rather than an idealised spherical model may circumvent this concern \citep{2014Sci...345.1330A}. 

Nonetheless, this had made an alternative ``heavy" seed scenario more attractive \citep[e.g.,][]{Woods+Agarwal+Bromm+19PASA}. If a metal free halo is massive enough to host atomic cooling, i.e., $T_{\rm vir}\sim10^4$~K, and is exposed to strong Lyman-Werner radiation, H$_2$ cooling is completely suppressed, and the high temperature prevents fragmentation \citep{Omukai01ApJ546p635O}. This leads to the formation of a super-massive star, a.k.a. quasi-star \citep[e.g.,][]{Bromm+Loeb03,Inayoshi+Omukai+Tasker14}, which directly collapses into a BH with a mass of $\sim10^{4}$--$10^{6}$~$M_{\odot}$ \citep[e.g.,][]{Ferrara+Salvadori+Yue+14,Umeda+16,Woods+Heger+Haemmerle20}. These massive BH seeds also form in relatively massive halos with high-density gas, which enables further gas accretion onto the BH. Hence, the heavy seeds can grow rapidly enough to explain the massive SMBHs found at $z\sim7$. 

Although the heavy seed scenario is attractive for explaining SMBH at high redshift, they are considered to be rare, with a comoving number density of about $10^{-6}$--$10^{-4}$~Mpc$^{-3}$ \citep[e.g.,][]{Habouzit+Volonteri+Latif+16}. This is a smaller number density than that for the Milky Way-sized galaxies. It is an interesting question whether or not the SMBH of the Milky Way formed from a single massive direct collapse to a BH, mainly via accretion. Alternatively, the SMBH of the Milky Way could be formed from more abundant Pop~III origin BHs, via accretion and mergers, i.e., the light seed scenario. The light seeds from Pop~III stars can be as massive as 1000~$M_{\odot}$ \citep{Hirano+Hosokawa+Yoshida+14}, and using cosmological numerical simulations \citet{Taylor+Kobayashi14} demonstrated that such light seeds can reproduce the observed BH mass$-$stellar velocity dispersion relation. In this scenario, BHs are expected to be found ubiquitously in galaxies of all masses, with intermediate-mass BHs \citep[IMBHs,][]{2018ApJ...863....1C, Greene+Strader+Ho19} of $M_{\rm BH}<10^5$--$10^6$~$M_{\odot}$ perhaps common in low-mass galaxies \citep[e.g.,][]{2021ApJ...923..246G}. 

Building off the promise of earlier discoveries of AGNs in local dwarf galaxies \cite[e.g.,][]{2003ApJ...588L..13F}, this predictive framework has fueled an intense search for IMBHs in dwarf galaxies, with some promising results \citep[e.g.,][]{2009Natur.460...73F,2009ApJ...704..439S,  Greene+Ho04,2013ApJ...775..116R,Secrest+12,Graham+Scott15,2015ApJ...798...38S,2015ApJ...809L..14B,
2016ApJ...818..172G,2018ApJ...869...49J,Baldassare+Dickey+Geha+Reines20,2021PASA...38...30D}, including dynamically estimated BH mass of NGC~205 having $M_{\rm BH}=6.8^{+95.6}_{-6.7}\times10^3$~$M_{\odot}$ \citep[3~$\sigma$,][]{Nguyen+Seth+Neumayer+19}. The mergers of BHs with $\sim10^4$--$10^7$~$M_{\odot}$ at the galactic center are a prime gravitational wave target of ESA's {\it Laser Interferometer Space Array} \citep[{\it LISA},][]{LISA+Amaro-Seoane+17}. These multi-messenger astronomy sources will provide strong constraints on the population and merger rates of IMBHs, ultimately revealing the formation mechanism of SMBHs \citep[e.g.,][]{Volonteri+Pfister+Beckman+20} and adding detail to the suspected accretion-driven origin of spiral galaxies \citep{Graham-S0}.

\jasmine{} can provide an independent test of the significance of the IMBH mergers for the formation of the SMBH of the Milky Way, because if the SMBH was built up by the mergers of IMBHs, these coalescences would heat up the older stars within 100~pc of the Galactic center, with the stellar profile becoming cored \citep{2014MNRAS.440..652T}. To demonstrate this, we set up an N-body simulation of a spherical bulge model, following \citet{2014MNRAS.440..652T}. Although, as discussed in the previous and the following sections, the Galactic center has a complex structure --- and there are an NSC (section~\ref{subsec: nsc}) and NSD (section~\ref{subsec:nsd}) --- here, for simplicity we consider a spherical isotropic bulge component only. Hence, this merely serves to show the potential capability of the \jasmine{} data with a simple model. More complex modelling studies are required to address how the different populations of the Galactic center stars are affected by the mergers of IMBHs, and how gas may spare the ejection of stars.

\begin{figure}
 \begin{center}
   \includegraphics[width=1.0\linewidth]{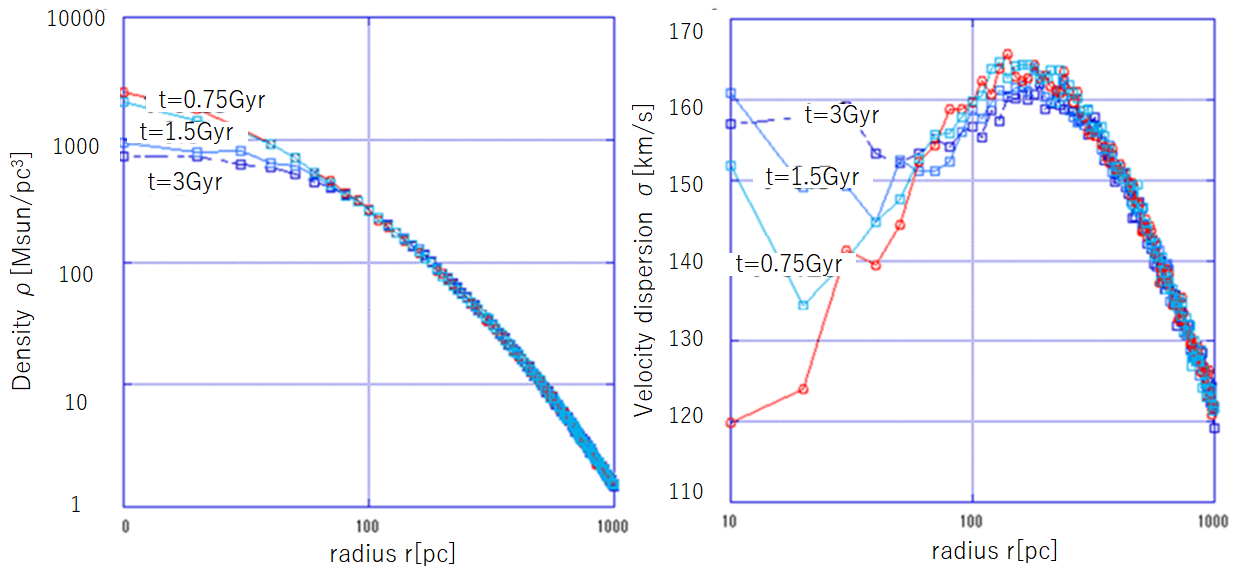}
 \end{center}
 \caption{Density (left) and velocity dispersion distribution (right) profiles of the stellar system. Red line shows the original density and velocity dispersion profile. The blue lines indicate the profiles at different times, as shown in the time labels, since the simulation of the mergers of the five BHs with $8\times10^5$~$M_{\odot}$ started.
 \label{fig:smbh-01}}
\end{figure}

We have first assumed a spherical stellar component following a Hernquist profile \citep{Hernquist90ApJ356p359} with the mass of $9\times10^9$~$M_{\odot}$ and half-mass radius of 1.1~kpc.\footnote{Our assumed bulge mass is slightly larger than the current upper limit of the bulge mass in the Milky Way from kinematics of the stars in the Galactic center region, e.g., about 10\% of the disk mass, i.e., $\sim5\times10^9$~$M_{\odot}$, suggested by \citet{2010ApJ...720L..72S}. However, for this simple numerical experiment, we consider a higher mass for increased stability, and as it would provide more conservative results for the dynamical response.} We use 524288~particles to describe this stellar system. We then add five $8\times10^5$~$M_{\odot}$ BHs, which follow the same distribution function as the stellar system. Because of dynamical friction, these BHs fall into the center of the system, and merge into a SMBH in the time scale of about one Gyr  \citep{2014MNRAS.440..652T}. The back reaction of dynamical friction heats up the stellar system in the central 100~pc. Figure~\ref{fig:smbh-01} shows the density profile (left) and velocity dispersion profile (right) evolution when the BHs merge into a central SMBH. The red lines of these figures show the initial condition of the Hernquist profile with the central density profile slope of $r^{-1}$. Blue lines show how the density and velocity dispersion profiles change as the BHs merge into the center of the system, and dynamical friction heats up the stellar system. After 3~Gyr, the density profile becomes significantly shallower with the slope of less than $r^{-0.5}$ within $r=100$~pc, and the velocity dispersion could be different from the initial velocity dispersion by as large as 30~km~s$^{-1}$ at $r\sim10$~pc. We ran the simulations with different masses of the seed BHs of $4\times10^5$~$M_{\odot}$ and $2\times10^5$~$M_{\odot}$, and found
that as long as the total mass is $4\times10^6$~$M_{\odot}$ like the SMBH of the Milky Way, the cored density distribution with a similar size and slope can be caused by the BH mergers, independent of the initial seed mass. A difference, depending on the mass of the seed BHs, is that lighter seed BHs take a longer time to merge into a single central SMBH. 

\jasmine{} is expected to observe about 6000 (68000) stars  with $H_\mathrm{w}<12.5$ (14.5)~mag and $J-H>2$~mag in the Galactic center, and measure their parallax and proper motion with accuracies of 25 (125)~\uas{} and 25 (125)~\uas{}~yr$^{-1}$ (about 1 (5)~km~s$^{-1}$ at $d=8.275$~kpc), respectively. The superb parallax accuracy of \jasmine{} for bright ($H_\mathrm{w}<12.5$~mag) stars will minimise the contamination of the foreground Galactic disk stars. Because the intrinsic NIR color of giants are  almost constant irrespective of their intrinsic luminosity, the color selection criterion for the Galactic center stars obtained from the parallax-color relation of the bright stars at the different positions of the sky can be applied to select the Galactic center stars for the fainter stars \citep[e.g.,][]{Nogueras-Lara+Schoedel+Neumayer21}.
Then, we can use the larger number of faint stars with similar colors, which enables us to observe the velocity dispersion profile within~100~pc to tell the difference in velocity dispersion by about 10~km~s$^{-1}$ as shown in figure~\ref{fig:smbh-01}.  Hence, \jasmine{} can detect the relic of possible BH mergers, if the SMBH of the Milky Way was built up via mergers of IMBHs. If such a heated velocity dispersion profile is not observed, it could be a sign that the SMBH started from a relatively massive BH and grew mainly via accretion, although there could be some alternative scenarios, such as an adiabatic contraction of the bulge stars due to the formation of an NSC and/or NSD.


\subsection{The Nuclear Stellar Disk: Structure} \label{subsec:nsd}

Extending out to $\sim200$~pc, the Galactic center hosts a NSD coincident with and of the same scale as the Central Molecular Zone \citep[CMZ,][]{2002A&A...384..112L}. The NSD has an exponential profile \citep{2013ApJ...769L..28N}, in accord with other galaxies \citep{2007ApJ...665.1084B, 2010MNRAS.407..969L, 2010A&A...518A..32M}. The radial extent of the disk is around 230 pc \citep{2016ARA&A..54..529B}, and the vertical scale-height is measured to be around 45~pc \citep{2013ApJ...769L..28N, 2016ARA&A..54..529B}. The total mass of the nuclear stellar disk is estimated to be around $1.4\times10^9$~$M_{\odot}$ \citep{2016ARA&A..54..529B}. The presence of classical Cepheids \citep{2011Natur.477..188M,2015ApJ...799...46M} reveals a thin disk of young stars continuously formed over the past $\sim100$~Myr \citep{2015ApJ...812L..29D}. While line-of-sight velocity studies have been done to understand the kinematics of the NSD \citep[e.g.,][]{2015ApJ...812L..21S}, a full 6D phase-space characterization awaits a high-precision astrometric mission for the Galactic center stars, which \jasmine{} will provide, as demonstrated with the current existing dataset in \citet{Sormani+Sanders+Fritz+22}.

\subsubsection{Non-axisymmetry}

An interesting question is whether there is a connection between the NSD and the nuclear bar as seen in simulations \citep[e.g.,][]{2002MNRAS.330...35A, Athanassoula2005, Cole+2014} and inferred in
NIR photometric campaigns \citep{2001A&A...379L..44A,2005ApJ...621L.105N,2008A&A...489..115R,2011A&A...534L..14G}. Nuclear bars are frequently found in earlier galaxy types \citep{1989Natur.338...45S,1993A&A...277...27F,2002ApJ...567...97L,2004A&A...415..941E}, and appear to be a distinct kinematical structure from the main outer bar. Using the VVV survey data, \citet{2011A&A...534L..14G} suggested that the Milky Way has a nuclear bar with a size of about 500~pc that is misaligned with the long bar. Such a nuclear bar, if it exists, can drive the inflow of the gas from the size of the CMZ, from about $\sim200$ to 10 pc, without destroying the NSD or CMZ's disk (or ring) structure \citep{Namekara+09}. However, \citet{Gerhard+Martinez-Valpuesta12} suggested that the nuclear bar observed in the star counts can be explained by the projection effect of the inner boxy bar \citep[see also][]{Fujii+2019}, thus casting doubt on the nuclear bar's existence. 

However, the NSD may still have a non-axisymmetric structure, and the strength of non-axisymmetry could determine the efficiency of the gas inflow. The precise motion of the stars in the NSD provided by \jasmine{} can assess non-axisymmetry of the NSD, because the kinematics of the NSD stars should be affected by non-axisymmetry, and will provide a stronger constraint than the star count observations. 

\subsubsection{Young star clusters}

There are two young star clusters, the Arches \citep{Nagata+95} and Quintuplet \citep{Kobayashi+83,Nagata+90,Okuda+90}, within about 30~pc from the Galactic center. The stellar mass of Arches is estimated to have an upper limit of $7\times10^4$~$M_{\odot}$ \citep{Figer+02} and the age is about 3.5~Myr \citep{Schneider+Izzard+deMink+14}. The line-of-sight velocity and the proper motion of the Arches cluster are estimated to be $v_{\rm LOS}=95\pm8$~km~s$^{-1}$ \citep{Figer+02} and 
$v_{\rm pm}=172\pm15$~km~s$^{-1}$ \citep{Clarkson+12}, respectively. 
The stellar mass and age of the Quintuplet are 10$^4$~$M_{\odot}$ \citep{Figer+McLean+Morris99} and 4.8~Myr \citep{Nogueras-Lara+2020}, respectively. The line of sight velocity and proper motion of the Quintuplet cluster are estimated to be $v_{\rm LOS}=102\pm2$~km~s$^{-1}$ \citep{Stole+14}
and $v_{\rm pm}=132\pm15$~km~s$^{-1}$ \citep{Stole+14}.

Based on high-resolution numerical simulations of a barred Milky Way-like galaxy that resolve the star formation in the CMZ, \citet{Sormani+Tress+Glober+20} argued that the current position and estimated velocity for the Arches and Quintuplet clusters indicate that both clusters formed at the location where the gas inflow from the bar collides with the CMZ. This is consistent with what is also discussed in \citet{Stole+14}, who suggested that the young star clusters formed at the transition place between the $x_1$-orbits (the gas in the leading side of the bar) and $x_2$-orbits (CMZ).

Recently, \citet{Hosek+Do+Lu+22} measured the absolute proper motion of these clusters using multi-epoch {\it Hubble Space Telescope} data and the {\it Gaia}~EDR3 reference frame, and obtained $(\mu_\mathrm{l}\cos\mathrm{b}, \mu_\mathrm{b}) = (-2.03\pm0.025, -0.30\pm0.029)$~mas~yr$^{-1}$ for Arches and $(-2.45\pm0.026, -0.37\pm0.029)$~mas~yr$^{-1}$ for Quintuplet, where $(\mu_\mathrm{l}\cos\mathrm{b}, \mu_\mathrm{b})$ are the proper motion in the Galactic longitude and latitude directions. These results indicate that these clusters are moving parallel to the Galactic mid-plane. 

\jasmine{} is expected to observe about 10 stars in Arches and about 20 stars in Quintuplet with 0.125~mas~yr$^{-1}$ proper motion accuracy. The accuracy of the proper motion measurements of these clusters are comparable to previous studies, but \jasmine{} will have a longer time baseline and an improved absolute reference frame. The combination of the \jasmine{} data and that of existing literature will further improve the proper motion measurements. 
The enhanced proper motions will test the above mentioned scenario and pinpoint the formation location of these clusters. 

\jasmine{} is also capable of detecting the presence of new young star clusters (see also section~\ref{subsec:bhhunt}). Revealing the formation location of the hidden young clusters in addition to the Arches and Quintuplet clusters will help to understand the star formation mechanism in the CMZ and their relation to the NSD, by comparing with observations of molecular gas in the Galactic center region \citep[e.g.,][see also the next section]{Henshaw+Longmore+Kruijssen+16,Sormani+Tress+Glober+20}. The recent ability to make precise measurements of the proper motion of star forming regions by Very Long Baseline Interferometry (VLBI) observations, such as the VLBI Exploration of Radio Astrometry (VERA), would provide further information on the star formation in the Galactic center \citep[e.g.,][]{Sakai+Oyama+Nagayama+23}. \jasmine{} data will be complementary to the VLBI astrometry data to further study the formation and evolution mechanism of the NSD.


\subsection{Nuclear Stellar Disk: The formation epoch of the Galactic Bar} \label{subsec:nsd-bar}

The central few kpc region of the Milky Way shows a prominent bar structure \citep[e.g.,][]{Binney+97,2016ARA&A..54..529B}. The shape of the inner region of the bar/bulge has been found to display a so-called boxy/peanut morphology \citep{McWilliam+Zoccali10,Wegg+Gerhard13,2017MNRAS.471.3988C}. The inner stellar kinematics shows a cylindrical rotation \citep{Howard+2008,Kunder+Koch+Rich+12,2011MNRAS.416L..60B,2016ApJ...819....2N}, which indicates that the inner bulge is dominated by the bar \citep{2010ApJ...720L..72S, 2011MNRAS.416L..60B}. \citet{2010ApJ...720L..72S} found that a classical bulge with a merger-origin makes up less than 8\% of the Galaxy's disk mass. The Galactic bulge is therefore primarily a disk-like structure built up with a more secular formation history \citep{1990A&A...233...82C,Kormendy+Kennicutt04,2016ASSL..418..391A}. 

The central bar components affect the radial and rotational velocity distribution of the Galactic disk stars. The presence of groups of stars moving with particular radial and rotational velocities in the solar neighborhood is considered to be due to the bar \citep{2000AJ....119..800D}. The Hercules stream, which is a group of stars rotating slower than the average local azimuthal velocity and moving outward in the disk, has been suggested to be caused by the outer Lindblad resonance of the bar being just inside of the Sun's orbital radius, allowing us to derive the pattern speed of the bar \citep{Dehnen99,2000AJ....119..800D,Ramos+Antoja+Figueras18,Fragkoudi+19}. However, recent studies demonstrate that such moving groups can also be caused by transient spiral arms \citep{DeSimone+2004,Quillen+11, Hunt+Hong+Bovy+18,Fujii+2019,Khoperskov+Gerhard22}, as well as the other resonances of the bar, including the co-rotation resonance \citep[e.g.,][]{Perez-Villegas+17,Monari+19a,Chiba+Friske+Schoenrich20,Chiba+Schoenrich21} and the 4:1 resonance \citep{Hunt+Bovy18}. 

On the other hand, NIR photometric surveys, such as VVV, and spectroscopic surveys, such as the Bulge Radial Velocity Assay \citep[BRAVA;][]{Kunder+Koch+Rich+12}, the Abundances and Radial velocity Galactic Origins Survey \citep[ARGOS;][]{Freeman+Ness+13,Ness+Freeman+Athanassoula+13}, the Pristine Inner Galaxy Survey \citep[PIGS;][]{Arestsen+Starkenburg+Martin+20} and APOGEE, have revealed the detailed stellar structure and line-of-sight velocities of the bar. These observations were directly compared with theoretical models \citep[e.g.,][]{2010ApJ...720L..72S,Portail+Wegg+Gerhard+15}, suggesting a slower pattern speed of the bar than what was inferred by assuming that the outer Lindblad resonance lay just inside of the solar radius. Recently, both the gas dynamics \citep{Sormani+Binney+Magorrian15,Li+Shen+Gerhard+Clarke22} and stellar dynamics from the Optical Gravitational Lensing Experiment \citep[OGLE;][]{Rattenbury+Mao+Debattista+07}, BRAVA, ARGOS, APOGEE, VVV and/or {\it Gaia}~DR2 data \citep{Portail+Gerhard+Wegg+Ness17,Sanders+Smith+Evans19,Bovy+Leung+Hunt+19,Clarke+Gerhard22} have converged to a value for the bar pattern speed of around 35--40~km~s$^{-1}$~kpc$^{-1}$.  

\citet{Bovy+Leung+Hunt+19} further analysed the age and stellar abundances of the stars within the bar region, and found it populated mainly by stars with older ages and higher levels of [$\mathrm{\alpha}$/Fe], akin to the properties of old thick disk stars. Bovy et al. further suggested that the Galactic bar formed when the old thick disk formed at an early epoch of the Milky Way's evolution. If the bar did form as early as 10~Gyr ago, it may have affected the chemical distribution of stars in the Galactic disk. For example, the Lindblad resonances due to the bar act as a barrier for stellar migration, and the stars formed inside and outside of the outer Lindblad resonance do not mix \citep{Halle+DiMatteo+Haywood+Combes15,Haywood+DiMatteo+Lehnert+18}. Hence, identifying the epoch of bar formation is an imperative task in understanding the evolution of the Galactic disk. Table \ref{tbl:BarAge} summarizes examples of observed estimations of the Galactic bar age in literature. It should be noted that the age of the stars in the Galactic bar does not tell us the formation epoch of the bar, because stars born both before and after the bar can be captured by it \citep[][]{Wozniak2007,Baba+Kawata+Schoenrich22}. Hence, the creation epoch of the bar remains a challenging question even in the post-{\it Gaia} era.

\begin{table*}
\begin{center}
\begin{tabular}{lll}
 \hline
 Age [Gyr]      & Method/Tracer & Literature\\
 \hline
 $\sim 3$--$4$  & $N$-body sim., pattern speed of the bar   & \citet{Tepper-Garcia+2021}\\
 $\sim 3$--$6$  & Spatial distribution / infrared Carbon stars in the outer bar & \citet{ColeWeinberg2002}\\
 $\sim 6$--$8$  & Spatial distribution / red giants in the inner ring   & \citet{Wylie+2021}\\
 $\sim 8$--$9$  & Spatial distribution / O-rich Miras in the BP-shaped bulge & \citet{Grady+2020}\\
 $\gtrsim 8$    & Spatial distribution / red clumps in the NSD & \citet{Nogueras-Lara+2020}\\
 $\sim 10$      & Kinematics and Spatial / red giants in the outer bar & \citet{Bovy+Leung+Hunt+19}\\
 \hline
\end{tabular}
\end{center}
\caption{
A selection of estimations of the age of the Galactic bar based on various observations.
\label{tbl:BarAge}
}
\end{table*}

\begin{figure}
 \begin{center}
   \includegraphics[width=1.0\linewidth]{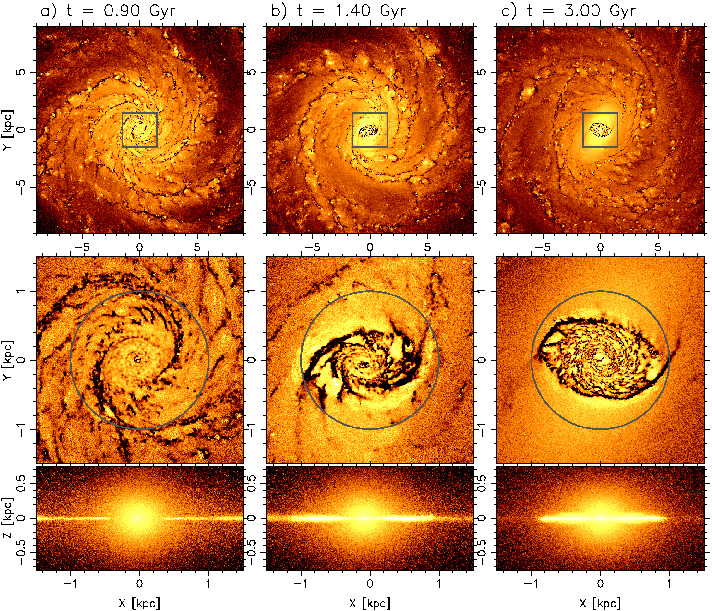}
 \end{center}
 \caption{
 Morphological evolution of a simulated Milky Way-sized galaxy \citep{Baba+Kawata20} in face-on views of the whole galaxy scale (top row), the central region (middle row), and in edge-on view of the central region (bottom) at the times (a) prior to, (b) during, and (c) after bar formation. Orange colors indicate surface density of stars on a logarithmic scale, and the dark filamentary structures in the face-on views (top and middle panels) indicate the cold dense gas. After the bar has formed, the major axis of the bar is orientated such that its major axis lies $25^\circ$ from the $y$-axis. 
 \label{fig:bar_formation_snapshot}
 }
\end{figure}

It is known that bar formation induces gas inflow to the central sub-kpc region of the host Galaxy, which leads to the formation of a compact and kinematically cold rotating nuclear gas disk \citep[e.g.,][]{Athanassoula1992b,Sormani+2018b}. Subsequently, the stars formed from the nuclear gas disk build up the NSD \citep[e.g.,][]{FriedliBenz1993,Athanassoula2005,WozniakMichel-Dansac2009,Martel+2013,Cole+2014,Ness+2014,Debattista+2015,Seo+2019}. Using an $N$-body/hydrodynamics simulation of a Milky Way-sized disk galaxy, \citet{Baba+Kawata20} demonstrated that bar formation triggers an intense burst of star formation in the central region, which creates the NSD (figure~\ref{fig:bar_formation_snapshot}). Consequently, the oldest age limit of the NSD stars displays a relatively sharp cut-off, and it tells us the age of the bar. In this way, the age distribution of the NSD in the Milky Way can tell us the formation epoch of the Galactic bar \citep[see also][]{Fragkoudi+2020}.

\citet{Baba+Kawata20} showed that the NSD is kinematically colder than the other stellar components, which is consistent with the recent observation mentioned above \citep{Schultheis+Fritz+Nandakumar+21}. Hence, the NSD stars can be identified kinematically. \citet{Baba+Kawata20} also suggested that stellar proper motions in addition to line-of-sight velocities are crucial to reducing contamination from other, kinematically hotter stars, such as those from the bulge, bar and Galactic disk. The superb astrometric accuracy of \jasmine{} will enable us to identify the NSD stars and reduce the contamination from the other stellar components. 

To obtain the age of the NSD stars, we will use Mira variables, because they are known to follow an age-period relation \citep{FeastWhitelock1997, Grady+2019,Zhang+Sanders23}.  Mira variables are intrinsically bright, and there are expected to be enough Miras in the \jasmine{} GCS to enable a robust age distribution study of the NSD. From the number of observed Miras in a smaller region of the Galactic center in \citet{Matsunaga+2009}, it is expected that about 2000 Miras whose color is red enough to be at the distance of the NSD and with a luminosity brighter than $H_\mathrm{w}=14.5$~mag will be observed within the GCS region of \jasmine{}. The proper motion measurements by \jasmine{} as well as the line-of-sight velocity from ground-based spectroscopic follow up (see section~\ref{sec:other_proj} for potential instruments to be used) will enable us to pick out Miras in the NSD. 

Before the launch of \jasmine{}, we need to identify Miras in the \jasmine{} GCS field (figure~\ref{fig:jgcs_sirius}) with long-term time-series photometric surveys. Recently, \citet{Sanders+Matsunaga+Kawata+22} identified more Miras from the VVV survey. We will also use the upcoming data from the {\it Prime focus Infrared Microlensing Experiment} ({\it PRIME}, see section \ref{subsubsec:prime}), which is a joint Japan-U.S.-South Africa 1.8m NIR telescope (with 1.3~deg$^2$ FoV) built in South Africa. The {\it PRIME} data will allow us to identify brighter Miras that are saturated in the VVV data and be valuable to verify the VVV Miras. 

\begin{figure}
 \begin{center}
   \includegraphics[width=1.0\linewidth]{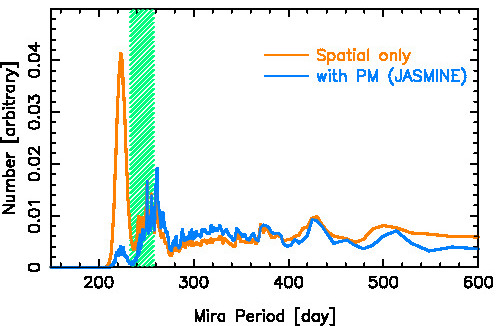}
 \end{center}
 \caption{
 Expected period distribution of the Mira variables in the NSD of the simulation shown in figure \ref{fig:bar_formation_snapshot}. The periods of Miras are derived from the age of star particles in the simulation snapshot at $t=5$~Gyr, using the empirical age--period relation in \citet{Grady+2019}. The orange line indicates the period distribution of Miras, when the star particles are chosen by the spatial information only, i.e., by the location corresponding to the Galactic longitude and latitude range of the NSD. The blue line shows the period distribution when the stars are selected by the Galactic longitude and latitude region of the NSD and the line-of-sight velocity and proper motion region corresponding to the NSD kinematics. In this simulation snapshot, the bar age is $3.5$ Gyr (green shaded region). Here, we assumed that there is a kinematically hot component, i.e., a classical bulge, which is 1 Gyr older than the formation age of the bar. The orange line shows the significant contamination from such a component, while the blue line with the additional kinematical selection of the NSD shows the sudden decrease of the population with shorter periods than about 250~days, which corresponds to the formation age of the bar. \label{fig:mira_period_dis_nsd}
 }
\end{figure}

Although the age-period relation of the Miras are still not well calibrated, we expect that it will be improved with $Gaia$ data \citep[e.g.,][]{Zhang+Sanders23}. Even without such a relation, the period of such Miras will be measured precisely, while identifying the shortest period of Miras in the NSD and comparing with the period of Miras in the other Galactic components will tell us the relative formation epoch of the Galactic bar (see figure~\ref{fig:mira_period_dis_nsd}), e.g., with respect to those for the thick and thin disks. This was demonstrated in \citet{Grady+2020}, who focused on the age of stars in the bar, but who could not reach the NSD because they studied Miras detected by {\it Gaia} \citep[see also][]{Catchpole+Whitelock+Feast+16}.

\subsubsection{Slowdown of the bar pattern speed}

Producing a sample of NSD stars with age information will also help to further uncover the slowdown history of the pattern speed of the Galactic bar. Recently, \citet{Chiba+Friske+Schoenrich20} and \citet{Chiba+Schoenrich21} suggested that the kinematics of local stars observed in $Gaia$~DR2 can be naturally explained if the pattern speed of the Galactic bar is slowing down. The slowdown of the bar pattern speed is often seen in $N$-body simulation of barred galaxies and is considered to be due to the transfer of the angular momentum of the bar to the dark halo \citep[e.g.,][]{Weinberg85}. The size of the NSD is likely affected by the location of the inner Lindblad resonance of the bar or the existence of $x_2$-orbits \citep[e.g.,][]{ReganTeuben2003,Li+2015}, which is likely affected by the pattern speed of the bar.
Hence, it is expected that the age distribution of the NSD disk as a function of radius and/or angular momentum could be sensitive to the slowdown history of the Galactic bar. Interestingly, a radial dependence of the age of the NSD stars is recently indicated by \citet{Nogueras-Lara+Schultheis+Najarro+23}, which motivates further studies.
\jasmine{} will be able to provide the age and kinematics of stars in the NSD, which will open up this new window to unravel the history of the Galactic bar pattern speed. How the NSD stellar age and kinematics are affected by the change in bar pattern speed is not well understood. Also, the bar pattern speed may not monotonically spin down, but may spin up, for example, due to gas infall. This could cause a more complicated mix of the age-kinematics relation of NSD stars. More theoretical studies on this topic are also encouraged. 

\subsubsection{Impacts on radial migration and the Sun's birth radius}

Identifying the bar formation epoch and the evolution of the pattern speed is crucial to answering how radial migration shaped the current chemodynamical structure of the Galactic disk. Bar formation induces significant radial migration in stars of the Galactic disk \citep[e.g.,][]{DiMatteo+2013,Khoperskov+19a}, and the coupling between bar resonances and spiral arms resonances further induces strong radial migration of disk stars \citep{Minchev+Famaey10}. The slowing down of the bar pattern speed means the resonances of the bar sweep a large radial range of the Galactic disk, and affect the kinematics of the disk stars at these radii and enhance radial migration \citep{Halle+DiMatteo+Haywood+Combes15,Halle+2018,Chiba+Friske+Schoenrich20,Khoperskov+19a}. 

The Sun is considered to have been formed in the inner disk ($R\sim 5$--7~kpc), and to have radially migrated outward to reach its current radius \citep[e.g.,][]{Wielen+Fuchs+Dettbarn96,Sellwood+Binney02,Minchev+18,Feltzing+Bowers+Agertz20,TsujimotoBaba2020}. It is a fascinating question as to whether the Galactic bar formed before or after the formation of the Sun. If the Galactic bar is younger than the age of the Sun, the orbital history of the Sun is likely impacted by the Galactic bar formation, in addition to radial migration due to resonances, making this a key question in understanding the Sun's dynamical history. 

Recently, the accurately measured chemical abundance pattern of a large number of disk stars by APOGEE combined with the kinematics provided by {\it Gaia} has enabled  modeling of the history of the inside-out disk formation of the Galactic disk and the impact of the radial migration of stars \citep{Hayden2015,Frankel+18,Frankel+20, imig2023}. However, radial migration due to bar formation and evolution has not yet been taken into account, because of the complexity of such a model with additional parameters needed to fit the data and thus requiring additional priors. One such key prior is the formation epoch of the bar. As discussed in this section, \jasmine{} will provide this important piece of the information from the age distribution of the NSD.

\subsubsection{Comparison with the cosmic bar fraction and the past merger history of the Milky Way}

It would be also interesting to compare the bar formation epoch of the Milky Way with the cosmic bar fraction \citep[e.g.,][]{Seth+Elmegreen+Elmegreen+08}, to study how common the formation epoch of the Milky Way is. Identifying the bar formation epoch in the Milky Way allows us to study the physical process of the Galactic bar formation by comparing with the expected formation epochs of the bar with the different mechanisms from the cosmological simulations \citep[e.g.,][]{Cavanagh+Bekki+Groves+22,Izquierdo-Villalba+Bonoli+Rosas-Guevara+22}. Another intriguing question is if the formation epoch of the bar of the Milky Way is similar to the time of the merger events of the {\it Gaia}-Sausage-Enceladus and Sagittarius dwarf, whose impact could have induced the bar formation of the Milky Way. 

\subsection{Nuclear Stellar Disk: A link to high redshift progenitors} \label{subsec:nsd-highzgal}

The size of the Galactic nuclear disk is similar to the size of star forming galaxies at redshift, $z>6$ \citep{Shibuya+Ouchi+Harikane15,Kikuchihara+20}, as also recently seen by {\it JWST} \citep{Ono+Harikane+Ouchi+22}. The compact disks observed at high redshift display a surface brightness profile following an exponential law, like a rotating disk \citep{Shibuya+Ouchi+Harikane+Nakajima19}. The theoretical study of \citet{Bekki00ApJ545p753} suggested that the gas accretion to a massive BH can create an NSD-like compact disk at high redshift. It would be a fascinating question whether the Milky Way started its formation from a compact disk similar to those observed at high redshift, and, if so, whether a part or the majority of the NSD could be a relic of such a compact disk epoch. 

The age and metallicity distribution of the NSD disk stars identified from the proper motions measured by \jasmine{} will answer this question regarding the early structure of the Milky Way. If \jasmine{} finds an old and metal poor population in the NSD, in addition to the populations formed after bar formation, they could be a relic of the high redshift disk. A comparison of the kinematics and metallicity properties between the Milky Way's old disk and the high redshift disks in Milky Way progenitors will be a new window to connect Galactic archaeology to the observations of high redshift galaxies \citep[see also][]{Clarkson+Sahu+Anderson+08,Renzini+Gennaro+Zoccali18}, which will be further advanced by the advent of {\it JWST} and future 30-m~class ground based telescopes, such as {\it E-ELT}, {\it GMT} and {\it TMT}.


\subsection{Inner disk: spiral arms and Galactoseismology}
\label{subsec:inner_disk}

The foreground stars of the \jasmine{} GCS field will give access to Galactic disk kinematics from the solar radius to the Galactic center. 
The {\it Gaia} data revealed the striking ridge-like features in the stellar rotation velocity distribution as a function of radius, i.e., the $R_{\rm GC}-V_{\rm \phi}$ diagram \citep{Antoja+Helmi+RomeroGomez+18,Kawata+Baba+Ciuca+18,Ramos+Antoja+Figueras18}. These features also correlate with the radial velocity of the stars \citep{Fragkoudi+19}, and are considered to be due to bar resonances and effects of the spiral arms \citep[e.g.,][]{Hunt+Hong+Bovy+18,Hunt+Bub+Bovy+19,Friske+Schoenrich19,Asano+2020}. Because of the heavy dust extinction in the inner disk, the {\it Gaia} data can reach only up to about 3 kpc, i.e., $R_{\rm GC}\gtrsim 5$~kpc, in the mid-plane toward the Galactic center. The NIR astrometry of \jasmine{} can extend this analysis all of the way to the Galactic center. 

The nature of spiral arms, especially whether they are classical density wave-like features \citep[][]{Lin+Shu64} or transient features \citep[e.g.,][]{Sellwood11}, is currently hotly debated \citep[see][for reviews]{Dobbs+Baba14,SellwoodMaster2022ARAA}. Spiral arms could be related to the bar \citep[e.g.,][]{Athanassolua12} and induced by satellite galaxy interactions \citep[e.g.,][]{Purcell+Bullock+Tollerud+11,Pettitt+Tasker+Wadsley16}.
Also, the effects of spiral arms on radial migration depend on the nature of the spiral arms. Radial migration is much more significant if the spiral arms are transient features \citep[e.g.,][]{Sellwood+Binney02,Grand+Kawata+Cropper12}. \jasmine{} will provide detailed kinematics of the stars towards the Galacitc center, which covers the Sagittarius arm and the Scutum-Centaurus arm. The locations of the spiral arms of the Galaxy are still not measured confidently from the stellar kinematics, although they can be traced with the gas, star forming regions, and young stars \citep[e.g.,][]{Nakanishi+Sofue06,Reid+Menten+Bruthaler+19,Skowron+Skowron+Mroz+19,Colombo+Duarte-Cabral+Pettitt+22}. It is also debated whether the Sagittarius arm is the major arm, i.e., a stellar arm, or purely a gas arm \citep[e.g.,][]{Benjamin+05, Pettitt+Dobbs+Acreman+Price14}, although the recent {\it Gaia}~EDR3 data indicate an excess of young stars in the Sagittarius arm \citep{Poggio+Drimmel+Cantat-Gaudin+21}. A crucial piece of information regarding the nature of spiral arms and the strength of stellar spiral arms is the stellar kinematics both inside and outside of the arm  \citep{Baba+Saitoh+Wada13,Baba+Kawata+Matsunaga+18,Kawata+Hunt+Grand+14}. In addition to the GCS, we plan to run the Galactic mid-plane survey  (see section~\ref{sec:gmps} for more details). The combination of these data can provide the kinematics of the both sides of the Sagittarius arm and Scutum-Centaurus arm at different radii. Comparison between these data and $N$-body modelling of disk galaxies will be necessary to ultimately answer the long-standing question in Galactic astronomy of the nature of spiral arms.

The {\it Gaia} data also revealed a vertical corrugation in the vertical velocity, $v_{\rm z}$, of the disk stars as a function of Galactocentric radius \citep{Schoenrich+Dehnen18,Gaia+Katz+18,Kawata+Baba+Ciuca+18}, which is related to the phase spiral most clearly seen in the $z-v_{\rm z}$ diagram when colored by the radial velocity, $v_{\rm rad}$ \citep{Antoja+Helmi+RomeroGomez+18,Bland-Hawthorn+Sharma+Tepper-Garcia+19,Hunt+Price-Whelan+Johnston+Darragh-Ford22}. Hence, they are likely closely linked with in-plane kinematical features, such as moving groups, the $R_{\rm GC}-V_{\rm \phi}$ ridge-like features, and spiral arm features \citep[e.g.,][]{Laporte+Minchev+Johnston+Gomez19,Khanna+Sharma+Tepper-Garcia+19,Bland-Hawthorn+Tepper-Garca21,Hunt+Stelea+Johnston+21,Antoja+Ramos+Lopez-Guitart+22}.  These vertical kinematical features are considered to be due to the recent ($<$Gyr) perturbation from the Sagittarius dwarf galaxy \citep{Bland-Hawthorn+Sharma+Tepper-Garcia+19,Laporte+Minchev+Johnston+Gomez19} and/or the wake of the dark matter halo induced by an earlier phase of the infall of the Sagittarius dwarf a few Gyr ago \citep{Grand+Pakmor+Fragkoudi+22}. However, it is also possible to be induced by bar buckling, which may have happened recently and created the X-shape/boxy inner Milky Way bar \citep{Khoperskov+DiMatteo+Gerhard+19}. Tracing the vertical corrugation seen around the solar radius towards the Galactic center will be key to answering whether this dynamical phenomenon is ubiquitous in the inner disk. A comparison of the observed corrugation features as a function of radius with different models of the Sagittarius dwarf perturbation and the bar buckling model will allow us to reveal the origin of the corrugation. Both the impact of the dwarf perturbation and the bar buckling can also significantly affect radial migration. The Galactoseismic information of both in-plane and vertical kinematics as measured in the inner disk by \jasmine{} will provide crucial information regarding the recent dynamical evolution of the Galactic disk.

The foreground disk stars of the \jasmine{} GCS field up to $d\sim6$~kpc from the Sun are dominated by red giants. The time-series photometry of \jasmine{} with about 46 instances of $\sim12.5$~s exposures every $\sim$530~min for 3~years of the nominal lifetime of the telescope will be highly valuable for asteroseismology. It will allow for the measurement of precise stellar masses, which will provide relative ages of stars, crucial information for Galactic archaeology studies \citep[e.g.,][]{Chaplin+Miglio13, Miglio+Chiappini+Mackereth+21}. 

It is also likely that a few open clusters will be discovered in this field. There is at least one star cluster (UBC335) in the new {\it Gaia}~DR2 star cluster catalogue \citep{Castro-Ginard+Jordi+Luri+20} in the \jasmine{} GCS field. These cluster data would be useful for the calibration of asteroseismic ages, as proposed in an ESA M7 mission candidate, {\it HAYDN} \citep{Miglio+Ciradi+Grudhal+21}. \jasmine{} can provide a proof of concept study for {\it HAYDN} in terms of asteroseismology in dense stellar fields. 

The ages of giant stars can also be obtained from follow-up spectroscopic data, mainly from [C/N] \citep{Masseron+Gilmore15,2016MNRAS.456.3655M}, which can be calibrated by their asteroseismic ages. Machine learning techniques are often used to train a model to map the chemical abundance patterns to ages, where having the high-quality training set is crucial \cite[e.g.,][]{Ness+Hogg+Rix+16,Das+Sanders19,Mackereth+Bovy+Leung+19,Ciuca+Kawata+Miglio+21}. Asteroseismology from \jasmine{} combined with spectroscopic follow up data will provide the unique calibration information for the inner disk red giants needed for future machine learning applications. 

\section{Exoplanets Survey (EPS)} \label{sec: exoplanets}

\subsection{Transiting Planet Survey around Late-type Stars} \label{subsec: exojasmine}

\begin{figure}
 \begin{center}
   \includegraphics[width=1.0\linewidth]{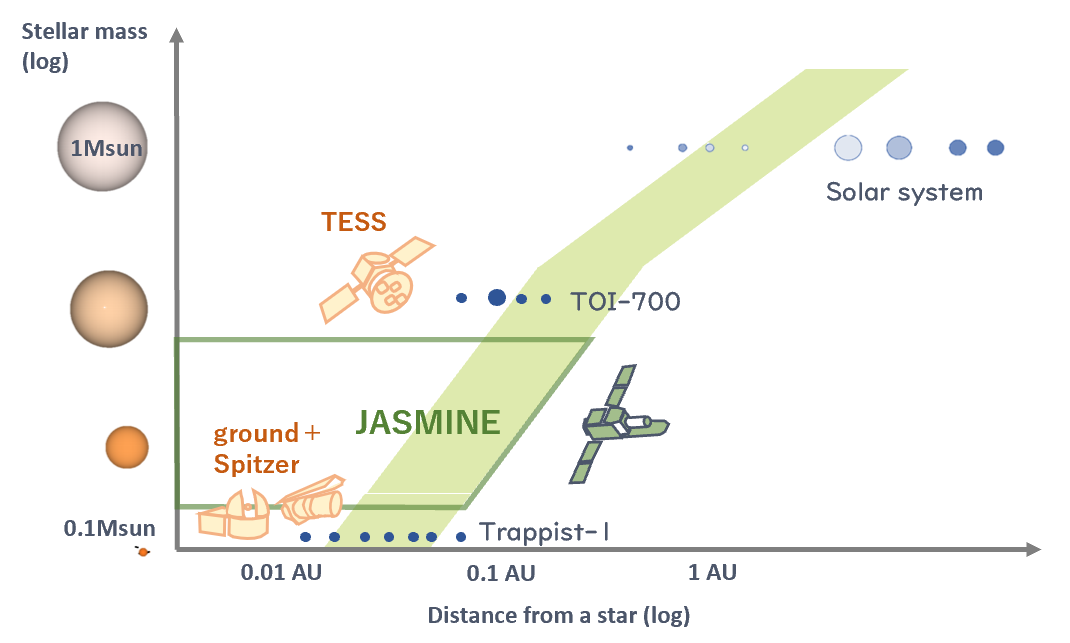}
 \end{center}
 \caption{Survey area for the \jasmine{} transit survey on the plane of star--planet distance versus stellar mass, contained within the green box.
 The HZ is shaded light green. \label{fig:exojasmine}}
\end{figure}

\begin{figure}
    \begin{center}
    \includegraphics[width=1.0\linewidth]{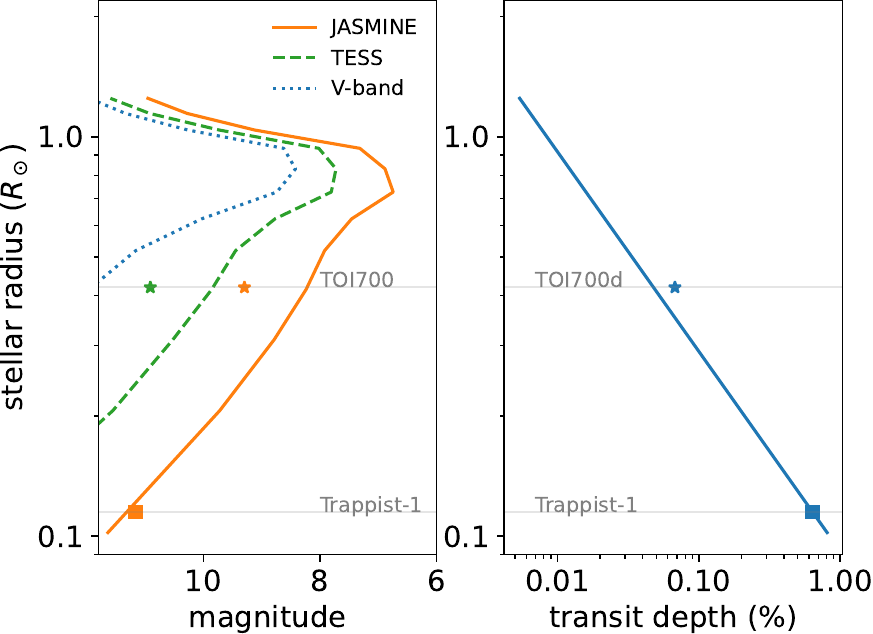}
    \end{center}
    \caption{Left: The lines show the apparent magnitudes of the 250th brightest star in the sky in each radius bin of width of $\Delta R$ = $0.1 R_\odot$, where the stars are from {\it TESS} candidate target list version 8 obtained via the Mikulski Archive for Space Telescope database. Solid, dashed, and dotted lines indicate the magnitudes in the \jasmine{} $H_\mathrm{w}$, {\it TESS}, and $V$-bands, respectively. The positions of TOI-700 and TRAPPIST-1 are indicated by stars and squares, respectively, and their color indicates the magnitude of \jasmine{} (orange) and {\it TESS} (green) magnitudes. Right: Transit depth due to an Earth-sized exoplanet as a function of stellar radius.}
    \label{fig:exo_basic}
\end{figure}

As described in the Introduction (section~\ref{sec:intro-exopl}) and shown in figure~\ref{fig:exojasmine}, terrestrial planets in the HZ around mid- to late-M dwarfs remain difficult to explore with both {\it TESS} and ground-based surveys. Searching for terrestrial planets around ultracool dwarfs like TRAPPIST-1 is challenging for {\it TESS} owing to its modest 10.5~cm aperture. In ground-based surveys, HZ terrestrial planets around earlier-type stars, like TOI-700,
exhibit transits that are too shallow and have too long orbital periods. In contrast, \jasmine{} has a 36~cm aperture and an NIR passband, and can also perform long continuous monitoring from space and suppress intrinsic variability due to stellar activity (see section \ref{ss:activity} for details). This makes \jasmine{} a unique probe of terrestrial planets in the HZ around mid- to late-M dwarfs that have orbital periods of several weeks, or semi-major axes of 0.03--0.3~AU. Figure \ref{fig:exo_basic} shows this trade-off relation between stellar brightness and transit depth. The left panel shows the apparent magnitudes of the 250th brightest star in a radius bin of width of 0.1$\,R_\odot$, where $R_\odot$ is the solar radius. 
For M-type stars with radii smaller than $0.5\,R_\odot$, smaller stars are apparently fainter both in visible and NIR bands. In contrast, the transit signals are larger for those smaller stars, as shown in the right panel. The sweet spot for precise NIR photometry with \jasmine{} therefore lies around $0.2$--$0.3\,R_\odot$.

\begin{figure*}
 \begin{center}
   \includegraphics[width=1.0\linewidth]{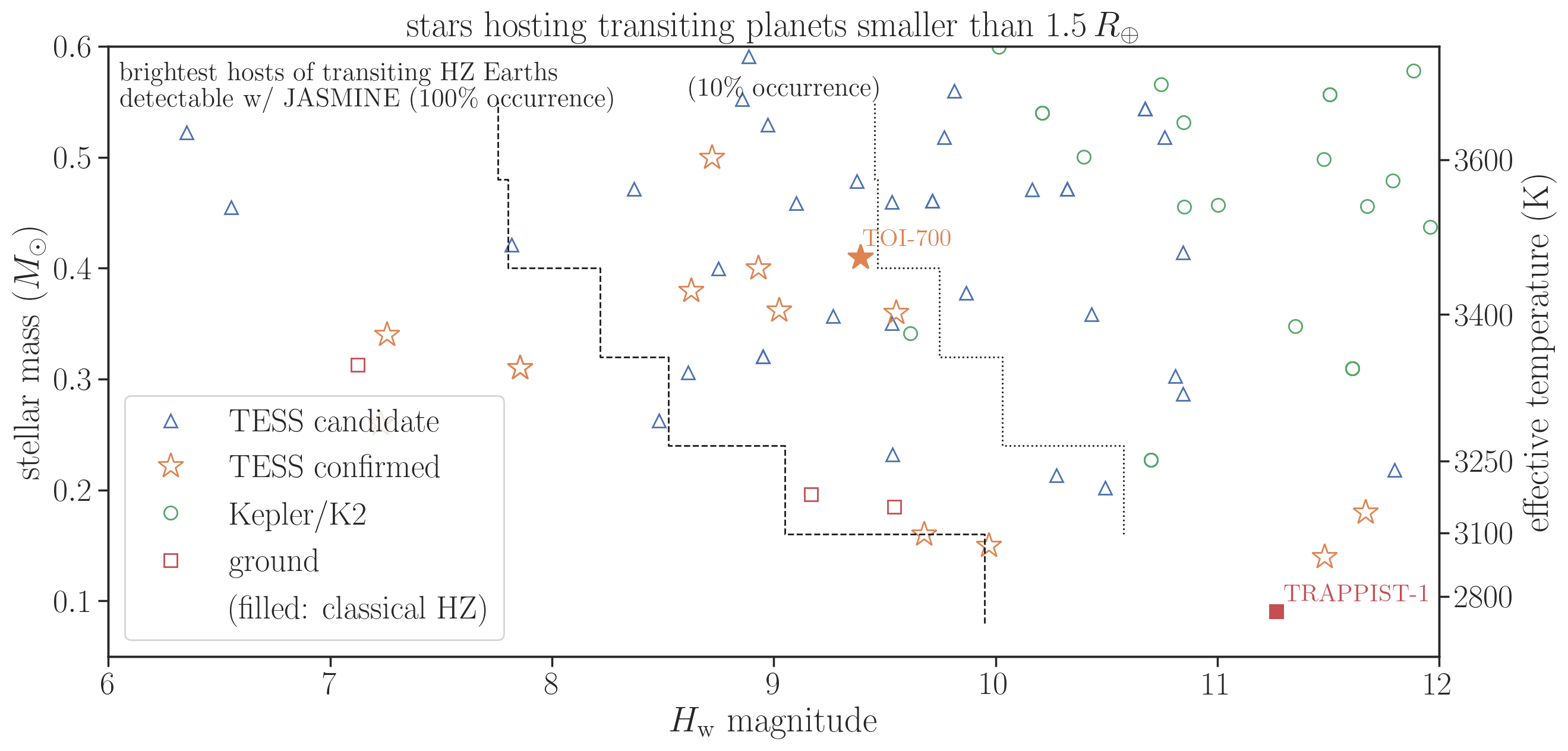}
 \end{center}
 \caption{
 \jasmine{}'s photometric precision is sufficient to find transiting Earth-sized planets in the HZ of the brightest mid-M dwarfs. The black dashed and dotted lines show the estimates of the $H_\mathrm{w}$ magnitudes of the brightest stars in the sky for which transit signals due to HZ Earths are large enough to be detected by \jasmine{} photometry (see text for details); here 100\% and 10\% of M dwarfs in the {\it TESS} candidate target lists harbor such planets, respectively. The open and filled symbols denote the masses and $H_\mathrm{w}$ magnitudes of stars with known transiting planets smaller than $1.5\,R_\oplus$. Most of these planets are outside the classical HZ except for TOI-700 and TRAPPIST-1, shown as filled star and square, respectively.
 \label{fig:target_exojasmine}
 }
\end{figure*}

Figure \ref{fig:target_exojasmine} assesses this statement more realistically. 
Here, we estimate the $H_\mathrm{w}$ magnitudes of the brightest M dwarfs in the entire sky that host transiting Earth-sized planets in the HZ whose transits are deep enough to be detected with \jasmine{}. The result is based on a Monte-Carlo simulation consisting of the following three steps:
(i) Assign Earth-sized planets at the inner edge of the classical HZ to certain fractions (here 100\% and 10\%) of nearby M dwarfs in the {\it TESS} candidate target list \citep{2019AJ....158..138S}.
(ii) Check which planets show transits, assuming that the orbits are isotropically oriented.
(iii) Check whether a single transit signal has a signal-to-noise ratio (S/N) greater than 7, where the signal is computed from the stellar radius, and the noise is evaluated for the timescale of expected transit duration using the noise model in section~\ref{subsec:NIRphoto}.
The dashed and dotted lines show the magnitudes of the brightest stars estimated from 10000 random realizations in each stellar mass bin, assuming that HZ Earths occur around 100\% and 10\% of the sample stars, respectively.
These estimates are compared with known transiting planets and planet candidates from ground- and space-based surveys (see legends), including the two Earth-sized planets ($<1.5\,R_\oplus$) in the HZ, TOI-700 and TRAPPIST-1, shown with filled symbols. 
The occurrence rate of HZ Earths around mid-M dwarfs is yet to be quantified; the Kepler sample suggests an occurrence rate of $\sim 30\%$ around earlier M dwarfs \citep[e.g.,][]{2020MNRAS.498.2249H, 2015ApJ...807...45D, 2013ApJ...767...95D}, and here the values of 100\% and 10\% are chosen to bracket this estimate. While it is unclear whether the estimate for earlier M dwarfs can be simply extrapolated to mid-M dwarfs, the fact that the two known planets lie close to the dotted line (i.e., the brightest planet hosts assuming the $10\%$ occurrence have already been found) appears to be consistent with the occurrence rate of $\gtrsim 10\%$ around later M dwarfs.
Our simulation results demonstrate the ability of \jasmine{} to identify Earth-sized transiting planets in the HZ of mid-M ($\sim0.2\,M_\odot$) dwarfs with $H_\mathrm{w}\lesssim10$~mag, if their occurrence rate is $\gtrsim \mathcal{O}(10\%)$ and if those stars are monitored over a sufficiently long time.

The habitability of exoplanets is often evaluated based on the concept of the habitable zone. For tidally locked exo-terrestrial planets, the inner edge of the habitable zone should be considered in relation to atmospheric dynamics, which are primarily affected by the planet's orbital period. Recent studies using 3D General Circulation Models (GCMs) suggest that a tidally locked terrestrial planet near the inner edge of the habitable zone of a star, with an effective temperature exceeding 3300~K, is likely to exhibit atmospheric circulation characterized by strong convective motions beneath the substellar point and a thermally direct circulation between the day and night hemispheres \citep[e.g.,][]{Showman_2013,Haqq-Misra_2018}. This strong convection facilitates the formation of thick clouds, which envelop the planet's dayside hemisphere, thereby increasing its albedo \citep{Yang_2013}. As a result, this cloud feedback mechanism can render a tidally locked terrestrial planet habitable even if a planet exists closer to the star than the inner boundary of the classical habitable zone, corresponding to the inner edge of $1.3$ solar insolation compared with $0.95$ solar insolation of the classical one for 3300~K. Conversely, the inner edge of the habitable zone for a star with an effective temperature below $\sim$3000~K gets closer to the classical definition. This is attributed to the dimmer nature of such stars, where clouds on the dayside are advected eastward off the substellar point due to rapid rotation \citep{Kopparapu_2017}. Hence, focusing on JASMINE targets around stars with effective temperatures near 3000~K is critical for understanding the boundary conditions affecting the habitability of tidally locked exo-terrestrial planets, as well as for elucidating their climate formation and verifying our theories regarding habitable planets.

\subsection{Follow-up Characterization from Space}\label{subsec: followup_transit} 

Photometry from space is important not only for finding new transiting planets, but also for characterizing them. For example, modeling of transit timing variations (TTVs) (i.e., variations in the orbital periods due to gravitational interactions between multiple planets) measured by {\it Spitzer} played a significant role in precisely weighing the planets of the TRAPPIST-1 system \citep[e.g.,][]{2018A&A...613A..68G,2021PSJ.....2....1A}. Such precise masses for Earth-mass planets are still difficult to obtain, even with state-of-the-art infrared spectrographs, but are crucial for understanding the composition of these planets and for interpreting their transmission spectra. Even if TTVs alone are not sufficient for obtaining precise masses, they still constrain mass ratios, and joint analyses with RV data improve mass measurements. Thus, follow-up observations of known multi-transit systems with \jasmine{} are valuable, even if no new transiting planet is found. These data contribute to the derivation of the best-constrained masses for terrestrial planets transiting late-type stars.

Long-term photometry from space is important even for single-transit planets without any known signature of dynamical interactions. Extending the baseline of transit measurements helps to pinpoint the ephemeris of transiting planets found from relatively short baseline observations. If follow-up spectroscopy is performed much later than the discovery, a small error in the ephemeris calculated from the discovery data could result in a transit prediction that is far from the actual value (stale ephemeris problem). Extending the transit baseline is essential to avoid wasting precious observation time, for example, with {\it JWST}. Checking for the presence or absence of TTVs is also important in this context.

The ability to perform high-cadence, high-precision, NIR photometry from space makes \jasmine{} a unique follow-up facility, similar to {\it Spitzer}, even for systems with shorter observing baselines. The detailed shapes of the transit light curves revealed from such measurements are important for ensuring that the system is not an eclipsing binary system (which tends to produce V-shaped dips). The chromaticity of the transit also helps to identify stellar eclipses if the candidates are obtained from optical observations. Better constraints on the transit shapes, in particular the durations of the transit ingress and egress, also improve the measurements of transit impact parameters, and hence the planetary radii \citep{2020AJ....160...89P}, as the two parameters are correlated through stellar limb darkening. This is important for learning about their compositions, given that the precision of stellar parameter measurements has now become comparable (or even better, depending on the instrument) to the precision of the radius ratio measurements from the transit light curve. NIR observations are particularly well suited for this task because the limb-darkening and stellar variabilities (i.e., the source of correlated noise) are both weaker than those in the optical regime. 

{\it Spitzer} had long been an important facility to perform such spaced-based follow-up observations until 2020. A similar role has now been fulfilled by the {\it CHaracterising ExOPlanet Satellite (CHEOPS)}. \jasmine{} could serve as a successor of these missions in late 2020s and address various other scientific questions as were covered by {\it Spitzer} \citep[cf.][]{2020NatAs...4..453D}. See table~\ref{tab:comparison_satellites} for comparisons of those space missions.

\begin{figure}
\begin{center}
  \includegraphics[width=1.0\linewidth]{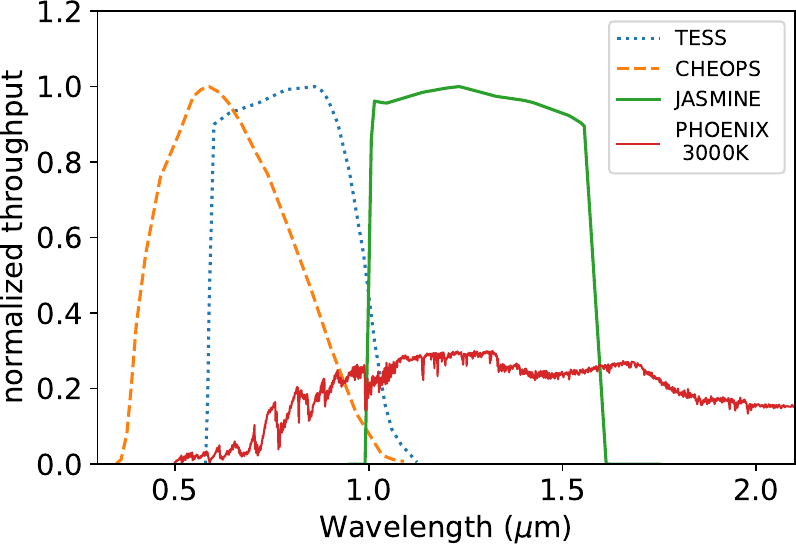}
\end{center}
\caption{Passbands of \jasmine{} (green solid), {\it TESS} (blue dotted), and {\it CHEOPS} (orange dashed), normalized by its each maximum throughput. The red solid line indicates the normalized photon flux of a M-type star (3000~K, $\log g=5$) by PHOENIX. \label{fig:comp}}
\end{figure}

\begin{figure}
\begin{center}
  \includegraphics[width=1.0\linewidth]{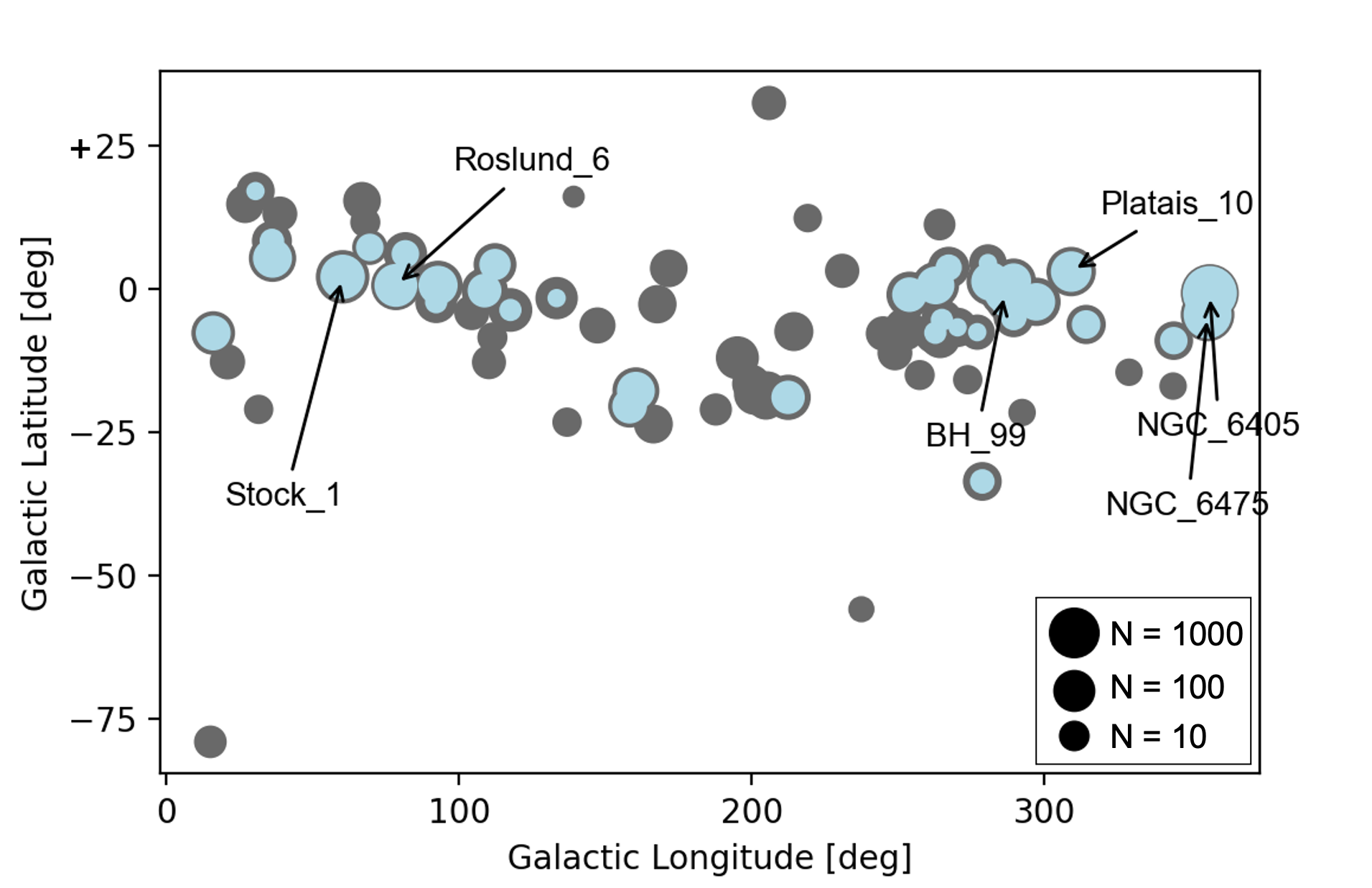}
\end{center}
\caption{Distribution of clusters that are closer than 500~pc from the Earth. The original list is cited from \citet{Cantat-Gaudin2018}. The sizes of the data points represent the number of detectable cool targets within the FoV of $0.55^{\circ} \times 0.55^{\circ}$, when transiting planets exist. The blue and gray colors correspond to assumed planetary radii of Earth-size and Neptunian-size, respectively. Details are described in section \ref{sec:youngplanet}. \label{fig:cluster}}
\end{figure}

\begin{table*}[ht]
\begin{center}
  \begin{tabular}{lcccccc}
  \hline\hline
  Mission & Telescope & Passband & Pixel size & FOV & Relative modulation amplitude & Status \\
  & & & & for late-type stars & by stellar activity$^\dagger$ &\\
\hline
\jasmine{} & 36~cm & 1.0--1.6 $\mathrm{\mu}$m & 0.4 arcsec & $0.55^\circ\times 0.55^\circ$ & $\sim 0.4$ & 2028 (planned)\\
TESS & 10.5 cm & 0.6--1.1 $\mathrm{\mu}$m  & 21 arcsec& $24^\circ\times24^\circ \times 4$ & $\sim 0.8$ & 2018-- \\
CHEOPS & 32 cm & 0.4--1.1 $\mathrm{\mu}$m  & 1 arcsec & $0.32^\circ\times 0.32^\circ$ & $\sim 0.9$ & 2019-- \\
Kepler & 95 cm & 0.4--0.9 $\mathrm{\mu}$m  & 4 arcsec & $12^\circ\times12^\circ$ & 1 & retired \\
Spitzer & 85 cm & 3.6--8 $\mathrm{\mu}$m  & 1--8 arcsec & $5^\prime\times5^\prime$& & retired \\
\hline
\end{tabular}
\end{center}
  \caption{Capability of performing photometry from space. ($^\dagger$ See section \ref{ss:activity} for further details). \label{tab:comparison_satellites}}
\end{table*}

\subsection{Searching for Young Planets}\label{sec:youngplanet} 
Other potential targets for the pointing--individual--target (PIT) type transit survey by \jasmine{} are stars in star clusters, young moving groups, and younger star-forming regions. Recently, transit surveys from space have unveiled a population of young close-in planets \citep[for example,][]{Mann2017, Mann2018, 2020Natur.582..497P}. 
These planets are often reported to have inflated radii for their insolation flux level received from host stars \citep{2016AJ....152...61M}, in comparison with their older counterparts. Some of super-Neptune-sized young planets were discovered as multi-planet systems, and those could be progenitors of ``compact-multi'' super-Earths \citep{David2019}, most commonly found around main-sequence solar-type stars. Hypotheses can be tested in relation to the formations of the super-Earth ``radius gaps'' \citep{2017AJ....154..109F} and the desert of the close-in Neptunian planets \citep{2016NatCo...711201L}. 
Measurements of eccentricities and inclinations (spin-orbit angles) of young planets would also allow the corroboration or refutation of planet migration theories for close-in large planets \citep[for example,][]{2020ApJ...890L..27H, 2020ApJ...899L..13H}. 
Overall, young exoplanet systems provide an ideal setting for testing hypotheses on planet formation and evolution. 

Younger systems allow two possible directions for \jasmine{} observations. One is a blind transit survey for stellar clusters, stars in star-forming regions, etc. Although the FoV of \jasmine{} is much smaller than that of ordinary transit surveys (i.e., often larger than $10^\circ\times 10^\circ$), many stars in the clusters (of the order of 10--100 for, e.g., the Beehive open cluster) are simultaneously observed, even with \jasmine{}'s FoV ($\approx 0.55^\circ\times 0.55^\circ$). This simultaneous photometry for multiple stars would significantly enhance the probability of transit detection compared with targeting field stars. 
For reference, the distribution of the clusters within 500~pc from the Earth, i.e., close enough to detect planet transits, is shown in figure \ref{fig:cluster}. The size of data points represents the number of cool stars which may show a larger transit signal than 4 $\sigma$ when transiting a planetary candidate's orbit. The blue and gray colors correspond to cases when the transiting planetary radii are Earth- and Neptune- sizes, respectively. Here, we computed stellar radii using relations between the {\it Gaia} {\it BP-RP} colors and effective temperatures in \citet{Mann2015}. The transit depth was derived as $(R_p/R_\star)^2 $ and the photometric noise follows figure \ref{fig:noise_exo}. There are a good number of clusters that contain $\sim 100$ of effective targets for the blind survey of transiting planets along the Galactic longitude. 
The other direction is the follow-up photometry of planet-hosts or planet-candidate hosting young stars for an additional planet(s) search, as in the case of M dwarf campaigns described in the previous sections. These follow-up observations target young stars hosting transiting close-in planets, similar to K2-25b and AU Mic b. It should be stressed that, for these targets, space photometry by {\it Spitzer} has played a significant role in confirming the planetary nature and refining transit ephemerides \citep{2020AJ....159...32T, 2020Natur.582..497P}.  
Reduced flux modulations resulting from stellar activity in the NIR region \citep{2021AJ....162..104M} make \jasmine{} optimal for these types of transit surveys from space (see section \ref{ss:activity}). The HZ around young stars (even M stars) tends to be further out in orbit ($P>50$ days) because of their inflated radii, but the HZ shrinks in the orbital distance as the system ages and the central star becomes smaller. Finding ``future-habitable'' planets would be an intriguing topic. 

\subsection{Stellar spin-down relations from young cluster observations}\label{subsec:Age-Rotation}
Photometric observations of stars in young clusters described in section \ref{sec:youngplanet} would usher in new knowledge on stellar rotation evolution especially young mid-to-late M dwarfs.  It has been known that stars spin down over time via magnetic braking (angular momentum loss) since \citet{1972ApJ...171..565S}. This has led to the development of the gyrochronology relation, which uses stellar rotation periods ($P_{\rm rot}$) as an indicator of stellar ages (e.g., \citealt{2003ApJ...586..464B}). 
This gyrochronology relation is important for considering the effects of stellar magnetic activities (e.g., X-ray/EUV emissions) on the planetary atmosphere over various ages (e.g., \citealt{2015A&A...577L...3T}; \citealt{2021A&A...649A..96J}). 

Recent \textit{K2} mission data have provided us the measurements of $P_{\rm rot}$ in benchmark open clusters over various ages from pre-main-sequence age ($\sim$3 Myr) to intermediate age ($\sim$2.7 Gyr) 
(\citealt{2020ApJ...904..140C}; \citealt{2020AJ....159..273R}; \citealt{2021A&A...649A..96J}). 
They showed that the formula describing the process of stellar spin-down cannot be as simple as first assumed.  
\citet{2020ApJ...904..140C} showed that low-mass stars show temporal stalling of spin down in medium ages. \citet{2020AJ....159..273R} reported on the rotation of stars in younger clusters (from 1 Myr to 790 Myr), and properties of rotation evolution differ greatly depending on spectral types. This would also be closely related with star-disk interactions in the pre-main sequence phase of stellar evolution (e.g., \citealt{2014prpl.conf..433B}). These recent studies showed the possibility that the rotation evolution of M dwarfs could be somewhat different (e.g., large scatter of M dwarfs at $\sim$790 Myr in figures 9 of \citealt{2020AJ....159..273R}) from that of G-dwarfs, and this difference is yet to be completely understood.  Moreover, in the previous studies using \textit{K2} \& \textit{TESS} data, there are very few measurements of mid-to-late M dwarfs. For example, in the \textit{K2} optical observations, most of mid-to-late M dwarfs were not selected as targets. 
Then NIR photometric observations using \jasmine{}
would help to fill in this “blank region” by measuring the rotation period of more mid-to-late M dwarfs in young stellar clusters, as  byproducts of young cluster observations described in section \ref{sec:youngplanet}. 

As discussed in the above sections, mid-to-late M dwarfs are bright in NIR wavelengths (figure \ref{fig:comp}), and the precision of \textit{JASMINE} photometric observations for  mid-to-late M dwarfs can be better than \textit{K2} \& \textit{TESS}.
In addition, the pixel size of \textit{JASMINE}, which is more than order of magnitude better than those of \textit{Kepler} and \textit{TESS} (table \ref{tab:comparison_satellites}), can be beneficial for such studies due to the capability to remove visual binary stars.

\subsection{Photometric Variability of Brown Dwarfs}\label{subsec:BDvar}

Brown dwarfs are intermediate objects between stars and planets with temperatures below $\sim 2400$~K. This temperature range resembles that of the primary targets for current observations of exoplanet atmospheres. Since brown dwarfs share many physical and chemical processes with gas giant exoplanets, understanding their atmospheres is also essential for studying these exoplanet atmospheres.

Many brown dwarfs reportedly show large photometric variations of typically up to a few percent \citep[see][for review]{2017AstRv..13....1B, 2018haex.bookE..94A} with timescales similar to their rotation periods (ranging from a few hours to a day).
Observed significant photometric variations in brown dwarfs are mainly attributed to the inhomogeneities of cloud opacity over their surface. 
Using the general circulation model (GCM), \citet{2021MNRAS.tmp..140T} recently demonstrated that cloud radiative feedback drives vigorous atmospheric circulation, producing inhomogeneities of cloud distributions over the surface.
Their simulations predict more prominent variability when viewed as equator-on rather than pole-on, consistent with the tentative trend of current observations \citep{2020AJ....160...38V}, pending further confirmation.
The dependence of the variability on other parameters such as the spectral type and gravity remains highly uncertain because of the lack of samples with precise photometric monitoring. This constitutes an obstacle to obtaining detailed insights into the cause of the variability, and further understanding physical and chemical processes in the brown dwarf atmospheres.
For example, if silicate clouds alone are responsible for the observed variability of brown dwarfs, one would expect the greatest variability at the L-type/T-type spectral transition. This is because those clouds start to form below the photosphere in the cold atmospheres of T-type dwarfs.
On the other hand, if some other types of clouds, such as $\mathrm{Na_2S}$ and $\mathrm{KCl}$, instead form in such cooler atmospheres \citep{2012ApJ...756..172M},
the trend of variability over the spectral type should be more complex (the variability amplitude should be maximized at the spectral type whose photospheres have the temperatures close to the condensation temperatures of each forming clouds).

In this context, the high photometric precision of \jasmine{} makes it suitable for systematically performing variability observations for several brown dwarfs with different fundamental parameters (spectral type, inclination angle, gravity, and age) to reveal the trend of variability against those parameters.
A recently developed dynamic mapping technique \citep{2020ApJ...900...48K} makes it possible to explore the origin of the variability, namely time-varying cloud distributions, and further understand atmospheric circulation and cloud formation mechanisms in brown dwarf atmospheres. 
Here, we note that, owing to the rapid rotation of brown dwarfs with rotation periods of the order of hours \citep{2017AstRv..13....1B, 2018haex.bookE..94A}, the observation time required to monitor the global surface of a brown dwarf is relatively short.
Whereas the variability observation campaign for 44 brown dwarfs was conducted using Channels 1 and 2 (3.6 and 4.5 $\mu$m) of the \textit{Spitzer} Space Telescope \citep{2015ApJ...799..154M}, \jasmine{} can perform these observations at different wavelengths ($1.0$--$1.6$~$\mu$m). Note that some of those brown dwarfs have been observed with the Wide Field Camera 3 (WFC3; $1.1$--$1.7$~$\mu$m) of the \textit{Hubble Space Telescope} \citep[e.g.,][]{2014ApJ...782...77B}, including the investigation of the spectral variability for a few targets \citep[e.g.,][]{2013ApJ...768..121A, 2015ApJ...798L..13Y}. 
Because different wavelengths can probe different depths in the atmosphere, combining the observations by \textit{Spitzer} and \jasmine{} will provide comprehensive insight into the vertical structure of the atmosphere, including clouds.

\begin{figure}
    \includegraphics[width=\linewidth]{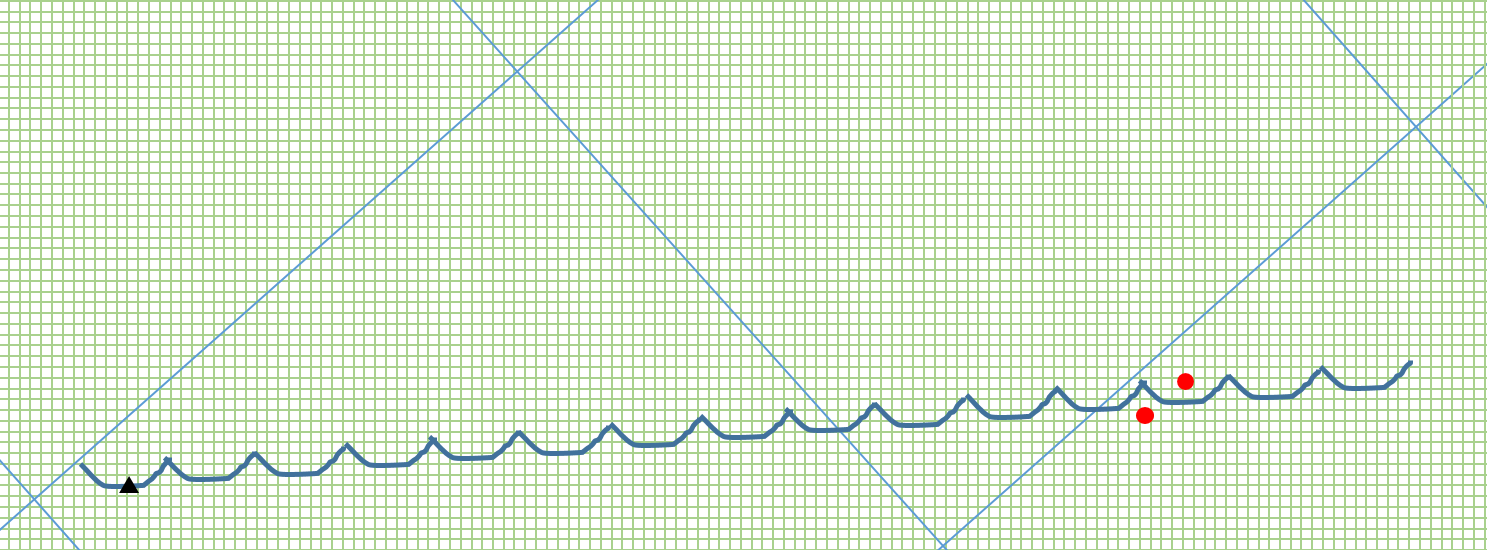}
    \caption{Luhman 16 orbit on the TESS (cyan , 21" pitch) and \jasmine{} (green, 0.4" pitch) detector pixel grids. The red dots correspond to a projected separation of the semimajor axis of Luhman 16AB. The triangle indicates the position, as observed by TESS in 2019 \citep{2021ApJ...906...64A}.}
    \label{fig:luhman16}
\end{figure}

In particular, \jasmine{} has an excellent capability for observing brown dwarf binaries.
Recently, \cite{2021ApJ...906...64A} observed the precise light curves of a nearby bright brown dwarf binary, Luhman~16 AB, using {\it TESS}.
Thanks to their long-term observation covering about 100 estimated rotation periods of Luhman 16B, they succeeded in extracting several minor modulation periods of 2.5, 6.94, and 90.8~hr in the observed light curve in addition to the dominant period of 5.28~hr.
They concluded that the 2.5 and 5.28~hr periods emerge from Luhman 16B possibly due to the atmospheric waves with half and one rotation periods while the 6.94~hr peak is likely the rotation period of Luhman~16 A.
As for the 90.8~hr period, they could not determine  which component it is originated from, but they tentatively attributed that modulation to the vortex in the polar regions.
\jasmine{} can resolve the semi-major axis of this system corresponding to 1.8 arcsec, while {\it TESS} was unable to separate them (figure \ref{fig:luhman16}).
Thus, \jasmine{} can confirm whether the 6.94~hr light curve modulation indeed corresponds to the rotation period of Luhman~16 A. In addition, if a long-term observation is possible, the component causing the 90.8~hr modulation will be identified, which will provide detailed insights into the atmospheric dynamics of the brown dwarfs.

Moreover, simultaneous observations with ground-based high-resolution spectrographs such as the Infrared Doppler \citep[IRD;][]{2018SPIE10702E..11K} mounted on the 8.2~m {\it Subaru} telescope for several best-suited targets will allow us to obtain more detailed surface maps of the temperatures and clouds for the observed brown dwarfs using the Doppler imaging technique, as done for Luhman~ 16 B \citep{2014A&A...566A.130C}.

\subsection{NIR Photometry by \jasmine{}} \label{subsec:NIRphoto}

Sections \ref{subsec: exojasmine}--\ref{subsec:BDvar} described the science cases that use precision NIR photometry by \jasmine{}. Here, we show how the required precision is attained. First, the photometric precision required to detect Earth-sized planets around M dwarf stars (\S 3.1) is $\sim 0.1\%$, which is less stringent than that achieved by {\it Kepler} and {\it TESS}. Detecting variability in young planets and brown dwarfs requires a similar photometric precision of 0.1\% (or milder). We first describe the statistical noise, including the shot noise, dark current, and readout noise of \jasmine{}. Next, we discuss the systematic noise caused by inter- and intra-pixel fluctuations of the detector sensitivity. Finally, we consider the extent to which astrophysical noise due to stellar activity can be suppressed using an NIR passband.

\subsubsection{Statistical Noise}

\begin{figure}
    \includegraphics[width=\linewidth]{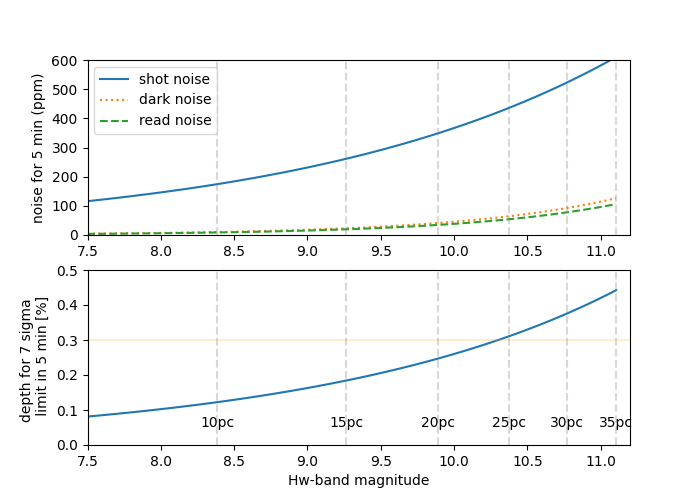}
    \caption{Statistical noise as a function of $H_\mathrm{w}$ magnitude. Top panel: 1 $\sigma$ level of shot, dark, and readout noises. Bottom: Transit depth corresponding to 7$\sigma$ detection during a 5-minute exposure (in terms of being able to measure the transit shape, the S/N at 5 minutes is shown). A telescope diameter of $36\,\mathrm{cm}$ was assumed (secondary telescope is 12.6 cm). The dark and read out noise are assumed to be 25 $\mathrm{e^-/s}$ and 15 $\mathrm{e^-}$/read. The limiting distances are computed for a $0.2 R_\odot$ star with a $3000\,\mathrm{K}$ blackbody spectrum.
   The horizontal yellow line indicates the typical depth of Earth-sized planets around a mid-M type star.
    \label{fig:noise_exo}
    }
\end{figure}

The top panel in figure \ref{fig:noise_exo} shows three statistical noise components: shot noise, dark current, and readout noise, as functions of the $H_\mathrm{w}$ magnitude. 
The stellar shot noise is most significant for exoplanet photometry targets.
The bottom panel shows the transit depth corresponding to a detection level of 7$\sigma$ in 5~minutes as a function of $H_\mathrm{w}$. 
In reality, a typical transit we search for lasts about $30$~minutes instead of 5~minutes, and so the limiting depth will be a few times smaller.
Consequently, one can search for a $\sim 0.2\%$ signal from a terrestrial planet around a $R_\star=0.2 R_\odot$ star that is brighter than $H_\mathrm{w}=11.5$~mag.  

\subsubsection{Impact of inter- and intrapixel sensitivity fluctuations on photometric precision}\label{ss:sensitivity} 

One main source of error in space-based precise photometry is an inhomogeneous detector sensitivity. {\it Kepler} and {\it TESS} avoid this systematic error by providing accurate attitude control for satellites. After its reaction wheel broke, {\it Kepler} lost precision in the attitude control to pin the stellar position in a single detector pixel. This resulted in $\sim1\%$ systematic error of the aperture photometry according to the satellite drift. These systematic errors can be corrected using several techniques, e.g., pixel-level decorrelation \citep{2016AJ....152..100L}. 
 To investigate the systematic errors in the detector sensitivity, we developed a detector simulator dedicated to \jasmine{}, called the \texttt{\jasmine{}-Image Simulator} (\texttt{JIS}) (Kamizuka et al. in prep). \texttt{JIS} simulates pixel images by incorporating both intra- and interpixel sensitivity models with attitude fluctuations. In the case of \jasmine{}, the point spread function (PSF) size was approximately a few times greater than the pixel size. This reduces the effect of the intrapixel fluctuation of the sensitivity, whereas the interpixel sensitivity is expected to be $\sim$1\% and it remains as a systematic error. We found that the systematic errors can be suppressed using a sufficiently sensitive detector to detect an Earth-sized planet around an M-type star after the flat correction of the interpixel sensitivity, as shown in section~\ref{subsec:eps_strategy}. In addition to precise measurements of the detector sensitivity prior to launch, we plan to have a single-mode fiber-type light source for the flat correction on board (Kotani et al. in prep). 

\subsubsection{Suppression of Flux Modulations by Stellar Activity}\label{ss:activity}

Enhanced stellar activity poses a challenge in the search for transiting planets around young stars. Most young stars are magnetically active, often creating dark spots and plages on their surface, which are revealed as strong periodic flux modulations in the light curves. For instance, the flux variation amplitude of the young M dwarf planet-host K2-33 is as large as $\approx 2\%$ \citep{2016AJ....152...61M} in the {\it Kepler} band. (In contrast, the transit depth of K2-33b is only $\approx 0.25\,\%$.) In the presence of such large flux modulations, together with the inflated radii of young stars, detecting ``small'' transiting planets (producing small transit depths) is not straightforward for young systems.

In addition, such variability can be an additional noise source, even for transit-signal detection in mature late-type stars. For the {\it TESS} passband (0.6--1.1 $\mu$m), there are approximately stellar variabilities of 400, 1000, and 4000 ppm on average for early to late M stars, corresponding to 0.35 $R_\odot$, 0.2 $R_\odot$, and 0.1 $R_\odot$ \citep{2021arXiv211208337B}, respectively. Considering the typical signal levels of terrestrial planets of 1200, 2500, and 10000 ppm for early to late M stars, flux modulation due to stellar activity can be a major source of noise when searching for the transit signal of terrestrial planets in the {\it TESS} band.

Photometric monitoring in the NIR region has an advantage over optical observations in terms of the reduced flux contrast of active surface regions (spots and etc.). Recent radial velocity measurements \citep[for example][]{2012ApJ...761..164C, 2019RNAAS...3...89B}, as well as photometric observations \citep{2021AJ....162..104M}, revealed that the contrast of typical surface spots on young active stars is mitigated by a factor of 2--3 in the NIR region. 
For reference, the expected relative modulation amplitudes by the stellar activity are listed in table~\ref{tab:comparison_satellites}, where {\it Kepler} passband case is used as a reference and set to be 1. The contrast was calculated by combining the PHOENIX atmospheric model and the response function for each passband in the same manner as in \citet{2021AJ....162..104M} when the stellar surface temperature and relative starspot temperature are 3500 K and 0.95, respectively. 
Miyakawa et al. (in prep.) implemented detailed simulations of the detection of close-in transiting planets around young cool stars (primarily M dwarfs). Their injection-recovery tests showed that the injected planets were recovered with enhanced rates in the NIR for all cases, and the recovery rate was significantly improved for more rapidly rotating stars ($P_\mathrm{rot}<$ a few days) with larger activity-induced modulations. 
This highlights the remarkable benefit of performing transit surveys by NIR photometry to identify small transiting planets around particularly young active stars, such as those in the Pleiades cluster ($\sim 100$ Myr) and star-forming regions ($\lesssim 10$ Myr). 

\subsection{Exoplanet Microlensing}
\label{sec:microlens}

Gas-giant planets, such as Jupiter and Saturn, are thought to have formed just outside the snow line, where icy materials are abundant. In this circumstance, a protoplanetary core grows quickly and starts accumulating the surrounding gas within the lifetime of the protoplanetary disk \citep[for example,][]{2004ApJ...604..388I}. However, the detailed process of gas-giant planet formation remains unclear. Unveiling the mass distribution of exoplanets outside the snowline is of particular importance in understanding this process. However, existing exoplanet detection techniques (except for microlensing) are insufficiently sensitive for detecting planets in this orbital region.

Microlensing detects planetary signals by observing the characteristic light curve produced by a background source star and altered by the gravitational lensing effect of the foreground planetary system. By monitoring hundreds of millions of stars in the Galactic bulge region, more than a hundred exoplanets have been detected using this technique. However, the ultrahigh stellar density of the field and the large distance to the planetary systems has made it difficult to further characterize each planetary systems. Although the {\it Nancy Grace Roman Space Telescope} (see section~\ref{sec:rst}) aims to improve the demographics of exoplanets substantially in the late 2020s by monitoring the Galactic bulge region \citep{2019ApJS..241....3P}, the difficulty of performing follow-up observations of each system remains.

This difficulty can be resolved if planetary systems are detected at close distances by microlensing. Although the event rate of such nearby planetary microlensing events is expected to be relatively small, the first such event was serendipitously discovered in 2017 \citep{2018MNRAS.476.2962N,2019AJ....158..206F}. High-cadence, large-sky-area surveys such as {\it All Sky Automated Survey for SuperNovae (ASAS-SN)}, {\it Zwicky Transient Facility (ZTF)}, {\it Tomo-e Gozen}, and {\it Vera C. Rubin Observatory's Legacy Survey of Space and Time (LSST)} have the potential to find more such events.

\jasmine{} can play an invaluable role in following up such nearby planetary microlensing event observations from the ground. Firstly, the high astrometric capability of \jasmine{} allows the centroid shift of the source star to be measured, thanks to the lensing effect (the signal is typically $\sim$1 mas). 
This will help to solve the degeneracy between the mass and distance of the lens system. Secondly, by simultaneously observing the same event from \jasmine{} and from the ground, one can measure the parallax effect in the microlensing light curve, which will also help solve for degeneracy. Thirdly, the NIR light curve from \jasmine{} allows the luminosity of the lens (host) star to be measured, thereby providing additional information about the host star in addition to its mass and distance (e.g., temperature). We note that although the GCS data will automatically provide all the required data, if events happen in the \jasmine{} GCS field, a full utilization of the advantages of \jasmine{} requires a target-of-opportunity (ToO) mode that can respond to a trigger within a few days. 

\subsection{Astrometric Planet Survey} \label{subsec: astrometric planet survey}

The astrometric detection of exoplanets has two key advantages over other exoplanet detection methods. First, the astrometric signal increases for planets more distant from their host stars, making this method complementary to the radial velocity and transit methods, which are sensitive only to short-period planets. Second, the two-dimensional motion of a star measured by astrometry, combined with Kepler's laws and an estimated stellar mass, allows a solution of both the absolute mass and the complete planetary orbit. This is generally not possible for exoplanets discovered based on radial velocities or microlensing. 

The astrometric signal (maximum shift of the stellar position by a planet) is given by 
\[
a_s \sim 30 (M_p/M_{\rm Jup}) (M_s/M_\odot)^{-1} (a / 3 {\rm au}) (d/100 {\rm pc})^{-1} \mu\mathrm{as},
\]
where $M_p$ denotes the planetary mass, $M_s$ the stellar mass, $a$ the semi-major axis, and $d$ the distance to the planetary system.
Although few exoplanets have been detected to date by this technique alone, {\it Gaia} is expected to detect tens of thousands of Jovian planets at a few au around solar-type stars using this technique \citep{2014ApJ...797...14P} because of its ultra-high astrometric precision at optical wavelengths ($\sim$25 \uas{} for stars with $G <$ 15~mag).  However, it remains difficult for {\it Gaia} to detect planets around ultra-cool dwarfs and/or long-period planets because of their lack of sensitivity. \jasmine{} can complement {\it Gaia} exoplanet exploration with its NIR capability and the long time baseline between {\it Gaia} and \jasmine{}. In the following subsections, we describe several scientific cases that \jasmine{} can pursue.

\subsubsection{Planets around ultra-cool dwarfs}

Core-accretion theories predict that massive planets are less abundant around lower-mass stars because of the lack of materials in the surrounding protoplanetary disks \citep[for example,][]{2005ApJ...626.1045I}. However, some Jovian planets have been discovered around mid-to-early M dwarfs, challenging current planet formation theories \citep[for example,][]{2019Sci...365.1441M}.
To further address this problem, it is important to 
unveil the planet population around further lower-mass stars or ultra-cool dwarfs (UCDs; $T_{\rm eff} \lesssim 3000$~K).
So far, only a limited number of planets have been discovered around UCDs owing to their faintness at optical wavelengths. 

\begin{figure}
    \begin{center}
    \includegraphics[width=8cm]{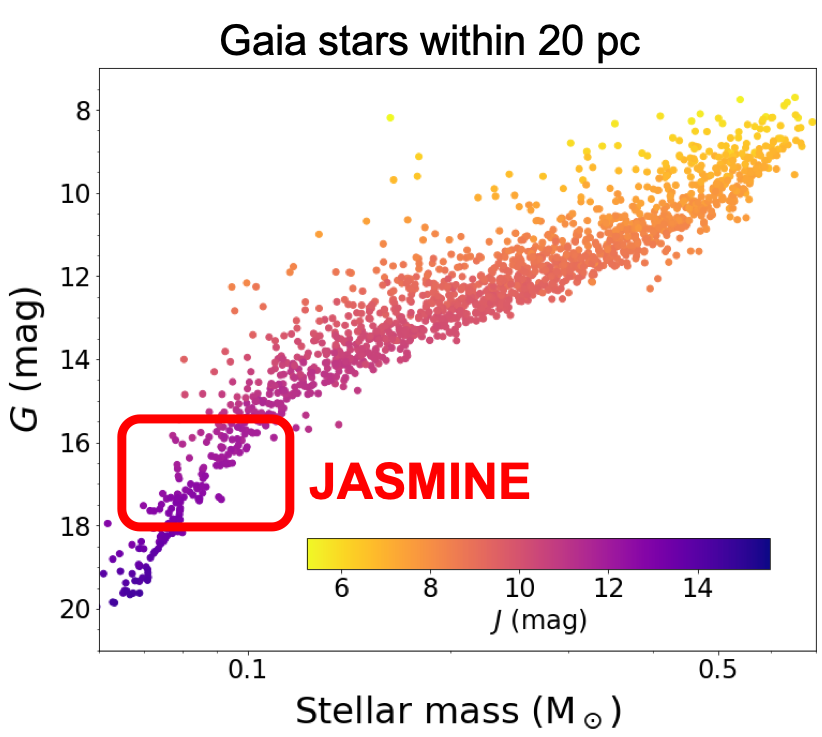}
    \end{center}
    \caption{The {\it Gaia} $G$-band ($y$ axis) and 2MASS $J$-band (colors) magnitudes of the stars within 20~pc as a function of the stellar mass ($x$ axis).  The UCD region to which \jasmine{} is sensitive to planets comparable to or higher than {\it Gaia} is indicated by the red box.
    \label{fig:Ms_vs_G}}
\end{figure}

The astrometry technique is well sensitive to distant planets around nearby UCDs because the astrometric signal increases inversely with the stellar mass. For example, the astrometric signal on a 0.1~$M_\odot$ star at a distance of 20~pc caused by a planet with a mass of 0.1~$M_{\rm jupiter}$ and an orbital period of 20~years reaches $\sim$500~$\mu$as. Because {\it Gaia} alone is likely difficult to firmly detect such planets owing to the faintness of nearby UCDs in optical ($G > $16 mag) and the limited time baseline ($\sim$10~years), \jasmine{} could play an essential role in confirming the candidates of such planets that will be detected by {\it Gaia} due to the NIR brightness of nearby UCDs ($J \sim$12--14 mag; see figure~\ref{fig:Ms_vs_G}), and the time baseline between {\it Gaia} and \jasmine{} (up to $\sim$18~years).

\subsubsection{Planet search in the Galactic Center Survey}

The data obtained by intensive observation with \jasmine{} toward the Galactic center (section~\ref{sec:gca}) will also be useful when searching for exoplanets using the astrometric technique. Despite the limited number of nearby stars within the \jasmine{} GCS area that are suitable for this planetary search, the expected astrometric precision toward this region (25~$\mu$as) will provide a good sensitivity to low-mass planets (down to $\sim$Neptune-mass) around M dwarfs. Potentially, about a thousand of M and K dwarfs that are bright enough for \jasmine{} are located within the GCS region (most within 1~kpc), providing good targets for astrometric planetary search.

\subsubsection{Synergy with radial velocity and direct imaging}

Astrometry has good synergy with radial velocity and direct imaging  techniques. Although the radial velocity technique is sensitive to long-period (tens of years) planets, this method alone cannot measure the orbital inclination. Thus, it cannot measure the true value but only the lower limit of the planetary mass. By measuring the astrometric signal of the host star of such a planet, one can determine the complete orbit and mass of the planet \citep[for example,][]{2019MNRAS.490.5002F}. \jasmine{} can play an important role in measuring the true mass of long-period planets (planetary candidates) around nearby M dwarfs discovered by radial velocity surveys, such as the one that is ongoing with the {\it Subaru}/IRD. Astrometry can also play a crucial role in determining the formation scenario of young self-luminous planets discovered by direct imaging. Although direct imaging can measure the complete orbit and luminosity of a planet, it alone cannot measure the dynamic mass of the planet, which is key to distinguishing different formation scenarios, i.e., hot and cold start models. Recently, the accelerations of host stars of several direct imaging planets were measured by combining the {\it Hipparcos} and {\it Gaia} proper motions, constraining their formation scenarios \citep[for example,][]{2019ApJ...871L...4D,2022MNRAS.509.4411D}. \jasmine{} will be able to contribute to such studies in combination with {\it Gaia}.


\section{Mission and Instrument Concept}
\label{sec:mission}

In this section we summarize the current concept study of the \jasmine{} mission and the survey plan to achieve the above mentioned main science objectives of the Galactic Center Archaeology and the HZ Exoplanet search. The mission and instrument concepts are still under development. Hence, the specifications summarized in this section are subject to change during the further development phases of the mission. 

\begin{figure}
 \includegraphics[width=\hsize]{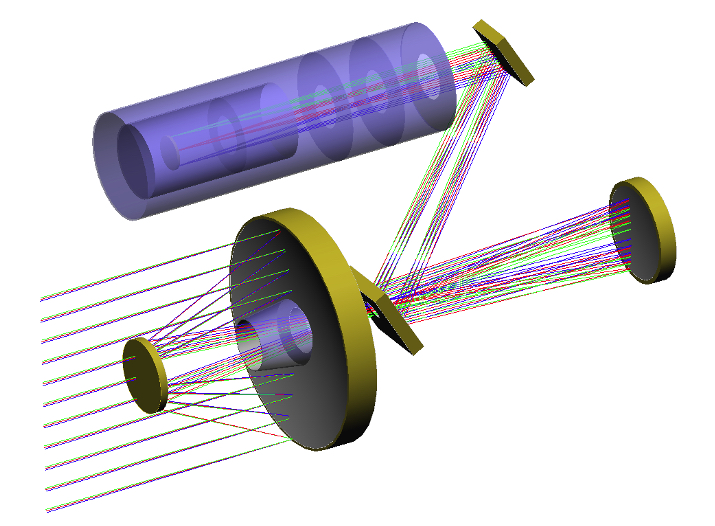}
    \caption{Preliminary optical design of \jasmine{}.}
    \label{fig:optics}
\end{figure}

\begin{figure}
 \includegraphics[width=\hsize]{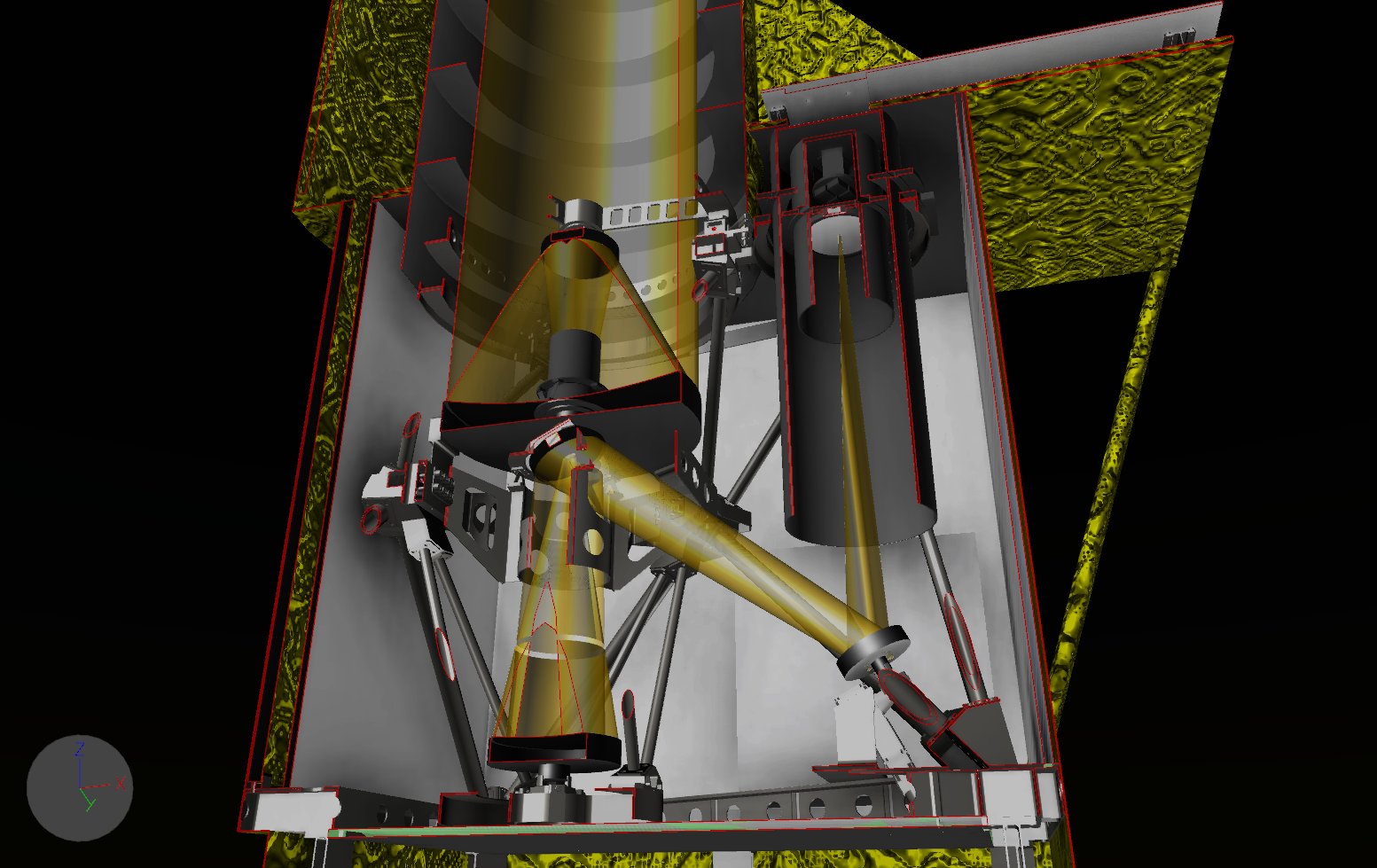}
    \caption{An example of schematic view of the payload layout.}
    \label{fig:payload}
\end{figure}

\subsection{Satellite and Payload Design}
\label{subsec:sat_payload}

The satellite system of \jasmine{} consists of a bus module and a mission module. The mission module includes a telescope and electronics box.
A large Sun shield and telescope hood are installed to prevent telescope from stray light and temperature change. The \jasmine{} satellite will be launched by JAXA Epsilon~S Launch Vehicle from the Uchinoura Space Center in Japan. \jasmine{} will be in a Sun-synchronous low Earth orbit at an altitude of about 600~km with a mission lifetime of 3~years. The weight of \jasmine{} is about 600~kg. 

\jasmine{} has a circular primary mirror with a diameter size of 36~cm. The current preliminary optical design adopts a modified Korsch system with three mirrors and two folding mirrors to fit the focal length (4.37~m) into the size of the payload (figure~\ref{fig:optics}). \jasmine{} is  planned to use CLEARCERAM\textsuperscript{\textregistered}-Z~EX for the mirror and Super-super invar alloy \citep{Ona+Sakaguchi+Ohno+Utsunomiya20} for the telescope structure. CLEARCERAM\textsuperscript{\textregistered}-Z~EX and the Super-super invar have extremely low ($0\pm1\times10^{-8}$ K$^{-1}$ for CLEARCERAM\textsuperscript{\textregistered}-Z~EX and  $0\pm5\times10^{-8}$ K$^{-1}$ for Super-super invar) coefficients of thermal expansion (CTE) at about 278~K, which is the operation temperature of the telescope. \jasmine{} uses four Hamamatsu Photonics InGaAs NIR detectors \citep[performance of a prototype model with a smaller format can be found in][]{Nakaya+Komiyama+Kashima+16}.
The FoV is $0.55^{\circ}\times0.55^{\circ}$. \jasmine{} uses an $H_\mathrm{w}$-filter which covers from 1.0~$\mu$m to 1.6~$\mu$m. Figure~\ref{fig:comp} shows the $H_\mathrm{w}$ passband.

The whole telescope of \jasmine{} is encased within the ``Telescope Panel Box" which insulates the telescope from outside (figure~\ref{fig:payload}). This enables precise control of the temperature variation of the telescope within 0.1~degree for 50 min, by keeping the temperature change inside the Telescope Panel Box within 1 degree. The operation temperature for the detector is required to be less than 173~K and the temperature variation is required to be kept within 1~degree level for 50 min. The structural thermal stability will keep the position of stars on the focal plane within 10~nm for 50~min. This is achieved by the above mentioned thermal control and the extremely low CTE of the CLEARCERAM\textsuperscript{\textregistered}-Z~EX mirror and Super-super invar telescope structure. The summary of the specifications of the satellite and telescope are shown in table~\ref{tab:spec}. 

\begin{table*}
\begin{center}
\begin{tabular}{ll}
 \hline
  Launch vehicle & JAXA Epsilon~S \\
 \hline
  Mass & $\sim600$~kg  \\
 \hline
  Volume & 1 m$^3$ \\
 \hline
  Power & 600 W \\
 \hline
  Orbit & 600~km Sun Synchronous Orbit \\  
 \hline 
  Downlink & Omnidirectional X-band \\
 \hline
   Mission lifetime & 3 years \\
 \hline
  Telescope & modified Korsch system \\
 \hline
  Aperture & 36~cm diameter \\
 \hline
  Focal length& 4.37~m \\
 \hline 
  Field of view & $0.55^{\circ}\times0.55^{\circ}$ \\
 \hline 
  Detector & Hamamatsu Photonics InGaAs 
  \\
 \hline 
  Number of detector & 4 \\
 \hline
  Number of pixels & $1952\times1952$ \\
 \hline
  Wavelength range & $H_\mathrm{w}$-band (1.0$-$1.6~$\mu m$) \\
 \hline 
  Pixel size & 472~mas \\
  \hline
\end{tabular}
\end{center}
\caption{
Summary of specifications of \jasmine{}.
\label{tab:spec}}
\end{table*}

\subsection{Galactic Center Survey Strategy}
\label{subsec:gcs_strategy}

The GCS is designed to achieve the science goals in section~\ref{sec:gca}. To achieve the precise astrometric accuracy, \jasmine{} will observe the same field repeatedly (about 60000 times). The GCS field covers a rectangular region of $-0.6^{\circ}<b<0.6^{\circ}$ and $-1.4^{\circ}<l<0.7^{\circ}$ (or $-0.7^{\circ}<l<1.4^{\circ}$) as described in section~\ref{sec:gca}.
With finite mission lifetime, the survey area on the sky is limited to ensure good astrometry, i.e., a sufficient number of repeat observations for each object. The \jasmine{} GCS is designed to cover the NSD (section~\ref{subsec:nsd}). The GCS covers both sides of the longitudes and latitudes of the central part of the NSD ($-0.6^{\circ}<b<0.6^{\circ}$ and $-0.7^{\circ}<l<0.7^{\circ}$). On the other hand, a compromise must be made for the outer part of the NSD, i.e., observing only one side of the disc, considering the apparent symmetry of the NSD.

\begin{figure}
    \includegraphics[width=0.95\hsize]{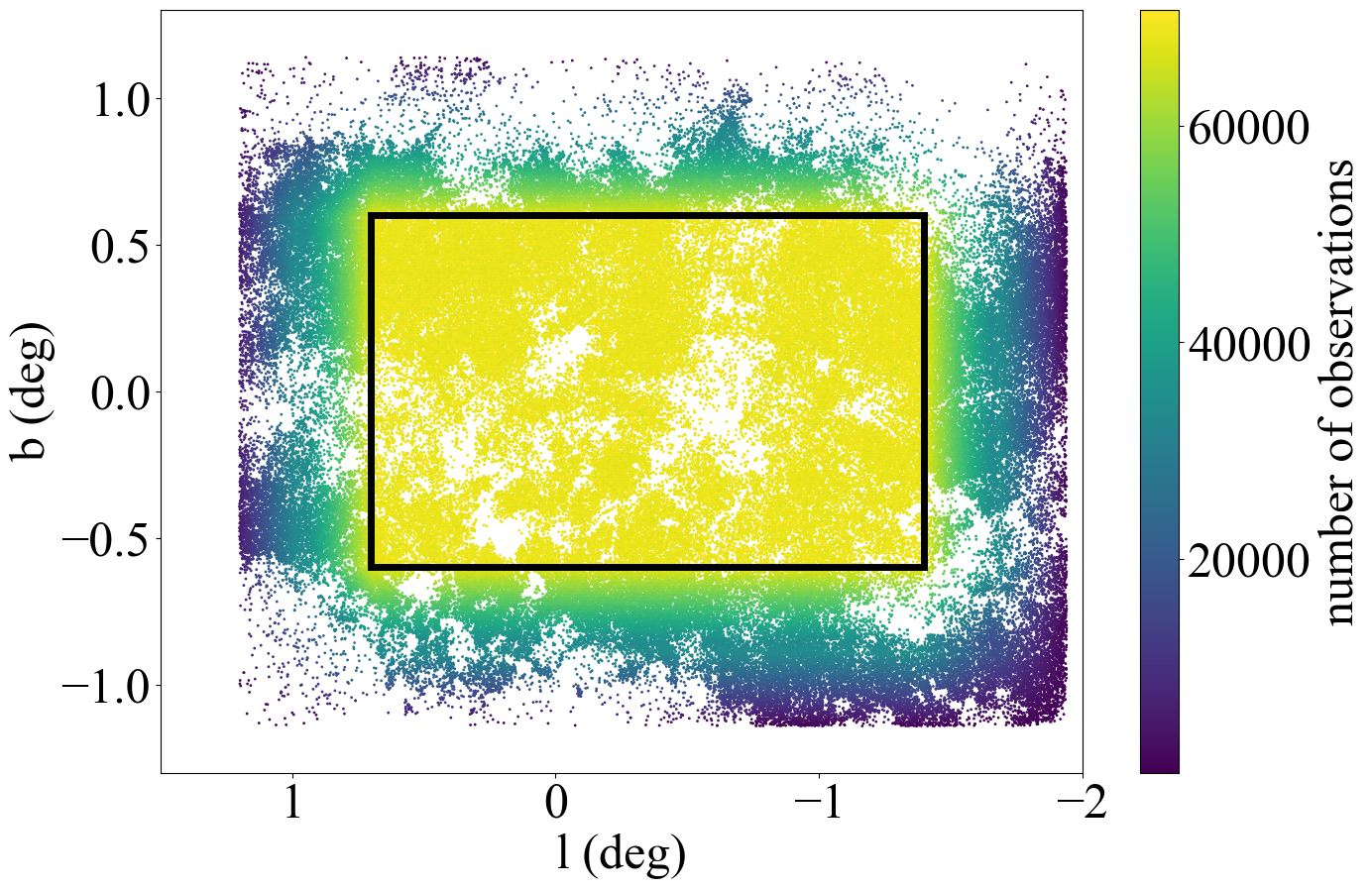}
    \caption{Number of the observations for each star observed towards the Galactic Center. The black rectangle highlights the \jasmine{} GCS field.}
    \label{fig:numobs}
\end{figure}

The whole GCS region will be mapped with a strategy to observe all the stars in this region for a similar number of times during the three years of the nominal operation period of \jasmine{} and detect each star at the different positions within the detector, to randomize the noise and reduce systematic biases. The expected number of observations for the stars in and around the GCS region is shown in figure~\ref{fig:numobs}. As can be seen from the figure, a number of stars surrounding the GCS region will be observed throughout the mission operation, though with a lower samping rate. Although the accuracy of the astrometry will be worse in this surrounding region, we will downlink the data for stars in this region and analyse their time-series astrometry and photometry. 

At each pointing of a single field of view of $0.55^{\circ}\times0.55^{\circ}$, \jasmine{} takes 12.5~sec exposure (plus 1~sec of read-out time) 46 times, which are combined to one frame which is hereafter defined as a ``small frame". The exposure of 12.5~sec corresponds to the saturation limit of $H_\mathrm{w}=9.7$~mag. Although the pixel resolution of each image is 0.472~arcsec, we will determine the centroid position of the stars with the accuracy of about 4~mas level for the stars brighter than $H_\mathrm{w}=12.5$~mag, using the effective Point Spread Function \citep[ePSF,][]{Anderson+King00}. 

The parallax accuracy of 25~\uas{} for the stars brighter than 12.5~mag is obtained by the repeated observation of about 60000 times in 3 years. In each orbit of about 97 min. \jasmine{} will observe 4 different small frames within 48\% of each orbit\footnote{The number of different fields observed per orbit is to be determined when the various trade-off studies are finalized.}. Two of these fields are overlapped about the half size of the field of view (either shifted vertically or horizontally, so that there will be two directions of the overlap) to correct the distortion of the images due to optical distortion and detector distortion. The observed fields in each orbit are chosen to map the main survey field homogeneously, but as random as possible. 
The distortions of the images are modelled with two dimensional 5th order polynomial functions by assuming the primary stars do not move in the short observational time in a single orbit. Here, the primary stars are defined as stars that do not have an intrinsic shift due to binaries or microlensing. As mentioned above, the telescope is stable enough to keep the position of stars on the focal plane to within 10~nm for 50~min, which allows us to ignore the time variation of the coefficients of the terms higher than the first order of the polynomial functions for the distortion correction, and maintain 10~$\mu$as stability of astrometric measurement. 
The time variation of the first order of the distortion, expansion and contraction, will be modelled with the stars whose parallax and proper motion is accurately measured with {\it Gaia}. 

\begin{figure}
 \includegraphics[width=\hsize]{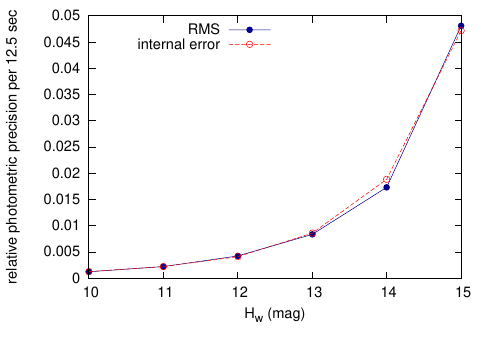}
 \caption{Estimation of the uncertainties per single exposure of 12.5~sec as a function of the magnitude of the stars. Red circles connected with the dashed line labeled as ``internal error" present the photometric errors computed from the Poisson noise and the read-out noise. The solid blue circle connected with the blue line labeled with ``RMS" show the photometric errors computed from the RMS error of the photometric measurement of 100 simulated images for each brightness of star created with JIS (Sec.~\ref{ss:sensitivity}).}
    \label{fig:gcs-photometry}
\end{figure}

The time-series photometry can be obtained from the GCS data. Figure~\ref{fig:gcs-photometry} shows the expected photometric uncertainties for a single exposure. 
The figure shows that the photometric accuracy is about 1~mmag for bright stars with $H_\mathrm{w}=10$~mag, but reaches to 40~mmag for the faint stars ($H_\mathrm{w}=14.5$~mag) in a small frame. Here, we assume that the inter- and intra-pixel sensitivity fluctuations are completely calibrated. The internal error in figure~\ref{fig:gcs-photometry} indicates the photometric errors computed from the Poisson noise and the read-out noise. ``RMS" shows the photometric errors computed from the RMS error of the photometric measurement of 100 simulated images created with \texttt{JIS} (section~\ref{ss:sensitivity}), which is almost consistent with the internal error. Each star is expected to be observed every $\sim$530~min. 

Because of the limited downlink capacity, only the data of $9\times9$ pixels around the target stars are planned to be donwlinked to the ground station. For these selected window regions, the data of every exposure, i.e., 46 images per small frame, will be downlinked, and will be used for astrometry and time-series photometry analysis.
However, we plan to downlink at least one full-frame image per each small frame, and make these data immediately available for the community. Such full-image data would be valuable for serendipitous discovery of any transient event. In addition, we can downlink all exposure data in a few additional pre-selected small regions. For example, \SgrA\ occasionally flares up in the near-infrared, and an extremely bright flare was observed in 2019. Converting from the reported magnitudes in $H$ and $K'$ bands in \citet{Do+Witzel+Gautam+19}, the observed peak magnitude of the 2019 flare corresponds to about $H_\mathrm{w}=18$~mag. This is out of reach for a single exposure of the \jasmine{} GCS. However, 46 stacked images per single pointing can detect such a flare. Depending on the scientific merit and the downlink capacity, we can downlink all exposure data of a small region around \SgrA\ and monitor the stacked flux.

\subsection{Exoplanet Survey Strategy}
\label{subsec:eps_strategy}

The field of view of \jasmine{} ($0.55^\circ \times 0.55^\circ$) makes it better suited to pointing--individual--target (PIT) observations than to a blind transit survey. Because the probability that an HZ planet transits a randomly chosen mid-M star is only $\mathcal{O}(1\%)$, and because the orbital period of the HZ planet is $\mathcal{O}(\mathrm{week})$, hundreds or more of such stars need to be monitored for at least a month to detect those planets in a blind transit survey. This is impractical with \jasmine{}. 

\begin{figure}
 \begin{center}
   \includegraphics[width=1.0\linewidth]{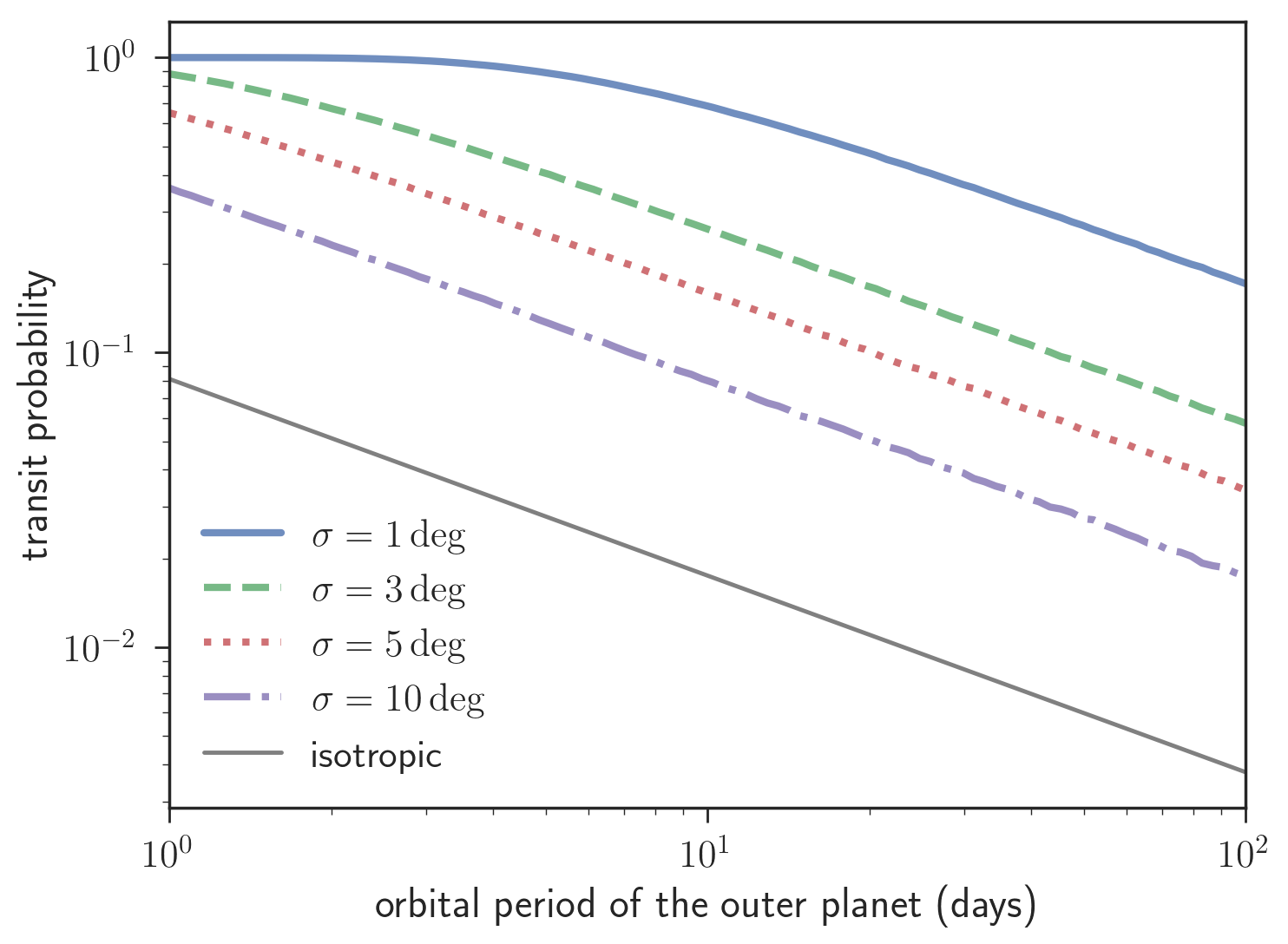}
 \end{center}
 \caption{Geometric transit probability of an outer planet, given the presence of an inner transiting planet, as a function of the different distributions of the mutual inclination of multiple planets. We assumed $M_\star=0.2 M_\odot$, $R_\star=0.2 R_\odot$, for Rayleigh distributions (with $\sigma$ as the mutual inclination distribution) and for an isotropic distribution. \label{fig:transitprobability}}
\end{figure}

Instead, \jasmine{} plans to focus on fewer ($\mathcal{O}(10)$) target stars that have a high prior probability of having transiting planets in the HZ. Specifically, we focus on stars with known transiting planets with orbital periods shorter than those in the HZ (i.e., $\lesssim 10\,\mathrm{days}$), which have been detected by ground- and/or space-based surveys prior to \jasmine{}. If these planets also have an outer planet in the HZ whose orbit is likely aligned with the inner transiting planet(s), the HZ planet is much more likely to transit than around a randomly chosen star: for mutual orbital inclinations of a few degrees, as inferred for multiplanetary systems from {\it Kepler} \citep{2014ApJ...790..146F}, the transit probability for the HZ planet conditioned on the presence of inner transiting planets is a few 10\% (figure \ref{fig:transitprobability}).
Such undetected HZ planets may be detected with a long-term follow-up monitoring using \jasmine{}.
This strategy makes it plausible to find HZ transiting planets by monitoring $\mathcal{O}(10)$ targets.
Indeed, the effectiveness of such a strategy has been demonstrated by {\it Spitzer} observations of TRAPPIST-1. They revealed five more terrestrial planets, d--g \citep{2017Natur.542..456G} on wider orbits covering the HZ than planets b and c, which were originally reported from a ground-based survey \citep{2016Natur.533..221G}.

\begin{figure}
\begin{center}
  \includegraphics[width=1.0\linewidth]{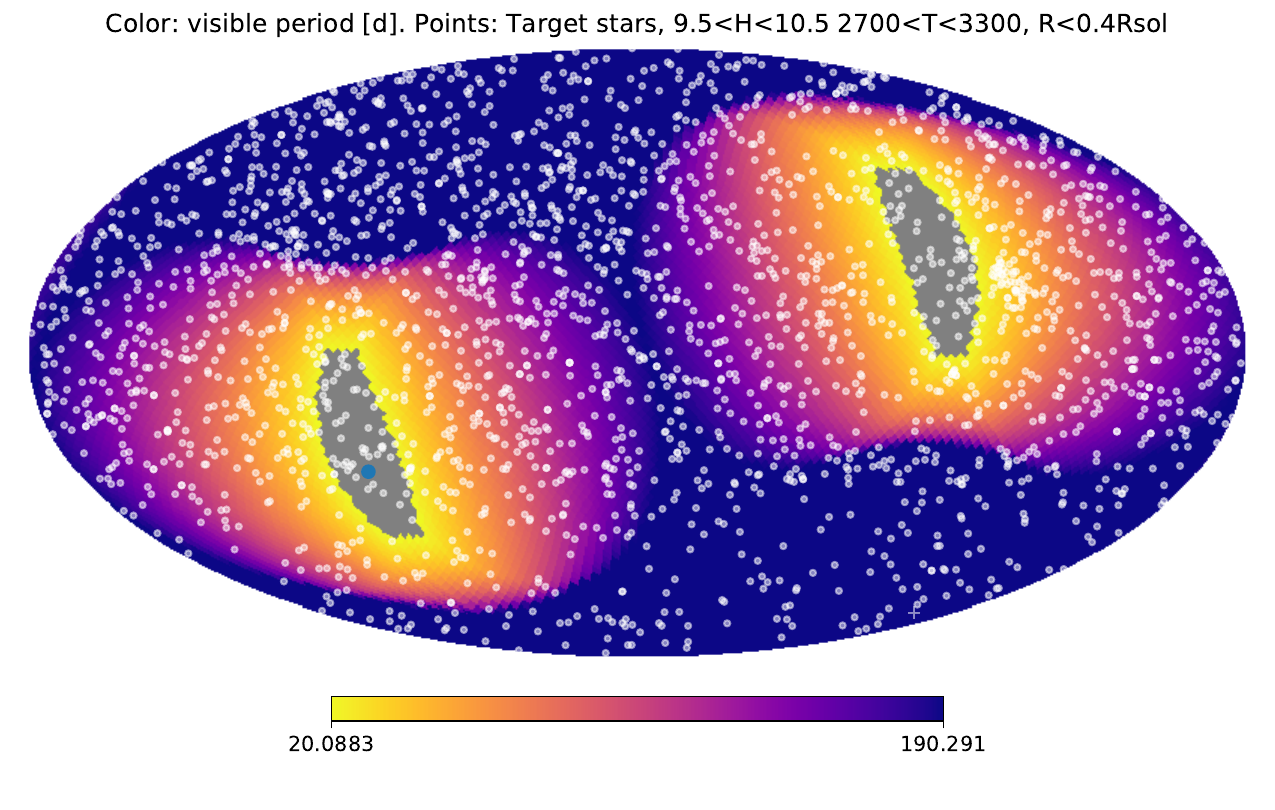}
\end{center}
\caption{Visibility of {JASMINE} on the equatorial coordinate system. The color indicates the total number of visible days per year. The small dots indicate the target M-type stars. The two gray areas correspond to the Galactic center and the anti-Galactic center, where \jasmine{} conducts an astrometric survey of the Galactic center. The blue dot indicates the Galactic center. \label{fig:vis}}
\end{figure}

To identify transiting planets in the HZ, \jasmine{} needs to monitor each star for at least few weeks in total, and the survey is planned to be performed during the periods when the Galactic center is not observable. 
Because \jasmine{} can observe the region within $45^\circ$ around the Sun, 
such observations are feasible for most stars in the sky (figure \ref{fig:vis}).
Here the white dots show the locations of potential target M dwarfs, the color corresponds to the total number of visible days per year, and the gray area corresponds to the region around the Galactic and anti-Galactic centers. Even excluding those latter regions, many mid- or late-M dwarfs can be observed with a sufficiently long baseline time.
In principle, the anti-Galactic center direction could also be observed for the exoplanet survey during the astrometric survey of the Galactic center. However, this may affect the thermal stability of the astrometric survey. A more detailed thermal stability analysis is needed to determine the visibility in the anti-Galactic direction. 

\begin{figure}
\begin{center}
  \includegraphics[width=1.0\linewidth]{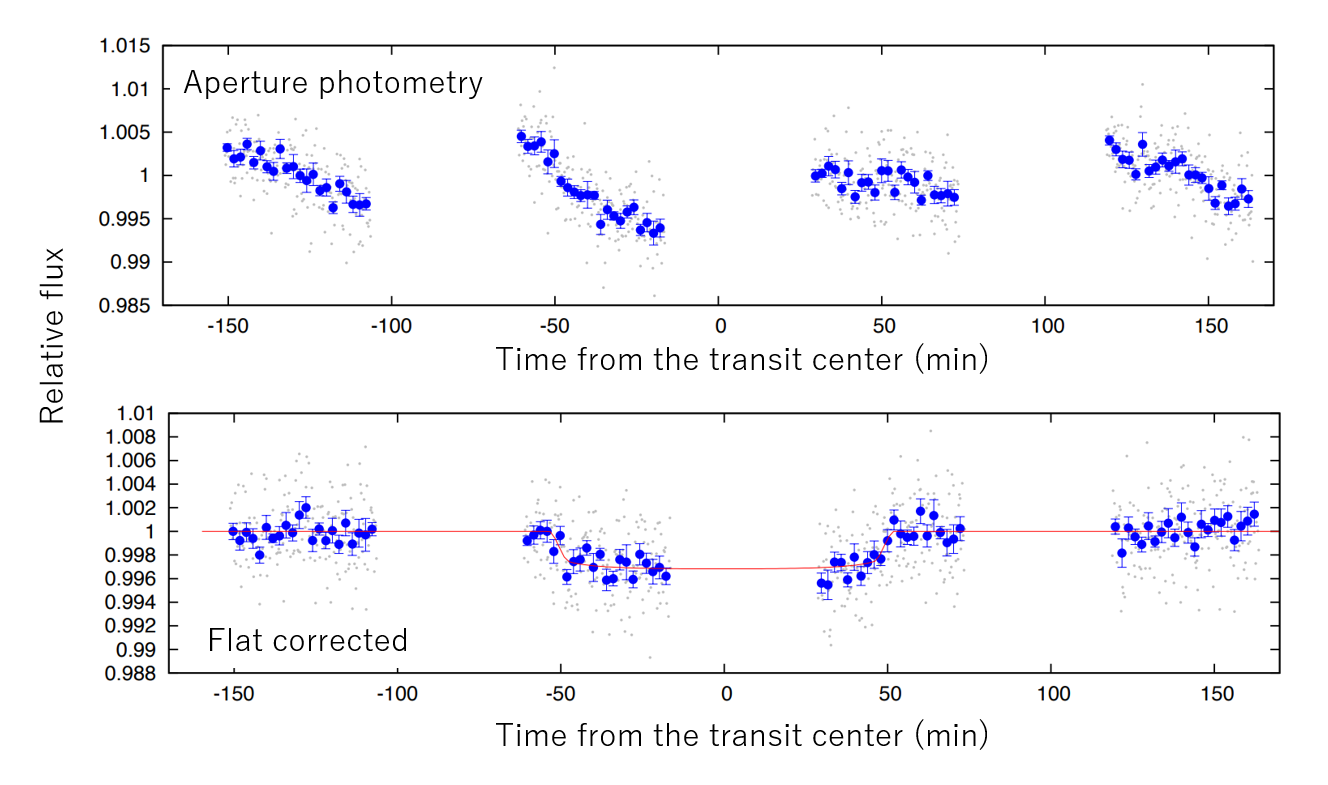}
\end{center}
\caption{Simulated transit signal assuming $R_\star=0.2 R_\odot$ and $R_p=1.2 R_\oplus$, $H_\mathrm{w}=10.5$~mag. The upper panel shows the light curve computed by simple aperture photometry of the simulated detector images. The detector images were computed using the \jasmine{} Image Simulator, which considers readout noise, dark current,  inter- and intrapixel sensitivity fluctuations, attitude control error, and pointing drift. The bottom panel shows a flat-corrected version of the top panel. Flat correction is applied by using an internal flat calibration ramp and/or self-calibrated flat using bright star cluster observations in orbit. \label{fig:exo_lc}}
\end{figure}

Figure \ref{fig:exo_lc} shows a simulated light curve of a mid- to late-M type star that includes a transit signal by a terrestrial planet. This simulation includes photon noise, dark current, readout noise, systematic error from intra- and interpixel sensitivity fluctuations, incorporating attitude control error of the satellite and a PSF of the optics with wavefront aberrations expressed by Zernike polynomials. An Earth-sized planet around a star with $R_\star = 0.2 R_\odot$ yields a transit signal of $\sim 0.2$--$0.3$\,\%, which can be detected in the simulated light curve of \jasmine{} after the flat correction of the interpixel sensitivity (see section \ref{ss:sensitivity} for details).

\section{Other Potential Science Cases}
\label{sec:other-science-cases}
\subsection{Galactic Mid-Plane Survey} \label{sec:gmps}

While the \textit{Gaia} mission is creating groundbreaking advances in the exploration of the structure and kinematics of the Galaxy as mentioned above, \textit{Gaia}'s contributions to understanding the large-scale dynamics of the youngest populations of the disk (``Extreme Population I'') will necessarily be limited, as even 1~$\deg$ of latitude is the scale height of these populations at a distance of about 5~kpc. Yet it is within this $\lesssim100$~pc from the Galactic mid-plane where is the main acting fields of the primary agents of disk dynamical evolution---e.g., Giant Molecular Clouds, the spiral arms, and the densest parts of the Galactic bar. To gather a complete understanding of processes such as secular orbital heating and radial migration that are driven by these perturbations requires contiguous kinematic information spanning the very youngest stars born in the Galactic midplane to their older siblings (well studied by other surveys) that have long since been scattered to dynamically hotter and/or radially migrated orbits.

Moreover, without accurate proper motions, even surveys like APOGEE, which is providing the first global view of the detailed chemistry and radial velocities of disk and bulge stars, are limited in their ability to place this information within a firm dynamical context; many thousands of mid-plane stars in the APOGEE database lack Gaia astrometry, so that 3D orbits are not possible to be inferred. Only a NIR astrometry facility can remedy this problem by supplying a similar global view of mid-plane stellar disk dynamics. The importance of accessing this ``optically hidden Milky Way'' has motivated discussions at ESA to follow-up {\it Gaia} with a NIR counterpart mission, {\it GaiaNIR} \citep[e.g.,][]{Hobbs+Brown+Hog+21}, but this likely will not be realized for a few decades.

\jasmine{} is well-suited to be a pathfinder for a proposed large flagship all-sky astrometric survey mission in the NIR, like the {\it GaiaNIR} concept, and is capable of having an immediate impact in the field of Galactic archaeology. In addition to the GCS in section~\ref{sec:gca}, to address the science objectives of the inner disk dynamics summarised in section~\ref{subsec:inner_disk}, tracing the dynamics and chemistry of the stars in the Galactic disk mid-plane at the various azimuthal angles of the disk is required. One potential targeting strategy is a small campaign of mosaicked \jasmine{} pointings, which are centered on the location of APOGEE $b = 0\deg$ fields, which are located every about $5~\deg$ in longitude around the entire sky. Such a strategy will instantly pay dividends through the value-added information on tens of thousands of stars with spectroscopy from the APOGEE and APOGEE-2 surveys, and their SDSS-V extension, \textit{Milky Way Mapper} (see section~\ref{subsec:sdss-v}), that will lack {\it Gaia}'s accurate astrometric data, because of the extreme foreground obscuration. These data will provide a systematic probe of the dynamics of the low latitude disk.

The required astrometric accuracy should not be as strict as the GCS. Although APOGEE spectroscopy yields $\sim0.1$~km~s$^{-1}$ accuracy radial velocities for stars, we do not require such accuracy in the transverse velocities. This is because we are primarily interested in measuring the velocity dispersions for young Population~I stars, which are of order 10--30~km~s$^{-1}$ per dimension.
For example,  200~$\mu$as~yr$^{-1}$ proper motions provided by \jasmine{} translates to a transverse velocity error of $< 5$~km~s$^{-1}$ for a star at 5~kpc (this is a bit higher than the approximate median distance of the APOGEE midplane stars, which are primarily red giants), and $< 10$~km~s$^{-1}$ for stars at 10~kpc. This relaxed requirement allows us to survey a large range of the Galactic longitude, when \jasmine{} is not observing primary science targets. 

\subsubsection{Star forming region} \label{subsec: sfregions}

A rigid adherence to the pointing strategy of the Galactic mid-plane survey is not necessary, and other interesting science problems--e.g., astrometric and photometric exploration of young stellar clusters and star forming regions (which naturally lie at low latitude) can also benefit from judicious placement of the \jasmine{} Galactic mid-plane Survey ``pickets''. For example, the APOGEE project has dedicated itself to intense spectroscopic probes of several star forming regions (e.g., the Orion Complex, the Perseus Molecular Cloud, and the Cygnus region), which may also be matched with \jasmine{} targeting.

Stars are mainly formed in Giant Molecular Clouds (GMCs). Moreover, star forming regions strongly concentrate into very compact sections within GMCs. Using the Two Micron All Sky Survey (2MASS) point source catalog, \citet{Carpenter00} estimated the fraction of young stellar populations contained within clusters to be 50\%--100\% for nearby cluster-forming GMCs, such as Perseus, Orion A, Orion B and MonR2. NIR surveys of young stellar populations using the {\it Spitzer Space Telescope} have confirmed that clustered star formation is the dominant mode of star formation in the Galaxy \citep[e.g.,][]{Poulton+Robitaille+Bonnell+08, Roman-Zunig+Elston+Ferreira+Lada08}.

This traditional picture is now challenged by the recent {\it Gaia} data releases, which are revealing more complex reality. \citet{Anders+Cantat-Gaudin+Quadrino-Lodoso+20} claimed that only about 16\% of the stars in the solar neighborhood formed in bound clusters, comparing the star formation rate in the solar neighborhood and the populations of the young star clusters. From the more small-scale kinematics of the OB associations, \citet{Ward+Kruijssen+Rix20} showed that the velocity fields of the OB association is highly substructured \citep[see also][]{Wright+Mamajek18}, which is not consistent with a monolithic scenario, where stars formed in the core of bound clouds and expanded subsequently due to the outflow of the gas caused by feedback. They discussed that these are consistent with a hierarchical star formation model, where stars formed in large-scale gravitationally unbound structures in molecular clouds.

Although the {\it Gaia} data is revolutionizing our understanding of star formation, these optical observations are unavoidably missing the information deep in the core of star forming regions where the dust extinction is too severe. The NIR astrometric observations with \jasmine{} and APOGEE spectroscopic data can unveil the whole picture of the star forming region hidden in the dust, which will complement what the {\it Gaia} data revealed and provide a more complete picture of these star forming regions.

\subsubsection{Milky Way Neighborhood Dwarf Galaxies} 
\label{subsec:mwdws}

The dwarf galaxy population around the Milky Way is diverse and new dwarfs are continuously being discovered \citep[e.g.,][]{2005ApJ...626L..85W, 2006ApJ...642L.137B,2006ApJ...643L.103Z}. The once troublesome “missing satellites problem” that plagued the $\Lambda$CDM cosmology theoretical framework is now steadily being refined and coming in line with observations of the dwarf galaxy populations around more massive galaxies such as the Milky Way \citep{2007ApJ...670..313S,2018PhRvL.121u1302K,2021ApJ...907...85M}. 
The ESA F-class mission {\it ARRAKIHS} (a planned 2030 lanuch) will further this inventory with unprecedented surface brightness levels in the {\it Euclid} VIS (0.550--0.900~\um{}), Y (0.920--1.230~\um{}) and J (1.169--1.590~\um{}) bands images of nearby galaxies.
One remarkable discovery that came from the recent Gaia DR2 is that of the giant dwarf galaxy Antlia~2 \citep{2019MNRAS.488.2743T}. The primary reason that this giant dwarf galaxy lurking in the dark matter halo of the Milky Way went undetected until now is that it lies only 11 degrees off the Galactic plane. Where Gaia has run into the limit of visual-wavelength extinction, \jasmine{} can go deeper into the Galactic mid-plane, as part of 
the Galactic mid-plane survey, 
allowing for serendipitous dwarf galaxy and globular/stellar cluster discoveries that cannot be made via any other method. 
With the anticipated astrometric precision and stellar content, we will be able to detect over-densities of stars that contain common proper motions indicative of low-latitude dwarf galaxies and/or clusters that are in the optical Zone of Avoidance, within a distance of about 10~kpc, as demonstrated for {\it Gaia} data in \citet{Antoja+Mateu+Aguilar+15} and \citet{Ciuca+Kawata+Ando+18}. So far, there is no dwarf galaxy found within 20~kpc, and Draco~2 at about 22~kpc is the closest \citep[e.g.,][]{McConnachie+Venn20}. Although the total field of view of the Galactic mid-plane survey is very limited, finding any galaxy within 20~kpc will be a significant discovery, and will be integral to studies of the survivability of dwarf galaxies within the inner Galaxy, their contribution to the bulge, and their impact on the Galactic disk \citep[e.g.,][]{DOnghia+Springel+Hernquist+10}.

\subsection{X-ray Binaries}
\label{sec: XRBs}

Another interesting target for which the precise astrometry with a short cadence observations of \jasmine{} can provide unique and impactful data is X-ray binaries, including gamma-ray binaries \citep{HESS+Aharonian+05AA442p1}. These are ideal laboratories for the study of high-energy astrophysics and prime future targets for multi-messenger astronomy, including continuous gravitational wave observations \citep[e.g.,][]{Middleton+Clearwater+Melatos+20}. Astrometric measurements of their companion stars enables us to measure the physical scale of the orbital parameters, and to unveil the mass of the compact object, whether it be a white dwarf, neutron star or BH \citep[][see also section~\ref{subsec:bhhunt_disk}]{Yamaguchi+Yano+Gouda18}. We listed below examples of X-ray binaries, which are particularly interesting targets for \jasmine{}, when \jasmine{} cannot observe the GCS field. This merely presents examples of potential targets, and is not an exhaustive list.

\smallskip

\noindent \textbf{$\mathrm{\gamma}$~Cassiopeia ($\mathrm{\gamma}$~Cas):} $\mathrm{\gamma}$~Cas is considered to be the first star identified as Be star (B~type star with emission lines) \citep{Secchi1866}.
However, now $\mathrm{\gamma}$~Cas is known to be a rare kind of Be star, which is characterised by a hard X-ray spectrum with a thermal X-ray emission, a high temperature ($>10$~keV) and a lack of strong variability \citep[e.g.,][]{2015ApJ...806..177M,Naze+Motch18,Tsujimoto+Morihana+Hayashi+18}. Despite the proximity of the object ($d\sim168$~pc) \citep{vanLeeuwen07} and several decades of observational and theoretical studies, X-ray emission mechanism and the nature of the lower mass secondary star are still debated \citep[e.g.,][]{Smoth19PASP131p4201,Langer+Baade+Bodensteiner+20}.
While the $\sim$204~day binary period with close to the circular orbit \citep{2002PASP..114.1226M,Nemravova+Harmanec+Koubsky+12} is known, the mass of smaller mass secondary star is not well measured and it is still debated if the secondary star is white dwarf or neutron star \citep{Langer+Baade+Bodensteiner+20}. 
With a visual magnitude of $\sim2$, $\mathrm{\gamma}$~Cas is too bright for {\it Gaia}. \jasmine{} can adjust the exposure time to observe such a bright object with a short cadence. Such time-series astrometric information will provide the precise orbital parameters and mass of the secondary star from astrometric observations \citep{Yamaguchi+Yano+Gouda18}, which will be a crucial to understanding the long-debated properties of $\mathrm{\gamma}$~Cas and the other similar systems ($\mathrm{\gamma}$~Cas analogs). 

\smallskip

\noindent \textbf{LSI +61 303 / HESS J0632+057:} These are both gamma-ray binaries, in which the source of gamma rays may be the impact of a relativistic pulsar wind on out-flowing protons in the disk of a Be star, UV photons from a massive main sequence star, or by the interaction of UV photons from such a star on the accretion disk of an X-ray binary counterpart \citep{2012Sci...335..175M}. Determining the masses of these companions can be achieved with high astrometric precision observations.

\smallskip

\noindent \textbf{X Per / V725 Tau:} Like $\gamma$~Cas, these are X-ray binaries with rapidly rotating B stars, and likely host neutron star companions. Neutron stars are characterized by their equation of state, which requires knowing their masses \citep[e.g.,][]{2010Natur.467.1081D}. This can be done with precise determination of orbital parameters \citep{2013arXiv1312.0029M}.

\subsection{Complementary Sciences with the Galactic Center Survey data} \label{sec:complementary}

The \jasmine{} GCS data will provide the accurate astrometry and time-series photometry for all the stars with $9.5~\mathrm{mag}<H_\mathrm{w}<14.5$~mag in the \jasmine{} GCS field. The GCS data should be valuable for the wide range of scientific studies, not just for the core science of \jasmine{} as shown in section~\ref{sec:gca}. In this section, we highlight some of these science cases. Note that the aim of this section is not to provide the comprehensive list, but merely list the potential science cases. We hope that many science cases that are not as premeditated will be developed by the wider science community. 

\subsubsection{Hunting Inner Disk BHs}
\label{subsec:bhhunt_disk}

Massive stars are expected to become BHs upon their demise \citep[e.g.,][]{Maeder92}. Therefore, it is expected that there are many stellar mass BHs floating around in the Milky Way \citep[e.g.,][]{Brown+Bethe94}. Stellar mass BHs are found in the Galaxy as X-ray binaries \citep[e.g.,][]{Ozel+10}, whose masses are around 5 to 20~$M_{\odot}$. Several gravitational wave detections of stellar mass BH binaries since the first detection of GW150914 by LIGO/VIRGO collaborations \citep{LIGOVIRGO+Abbott+16a} revealed that there are indeed stellar mass BHs, up to $M_{\rm BH}\sim150$~$M_{\odot}$ \citep{LIGOVIRGO+Abbott+20ApJ...900L..13A}. Two remaining questions are what is the mass function of this BH population and how they are spatially distributed in the Galaxy. These questions are also related to the origin of SMBHs, as discussed in section~\ref{sec:smbh}. 

One promising method to detect a large population of these stellar mass BHs is finding a binary system of a BH and a companion star bright enough to allow for its kinematics to be measured with astrometry and/or spectroscopic observations. These system does not require the companion star be interacting to the BH and be emit X-ray, i.e., non-interacting BH  \citep{Thompson+Kochanek+Stanek+19}. Such non-interacting BHs are expected to be observed by precise astrometry available with {\it Gaia} \citep[e.g.,][]{Penoyre+Belokurov+Evans22}. Recently \citet{Shikauchi+Tanikawa+Kawanaka22} estimated that {\it Gaia} will detect $\sim1.1$--46 non-interacting BH binaries \citep[see also][]{Kawanaka+Yamaguchi+Piran+17, Yamaguchi+Kawanaka+Bulik+Piran18}. In fact, so far from {\it Gaia}~DR3 and follow-up observations, two non-interacting BH binaries, {\it Gaia}~BH1 with a Sun-like star \citep{Chakrabarti+Simon+Craig+23,El-Badry+Rix+Quataert+Howard+Isaacson+23} and {\it Gaia}~BH2 with a red giant star \citep{El-Badry+Rix+Cendes+Conroy+Quataert+23,Tanikawa+Hattori+Kawanaka+22}, have been found.
\jasmine{} will offer similarly precise astrometry for stars in the inner disk from the Sun to the Galactic center, where {\it Gaia} cannot observe due to the high extinction. Therefore, \jasmine{} is expected to uncover the population of BHs in the inner Galaxy. 
According to a similar model of \citet{Shikauchi+Tanikawa+Kawanaka22}, in the \jasmine{} GCS region 100--1000 BH$-$star binaries are expected to exist. Further study of how many of such binaries can be detected by \jasmine{} is ongoing. 

Another way of hunting BHs with \jasmine{} is microlensing. \jasmine{} will offer the time-series photometry of the Galactic center region where the stellar density is very high. 
We expect that \jasmine{} will find about 3 microlensing events during the nominal operation of 3 years, which is an optimistic estimate based on the VVV microlensing survey results \citep{Navarro+Minniti+Pullen+Ramos20}. As suggested by \citet{Abrams+Takada20}, the long timescale ($>100$~days) microlensing events are expected to be dominated by a high mass ($\gtrsim30$~$M_{\odot}$) BH lens. Photometric microlensing itself does not give us the lens' mass. However, \jasmine{} can also detect astrometric microlensing from the centroid shift of the source \citep[e.g.,][]{Dominik+Sahu00,Belokurov+Evans02}. Astrometric microlensing enables us to measure the source mass if the mass and distance to the lens are optimum for the measurement. Recently, the first astrometric microlensing measurement has been reported for a microlensing event found by the ground-based observation and followed up by the {\it Hubble Space Telescope} for astrometry \citep{Lam+Lu+Udalski+22,Sahu+Anderson+Casertano+22}. However, the lens masses for the same event, OGLE-2011-BLG-0462/MOA-2011-BLG-191, reported by the two teams are so far quite different. \citet{Sahu+Anderson+Casertano+22} reported the lens mass of $7.1\pm1.3$~$M_{\odot}$, clearly indicating a BH, at distance of $1.58\pm0.18$~kpc, while \citet{Lam+Lu+Udalski+22} reported the lens mass of 1.6--4.2~$M_{\odot}$, which could be a neutron star or BH, at a distance of 0.69--1.37~kpc. This tension could be due to systematic uncertainty from the two independent measurements of photometry and astrometry.
The astrometric displacement of this event was larger than 1~mas and large enough to be clearly detected by \jasmine{}. \jasmine{} can provide both photometric and astrometric information, which may help to reduce the systematic uncertainty of such microlensing events. 
The chance of having such an event in the \jasmine{} GCS field in the lifetime of \jasmine{} could be slim, but one event of precise measurement of the high mass BH would still be an exciting outcome, because only few of such events are expected even in the {\it Gaia} data \citep{Rybicki+18}.

\subsubsection{Hunting IMBHs in the Galactic center}
\label{subsec:bhhunt}

A pressing mystery is the low number of confidently confirmed IMBHs $M_\mathrm{BH}=100$--$10^5$~$M_{\odot}$ \citep{2018ApJ...863....1C}.  Many candidates are, however, known \citep[e.g.,][]{2021ApJ...923..246G}, and they may form the low-mass ($M_{\rm bh}<10^5\,M_\odot$) extension to the quadratic BH/bulge mass scaling relation for disk galaxies \citep{Graham+Scott15}. 
The Galactic center is an attractive area to explore for these long-sought after IMBHs \citep[see][for a review]{Greene+Strader+Ho19}, especially if brought in through the capture of dwarf-mass galaxies.  Furthermore, 
\citet{PortegiesZwart+06} demonstrated that some massive star clusters formed in the central 100~pc undergo core collapse before the massive stars die, i.e., $\sim3$~Myr, which induces a runaway stellar merger and creates an IMBH. They estimated that within 10~pc from the Galactic center about 50 IMBHs may exist. Some of them could be still within the survived star clusters \citep[e.g.,][]{Fujii+Iwasawa+Funato+Makino08}, like the star clusters near the Galactic center, the Arches \citep{Figer+02} and the Quintuplet \citep{Figer+McLean+Morris99}. Detecting a star cluster in a high stellar density region like the Galactic center only with photometric data is difficult. However, the proper motion data from \jasmine{} will enable the detection of star clusters in the Galactic center (see also section~\ref{subsec:nsd}), which can inform follow-up studies of their cluster centers with X-ray and/or radio surveys \citep[e.g.,][]{Oka+Tsujimoto+Iwata+17, 2017ApJ...850L...5T}. 

Interestingly, so far five IMBH candidates have been discovered as high-velocity (velocity width $>50$~km~s$^{-1}$) compact ($<5$~pc) clouds in the Galactic center  \citep{Takekawa+Oka+Iwata+Tsujimoto+Nomura20}. The advent of the \textit{ALMA} enables the measurement of the detailed velocity structure of the compact clouds less than 0.1~pc from the center, which is consistent with Keplearian rotation around a massive object whose inferred mass between $10^4$ and $10^5$~$M_{\odot}$ \citep[e.g.,][]{Tsuboi+Kimura+Tsutsumi+19,Takekawa+Oka+Iwata+Tsujimoto+Nomura20}. 
Although further studies are required to prove that they are the true IMBHs, these observations may indicate that several IMBHs exist in the Galactic center. 

With \jasmine{}, IMBHs can be detected as a binary motion of bright stars around an IMBH or astrometric microlensing, as discussed in the previous section, if such systems exist or such an event occurs. For example, if a 1~$M_{\odot}$ AGB star is rotating around a 1000~$M_{\odot}$ BH with the orbital period of 3 years with zero eccentricity at a distance of 8~kpc, the semi-major axis of the orbit corresponds to 2.6~mas, which can be detected by \jasmine{}.

There will be an astrometric microlensing event if an IMBH crosses in front of a distant star. Following \citet{Toki+Takada21}, we can consider an event for a source star at 8~kpc, and a lens object of an IMBH with 1000~$M_{\odot}$ crossing at 7.5~kpc. The Einstein time-scale of this event is 713 days, and the maximum displacement due to the astrometric microlensing is 2.9~mas \citep{Toki+Takada21}. This can be detected by \jasmine{}, though such an event would be extremely rare. 


\subsubsection{Gravitational Waves}

Gravitational waves (GWs) have been successfully detected by the {\it Laser Interferometer Gravitational-Wave Observatory (LIGO)}, {\it Virgo} and the {\it Kamioka Gravitational Wave Detector (KAGRA)} collaborations. 
The sources for these events are merging compact objects such as BHs and neutron stars. 
It is of importance to detect gravitational waves from SMBH binaries to study the growth mechanism of SMBHs. These waves have much longer wavelengths than detectable by ground-based detectors. Astrometry could be a valuable resource to detect or constrain such low frequency gravitational waves \citep[e.g.,][]{Klioner18}.

Here, we estimate strain sensitivity of the \jasmine{} GCS. It is well known that the maximal magnitude of the astrometric effect of a gravitational wave is $h/2$ for $h$ being the strain. The astrometric accuracy of single observations of \jasmine{} being $\Delta\theta = 4$~mas for stars with magnitude $H_w<12.5$~mag and the uncertainty grows exponentially for fainter sources. Given that each star will be observed around $N_{\rm obs}=68000$ times, considering a realistically expected distribution of stars in $H_w$ magnitudes of the GCS and using the theoretical formulation developed for {\it Gaia}-like astrometry \citep{Klioner18}, one can conclude that the full sensitivity of \jasmine{} to the effects of a gravitational wave will be $h=3\times10^{-13}$. Here we assume that the instrument is ideally calibrated, so that the full accuracy scales as $N_{\rm obs}^{-1/2}$ for each source and the sensitivity is also accordingly computed from the combination of the contributions from individual sources.

However, the astrometric effect of a gravitationa wave is proportional to the sine of the angular separation $\chi$ between the directions of observations and that towards the gravitational wave source \citep{Book+Flanagan11,Klioner18}. Although \jasmine{} makes relative astrometry only, the variation of the astrometric effect within the observed field on the sky can be detected.  Therefore, the sensitivity quoted above should be scaled by $|\,\sin(\chi+f/2)-\sin(\chi-f/2)\,|=2\sin(f/2)\,|\,\cos\chi\,|$, where $f$ is the extension of the observed field, being $f\approx2^\circ$ for the \jasmine{} GCS.  Therefore, the theoretical sensitivity of \jasmine{} can be estimated $h=8.6\times10^{-12}\,|\,\cos\chi\,|$. Interestingly, the maximal sensitivity is reached for the gravitational sources approximately in the direction of observations or the opposite direction where $|\,\cos\chi\,| \approx1$.
This theoretical sensitivity is valid for the gravitational wave periods between the typical cadence of observations and the duration of observations by \jasmine{}.


\subsubsection{Ultra Light Dark Matter}
\label{uldm}

The precise measurement of the kinematics of stars revealed by \jasmine{} would enable us to reconstruct the mass distribution in the Galactic center \citep[e.g.,][see also sections~\ref{subsec: nsc} and \ref{subsec:nsd}]{Genzel+96,Chatzopoulos+15}. It is believed that in the central 100~pc of the Galaxy baryons dominate the mass profile, and the total mass measured from the dynamical model is consistent with what is expected from the stellar density profile \citep[e.g.,][]{2002A&A...384..112L,Fritz+16}. However, the Galactic center is attracting interests in testing for the existence of a particular dark matter candidate, namely Ultra Light Dark Matter (ULDM), including Axion-like ULDM particles \citep[e.g.,][for a review]{Ferreira+20}. Although ULDM behaves like conventional cold dark matter on the large scales, ULDM is expected to produce a soliton core in the galactic center in the de Broglie wavelength scale due to Bose-Einstein condensation. \citet{Schive+14} suggested that ULDM particle masses of $\sim8\times10^{-23}$~eV can explain the dynamical mass profile of the Fornax dwarf galaxy \citep[but see also][]{Safarzadeh+Spergel20}. \citet{Bar+18} suggested that the soliton core created from the dark matter particle mass less than $10^{-19}$~eV can influence the gravitational potential in the Galactic center significantly. \citet{DeMartino+20} showed that stellar velocity dispersion observed in the Galactic center implies a soliton core as massive as $\sim10^9$~$M_{\odot}$, expected from ULDM particles with $~10^{-22}$~eV. \citet{Maleki+20} also demonstrated that a soliton core corresponding to a particle mass of $\sim2.5\times10^{-21}$~eV explains the rotation curve of the Milky Way in the central region. \citet{Li+Shen+Schive20} showed that such a massive soliton core as suggested above can influence the kinematic properties of the nuclear gas disk on the scale of $\sim200$~pc. 

Recently, \citet{Toguz+Kawata+Seabroke+Read22} demonstrated that the kinematics of stars in the NSC can provide constraints on the particle mass range of ULDM. \citet{Toguz+Kawata+Seabroke+Read22} applied a simple isotropic dynamical model to the kinematics data of the NSC stars in \citet{Fritz+16}, and rejected the mass range of ULDM between $10^{-20.4}$~eV and $10^{-18.5}$~eV. \jasmine{} will provide the precise kinematics of the stars in the NSD (section~\ref{subsec:nsd}) which is the dominant stellar component from a few pc to $\sim200$~pc. This size corresponds to the size of the soliton core of $~10^{-19}$--$10^{-22}$~eV ULDM. Using the dynamical modelling of these stellar structures, the precise astrometric information of \jasmine{} may uncover indirect evidence of ULDM or provide stringent constraints on the existence of ULDM whose particle mass between $10^{-22}$~eV and $10^{-19}$~eV.


\subsubsection{Identifying disrupted globular cluster population}
\label{subsec:n-rich}

Recent observational studies of the Galactic bulge by APOGEE have discovered that a significant fraction of the bulge stars have unusually
high [N/Fe] \citep[e.g.,][]{Schiavon+Zamora+Carrera+17}. These N-rich stars are not found in the Galactic disk, but they are ubiquitous in globular clusters. Accordingly, one of the possible scenarios for the formation of the N-rich stars in the Galactic bulge is that the stars originate from globular clusters that had been completely destroyed by the strong tidal field of the Galactic bulge. Interestingly, these N-rich stars have been discovered in elliptical galaxies \citep[e.g.,][]{Schiavon07,vanDokkum+Conroy+Villaume+17}, which suggests that N-rich populations are common in galactic bulges and elliptical galaxies, i.e., not just in the Galactic bulge, in line with the indistinguishable properties of classical bulges and elliptical galaxies \citep[e.g.,][]{1999fgb..conf....9R,2009ApJS..182..216K,2010ApJ...716..942F}.

Globular clusters can spiral into the central region of the Galactic bulge due to dynamical friction \citep[e.g.,][]{Tremaine+Ostriker+Spitzer75}, and they can be more severely influenced by the tidal field of the bulge in the inner region. Accordingly, if such a globular cluster destruction scenario for the N-rich stars is correct, then stars from the destroyed globular clusters can be a  major population in the central region of the Galactic halo. In fact, using APOGEE DR16, \citet{Horta+Mackereth+Schiavon+21} estimated that N-rich stars contribute to about 17\% of the total halo stars at 1.5~kpc from the Galactic center \citep[see also][]{Fernandez-Tricado+Beers+Barbuy+22}. 
\jasmine{} will enable us to investigate the 3D spatial distributions and kinematics of N-rich (globular cluster origin) and N-normal halo stars in the central region through its superb accuracy in its proper motion measurement. Because the globular cluster origin stars could inherit unique kinematics different from the other halo stars, such 3D dynamics of N-rich stars will contribute to our understanding of the formation of the inner bulge. In APOGEE DR17 \citep{SDSSDR17+22}, there are 436 stars with the measured [N/Fe] and [Fe/H] 
in the \jasmine{} GCS field with good quality stars, i.e., STARFLAG$=0$, ASPCAPFLAG$=0$, SNR$>70$, 3250~K$<T_{\rm eff}<$4500~K and $\log g<3$ \citep{Kisku+Schiavon+Horta+21},
and 6 stars of them are N-rich stars ([N/Fe]$>0.5$, $-1.5<$[Fe/H]$<0.0$). All these stars are bright enough for \jasmine{} to observe. Therefore, it is promising that \jasmine{} will provide the proper motion of good number of N-rich stars in the Galactic center field in the combination with future high-resolution high-quality spectroscopic surveys of the Galactic center field, which will help to discover more N-rich stars. 

\subsubsection{Relics of Ancient Mergers} \label{subsec:ram}

An ancient galaxy merger of {\it Gaia}-Sausage-Enceladus discovered in the {\it Gaia} data (section~\ref{sec:intro}) leaves questions like ``where is the core of the remnant now?" and ``has the core of the progenitor galaxy reached to the Galactic center?". To assess the possibility of identifying such merger remnants in the \jasmine{} GCS, we again use APOGEE DR17, but apply a slightly different quality cut, i.e., STARFLAG$=0$, APPCAPFLAG$=0$, SNR$>70$, 3500~K$<T_{\rm eff}<$5500~K and $\log g<3.6$,
following \citet{Horta+Schiavon+mackereth+22} who used APOGEE DR17 to chemically characterise halo substructures of the likely accreted populations. We find that there are 284 APOGEE high-quality star data within the \jasmine{} GCS field. The {\it Gaia}-Sausage-Enceladus remnants occupies a distinct stellar abundance distribution in the [$\mathrm{\alpha}$/Fe]-[Fe/H] plane \citep[e.g.,][]{Haywood+DiMatteo+Lehnert+Snaith+Gomez18,Helmi+Babusiaux+Koppelman+18,Das+Hawkins+Jofre20}. Out of this sample, we find 4 stars within the abundances expected for the {\it Gaia}-Sausage-Enceladus remnants, i.e., stars with [Fe/H]$<-1.1$ and [Mg/Fe]$<-0.28$. All these stars are brighter than $H_\mathrm{w}=14.5$~mag. 

Note, however, that the APOGEE DR17 sample is not a complete sample up to $H_\mathrm{w}=14.5$~mag, but has a sample selection due to colors and/or the specific scientific targets. The \jasmine{} GCS will obtain the precise proper motion for about 1000 times more stars than present in the APOGEE data. Obtaining accurate proper motions and orbits of these potential remnant stars of the {\it Gaia}-Sausage-Enceladus interaction in the inner Galactic disk will allow studies to test the association with the already measured {\it Gaia}-Sausage-Enceladus remnants, which have so far been found exclusively in the solar neighborhood. 

\citet{Horta+Schiavon+Mackereth+21} found the Inner Galactic Structure (IGS) which has a similar chemical properties to the accreted components of the Galactic halo. They suggest that this could be a relic of an ancient accretion of another galaxy in the Milky Way earlier than the {\it Gaia}-Sausage-Enceladus merger and could be a more massive progenitor than {\it Gaia}-Sausage-Enceladus. Further studies with {\it Gaia}~DR3 and ground-based spectroscopic data \citep[e.g.,][]{Belokurov+Kravtsov22,Rix+Chandra+Andrae+22} argue that such centrally concentrated metal poor stars are relics of the ancient Milky Way proto-Galaxy, which could be mix of merger and in-situ populations from the early epoch of the Milky Way formation.  \jasmine{} can provide the proper motion of the stars in the Galactic center where {\it Gaia} cannot observe, and will help to identify the inner extension of the ancient populations.


\subsubsection{Origin of Hyper-velocity Stars}

\cite{1988Natur.331..687H} theoretically predicted 
that the SMBH at the Galactic center (\SgrA) ejects stars with extremely large velocities as a result of close encounter and disruption of stellar binaries near the SMBH. \cite{2003ApJ...599.1129Y} expanded upon the possible ejection mechanisms. The discoveries of young hyper-velocity stars (HVSs) in the halo \citep{Brown2005,Zheng2014,Huang2017,Brown2015ARAA, 2018AJ....156..265M, Koposov2019arXiv} confirmed this prediction. Among these discoveries, the most intriguing one is the A-type HVS dubbed S5-HVS1 \citep{Koposov2019arXiv}. Based on the astrometric data from {\it Gaia} and a follow-up spectroscopic observation, it turned out that this star was ejected from the Galactic center 4.8 Myr ago with the ejection velocity of $\sim 1800 \;{\mathrm{km\;s^{-1}}}$. Some numerical simulations suggests that the ejection rate of HVSs is around $10^{-5}$--$10^{-4}~\mathrm{yr^{-1}}$ \citep{Brown2015ARAA}. This ejection rate suggests that there are 1 to 10 HVSs within a sphere of radius $0.1$ kpc centered at the Galactic center, given their typical velocity $\sim 1000\;{\mathrm{km\;s^{-1}}}$.
Of course, what we can expect to observe with \jasmine{} is a tiny fraction of them, 
because they need to be bright enough to be detected by \jasmine{}. 
Given that the GCS area of \jasmine{} includes a square region of $\pm0.6^{\circ}$ around the Galactic center  ($0.6^{\circ}$ corresponds to about 0.09 kpc at the projected distance of 8.275~kpc), 
it is an enticing prospect to look for HVS candidates with \jasmine{}. If \jasmine{} discovers an HVS within $r<0.1$ kpc from the Galactic center, this will be very useful to understand the detailed mechanism of HVS ejection. For example, a HVS with a velocity of $1000 \;{\mathrm{km\;s^{-1}}}$ at $r=0.1$ kpc can be traced back to the Galactic center by integrating the orbit backward in time for just $0.1$ Myr. This means that we can probe the environment near the SMBH in the immediate past (just 0.1 Myr ago), 
such as the binary fraction near the SMBH or the orbital distribution near the SMBH. 
If \jasmine{} finds an HVS whose orbit is not consistent with the SMBH origin, it may indicate the existence of IMBH, which can produce an HVS, in the Galactic center.

\subsubsection{X-ray Sources and the Origin of the Galactic Ridge X-ray Emission}

The apparently extended hard ($\geq$ 2 keV) X-ray emission along the Galactic plane has been known as the Galactic Ridge X-ray Emission (GRXE) since the early 1980s \citep[e.g.,][]{Worrall1982}. This emission extends tens of degrees in Galactic longitude and a few degrees in Galactic latitude in $|l|<45^\circ$ and $|b|<1.5^\circ$. The GRXE has an integrated X-ray luminosity of $\sim$1$\times$10$^{38}$ ${{\rm erg}\ {\rm s}^{-1}}$ in the $2$--$10$~keV range \citep{Koyama1986, Valinia1998} at a distance of the Galactic center. The X-ray spectrum is described by a two-temperature thermal plasma ($\sim$1 and 5--10 keV) with the K shell emission lines \citep[e.g.,][]{Koyama1996, Yamauchi2009}; from neutral or lowly-ionized Fe at 6.4 keV (Fe\,${\rm{I}}$) as well as from highly-ionized Fe at 6.7~keV (Fe\,${\rm{XXV}}$) and 7.0 keV (Fe\,${\rm{XXVI}}$). 

It has been under intensive debate whether GRXE is a truly diffuse emission of a low surface brightness along the Galactic plane or a composition of discrete faint unresolved X-ray sources such as cataclysmic variables (CVs) and X-ray active stars \citep[e.g.,][]{Yuasa2012, Hong2012, Revnivtsev2006}. 
Many X-ray observations were carried out on this topic. In a Galactic bulge region ($l=0.08^\circ, b=-1.42^\circ$), $\sim$80\% of the diffuse X-ray emission was resolved into faint X-ray point sources using the deepest X-ray observation with the {\it Chandra} X-ray Observatory having an excellent spatial resolution of 0.5$\arcsec$. This indicates that the apparently diffuse emission in the Galactic bulge is primarily made of faint discrete X-ray sources \citep{Revnivtsev2009}. There are several candidates for such population of faint X-ray sources, including magnetic CVs \citep[e.g.,][]{Yuasa2012, Hong2012}, non-magnetic CVs \citep{Nobukawa2016} and X-ray active stars \citep[e.g.,][]{Revnivtsev2006}.

However, it is difficult to constrain the nature of these faint X-ray point sources from X-ray observations alone, because most of these sources are detected only with a limited number of X-ray photons (less than 10 photons) even with the deepest observations.
Thus, follow-up observations at longer wavelengths are needed. Because of the large interstellar absorption toward the Galactic plane, NIR observations are more
suited than optical observations. NIR identifications of X-ray point sources were performed along the Galactic plane \citep[e.g.,][]{Laycock2005, Morihana2016}, which provided clues to the nature of faint X-ray point source populations that make up the GRXE. The distance of these sources is unknown. Thus, classification of the sources is based on the X-ray to NIR flux ratio; high values suggest sources containing a compact object such as CVs and low values suggest sources otherwise such as stars. If the distance is obtained for many of these sources with \jasmine{}, we can discuss their nature based on the absolute luminosity both in X-ray and NIR bands and discriminate foreground contamination in the line of sight. A more robust classification of X-ray sources and their 3D distribution allow us to constrain the Galactic X-ray point source population for the different locations and components of our Galaxy, providing a hint to understanding the formation history of our Galaxy.

A large fraction of the observing fields of the GCS using \jasmine{} ($-0.6^{\circ}<b<0.6^{\circ}$ and $-1.4^{\circ}<l<0.7^{\circ}$ or $-0.7^{\circ}<l<1.4^{\circ}$ in section~\ref{sec:gca}),
was observed with {\it Chandra} \citep[the {\it Chandra} Multiwavelength Plane Survey; $-0.4^{\circ}<b<0.4^{\circ}$ and $-1.0^{\circ}<l<1.0^{\circ}$,][]{Grindlay2005}. A total of 9017 X-ray point sources were detected with a total exposure of 2.5 Ms \citep{Muno2009}. 
NIR identifications for these X-ray point sources were also made \citep{Mauerhan2009}. Based on this, we estimate that $\sim$600 X-ray sources will be identified in NIR brighter than 12.5 mag in the $H_\mathrm{w}$-band in the \jasmine{} GCS region. This is a significant improvement compared to the {\it Gaia}~DR3 optical identification and astrometric distances for $\sim$100 sources \cite[]{Gaia+Vallenari+23}, which are mostly foreground sources located within 2 kpc. This will be complemented with \jasmine{}.


\subsubsection{Observations of small solar system bodies}

As solar system bodies are moving objects, they are good targets for astrometry and time-series photometry. Precise astrometry improves the orbital elements of small solar system bodies such as comets and asteroids. It provides a solid foundation in several fields. Risks of minor bodies that threaten the Earth (potentially hazardous asteroids) can be precisely assessed. Non-gravitational effects such as the Yarkovsky effect can be quantitatively measured. An asteroid family, asteroids derived from the same parent body, can be identified. Astrometry of interstellar objects such as 1I/'Oumuamua and 2I/Borisov is essential to understand their origins. The rotation periods and shapes of minor bodies are derived from time-series photometry, leading to their internal structure estimates (bulk density). A binary system can be identified if it shows an eclipse or mutual event. Time-series photometry is also useful for tracking the brightness changes of active asteroids. Since \jasmine{} is in a Sun-synchronous polar orbit, \jasmine{}'s photometry will be complementary to ground-based observations. Finally, non-targeted, serendipitous surveys provide opportunities to discover new minor bodies.

The expected number of small solar system bodies via \jasmine{} is estimated as follows. Here, we focus on asteroids, the most abundant objects among minor bodies detectable by \jasmine{}. The spectral energy density of an asteroid is generally dominated by two components: reflected sunlight in optical wavelengths and thermal emission in infrared wavelengths. \jasmine{}'s $H_\mathrm{w}$-band is located at transitional wavelengths between the two components. Hence, unfortunately, the minor bodies are fainter in $H_\mathrm{w}$-band, and it is more challenging to detect them with \jasmine{}. To evaluate the observability of asteroids, we propagate the positions of known asteroids and check if they cross the \jasmine{} GCS observing region. The orbital elements of asteroids were retrieved from Lowell Minor Planet Services operated by Lowell Observatory on 26 August 2022. Objects with large uncertainties were removed. The total number of objects was 1192756, which includes 1148593 Main Belt Asteroids, 28829 Near Earth Asteroids, 11458 Jupiter Trojans, and 3876 Trans-Neptunian Objects. The topocentric (geocentric) coordinates were calculated from 1 January 2028 to 31 December 2031. For the sake of simplicity, \jasmine{}'s observing region was defined as a circle with a radius of 0.7 degrees centered at $(l, b) = (359.9^{\circ}, 0.0^{\circ})$ in Galactic coordinates. The defined region differs from the current baseline of \jasmine{} GCS, but the number of observable objects is not significantly affected. With the distances, absolute magnitudes, and slope parameters, the apparent magnitudes of the bodies in the $V$-band can be calculated \citep{Bowell+Hapke+Domingue+89}. We then assume that the $V-H_\mathrm{w}$ color for the objects is the same as the Sun, i.e., $(V-H_\mathrm{w})_{\solar}\sim1.21$, and convert the $V$-band to $H_\mathrm{w}$-band magnitude, assuming that all the asteroids have flat reflection spectra.

\begin{figure}    
    \includegraphics[width=.9\linewidth]{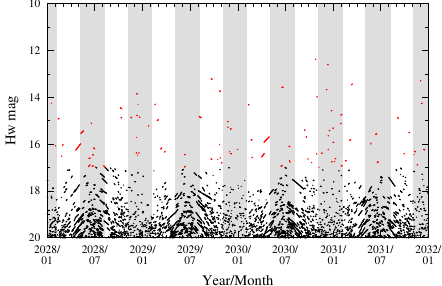}
    \caption{The brightnesses of asteroids crossing the \jasmine{} GCS observation field, which is approximated with a 0.7~deg radius circle region from $(l, b) = (359.9^{\circ}, 0.0^{\circ})$ . The asteroids brighter than $17.0$~mag are shown in red. The gray shaded regions show the seasons when the Galactic center is not accessible by \jasmine{}.}
    \label{fig:sssb:timeline}
\end{figure}

\begin{figure}
    \centering
    \includegraphics[width=.9\linewidth]{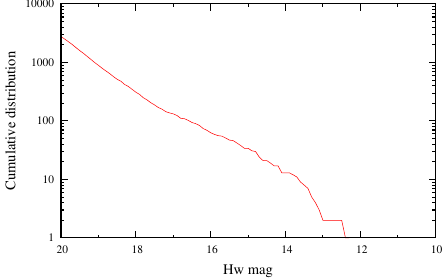}
    \caption{The cumulative distribution of the $H_\mathrm{w}$ magnitude for asteroids observable throughout the JASMINE operation period.}
    \label{fig:sssb:histogram}
\end{figure}

Figure \ref{fig:sssb:timeline} shows the brightnesses of asteroids crossing the  \jasmine{} GCS region at different epoch. Each segment shows an individual asteroid. Due to the operation constraint of the satellite, \jasmine{} does not observe the Galactic center in the gray-shaded seasons. The length of the segment represents the observable duration, which depends on the relative motion to \jasmine{}. A handful of asteroids can be observed at apparent magnitudes brighter than $H_\mathrm{w}=14.5$~mag (sufficient for astrometry) and $H_\mathrm{w}=17.0$~mag (for photometry). Figure \ref{fig:sssb:histogram} illustrates the cumulative histogram of the $H_\mathrm{w}$ magnitude for observable asteroids. The numbers of objects brighter than $H_\mathrm{w}=14.5$ and 17.0~mag are about 10 and 100, respectively, throughout the operation period. The expected number of the potential targets for astrometry in the \jasmine{} GCS is rather small. Thus, it is unlikely that there are serendipitous astrometric measurements with \jasmine{}, while targeted observations for known objects to provide additional astrometric information are preferred. 

Taking advantage of accurate photometry, \jasmine{} may observe occultation events by minor bodies. Occultation provides valuable information for shape modeling and binary search. Since the brightness of a minor body does not matter in an occultation observation, the number of potential targets significantly increases. Detecting an occultation event is feasible with accurate orbital elements and dedicated observation planning. Serendipitous observations of occultation events may detect new objects that are unreachable even by large telescopes with apertures of $\sim10$~m. \citet{Schlichting+Ofek+Wenz+09} claimed an occultation event by a Trans-Neptunian Object with a radius of 500~m at 45~au in archival data by the {\it Hubble Space Telescope}'s Fine Guidance Sensors. \citet{Arimatsu+Tsumura+Usui+19} detected an occultation event by a Trans-Neptunian Object with a radius of 1.3~km using a coordinated observation system of multiple low-cost commercial off-the-shelf 0.28~m aperture telescopes, the Organized Autotelescopes for Serendipitous Event Survey \citep[OASES,][]{Arimatsu+Tsumura+Ichikawa+17}. Occultation events sometimes reveal additional features of satellites and rings. Rings around (2060) Chiron and (10199) Chariklo were identified by ground-based occultation observations \citep[e.g.,][]{Ortiz+Duffard+Pinilla-Alonso+15,Braga-Ribas+Sicardy+Ortiz+14}. Recently, a ring beyond the Roche limit was discovered around (50000) Quaoar \citep{Morgado+Sicardy+Braga-Ribas+23}. High precision photometry with \jasmine{} has the potential to detect minor features in light curves \citep{2022A&A...664L..15M}. As for the serendipitous survey, careful assessment of false detection is required. OASES adopted simultaneous observations with multiple telescopes, to minimise the false detection rate. \jasmine{} may be able to detect occultation events after all anomalous signals are suppressed by careful calibration. 


\section{Synergies with the other projects}
\label{sec:other_proj}

In this section, we summarize the Galactic stellar survey projects complementary to \jasmine{} and planned to be operating in late 2020s. Although there are many projects relevant to the \jasmine{} science cases, here, we only list the projects and/or new instruments more relevant to the \jasmine{}'s main survey targets of the GCS and the Exoplanet Survey, targeting M dwarfs. 

\subsection{Ground-Based Surveys}

\subsubsection{\it PRIME}
\label{subsubsec:prime}

The Prime Focus Infrared Microlensing Experiment ({\it PRIME}) is a wide field (1.56~deg$^2$) 1.8~m telescope at the South African Astronomical Observatory, which is planned to operate from 2023. {\it PRIME} is jointly managed by Japan, the USA and South Africa. {\it PRIME} has $z, Y, J$ and $H$-band filters and several narrow-band filters. 
{\it PRIME} will observe the Galactic center region around $-3^{\circ}<l<3^{\circ}$ and $-2^{\circ}<b<2^{\circ}$, which covers the whole area of the \jasmine{} GCS field. The prime target of {\it PRIME} is a microlensing exoplanet search. 
 Joint observations with \jasmine{} can help to constrain the parameters of the exoplanet detection further with the additional accurate astrometry information of the source stars that \jasmine{} can provide. 
The time-series photometry of {\it PRIME} will also find many variable stars. As mentioned above, the {\it PRIME} data will be used to provide the catalog of Miras observable with \jasmine{}. 

\subsubsection{{\it Vera C. Rubin Observatory}/Legacy Survey of Space and Time (LSST)}

The {\it Vera C. Rubin Observatory} is located on the Cerro Panch\'on ridge in Chile, and will run 
the ten-year Legacy Survey of Space and Time (LSST) with an 8.4~m (6.5~m effective) Simonyi Survey Telescope \citep{Ivezic+19}. The {\it Rubin Observatory} LSST Camera will have a 3.5-degree field of view with about 32 gigapixels with 0.2~arcsec sampling pixel size. There will be six filters ($u, g, r, i, z$ and $y$) covering 320--1050~nm. The survey is planed to begin in 2024, and the main survey will observe 18000~deg$^2$ region of the sky about 800 times in the planed duration of 10~years. The co-added map will reach to $r\sim27.5$~mag, and it is anticipated to detect about 20 billion stars and a similar number of galaxies. The main science drivers of LSST are probing the properties of dark energy and dark matter, cataloging an inventory of the solar system, exploring the transient optical sky and mapping the Milky Way. The survey field covers the Galactic bulge and the Galactic center. The LSST is capable of providing the astrometric measurements for fainter stars than possible with {\it Gaia}. With the 10 year baseline, the expected uncertainties of parallax and proper motions are respectively $\sigma_{\rm \pi}=0.6$~mas and $\sigma_{\rm \mu}=0.2$~mas~yr$^{-1}$ for the stars brighter than $r=21$~mag, and $\sigma_{\rm \pi}=2.9$~mas and $\sigma_{\rm \mu}=1.0$~mas~yr$^{-1}$ for the stars brighter than $r=24$~mag.  In addition, the time-series photometry of the LSST will help to find many variable stars and microlensing events. The majority of them will be too faint for \jasmine{} to follow up. However, if they are bright enough and in the same field, \jasmine{} can provide more accurate astrometric information.

\subsubsection{SDSS-V}
\label{subsec:sdss-v}

The Sloan Digital Sky Survey (SDSS)-V \citep{Kollmeier+17} is an ambitious project to run an all-sky multi-epoch spectroscopic survey, utilising telescopes in both Northern and Southern hemispheres. The survey will provide optical and IR spectra covering 2500~deg$^2$, ultra-wide field for more than 6 million objects in five years (2020--2025). SDSS-V uses the telescopes at Apache Point Observatory (APO) in USA and and Las Campanas Observatory (LCO) in Chile. At APO, 2.5~m {\it Sloan telescope} will be continuously used for SDSS-V full-time. At LCO, more than 300 nights per year of telescope time of 2.5~m {\it du Pont telescope} will be dedicated to this survey. The survey will also use the smaller (1~m to 16~cm) telescopes at APO and LCO. The NIR APOGEE spectrograph (300 fibers, $R=22000$, $\lambda=1.5$--1.7~$\mu$m), the eBOSS optical spectrograph (500~fibers, $R\sim2000$, $\lambda=0.36$--1.0~$\mu$m) and the MaNGA multi-object IFU ($R\sim4000$, $\lambda=0.36$--1.0~$\mu$m) will be used at both APO and LCO. SDSS-V will run three surveys, the Milky Way Mapper, the Black Hole Mapper and the Local Volume Mapper. The most relevant survey to \jasmine{} is the Milky Way Mapper, which plans to observe 4--5 million stars in the Milky Way, with the NIR APOGEE spectrograph and/or the optical BOSS spectrograph. The Milky Way Mapper aims to understand the evolution of the Milky Way, the physics of the stars and the interstellar medium as well as multiple stars and exo-planetary systems. The Galactic Genesis Survey as a part of the Milky Way Mapper targets the stars with $H<11$~mag and $G-H>3.5$~mag, which are likely to overlap with the bright target stars of the \jasmine{} GCS fields, and will provide accurate radial velocity and abundance patterns. 

\subsubsection{{\it Subaru}/PFS}

{\it Subaru} Prime Focus Spectrograph \citep[PFS;][]{Takada+14} is the next generation instrument of the 8.2~m {\it Subaru} telescope at the summit of Maunakea, Hawai'i in the US, operated by National Astronomical Observatory of Japan. PFS is a joint instrument of the institutes in Japan, Taiwan, the USA, France, Brazil, Germany and China. PFS has $\sim1.38$~deg$^2$ field of view, and about 2400 science fibers. PFS consists of blue ($\lambda=0.38$--0.65~$\mu$m, $R\sim2300$), red (low resolution mode: $\lambda=0.63$--0.97~$\mathrm{\mu}$m, $R\sim3000$; medium resolution mode: $\lambda=0.71$--0.885~$\mathrm{\mu}$m, $R\sim5000$) and NIR ($\lambda=0.94$--1.26~$\mathrm{\mu}$m, $R\sim4300$) spectrographs. It is scheduled to start operating in 2024. About 300 nights over 5 years of {\it Subaru} time will be dedicated for the PFS survey for cosmology, galaxy evolution and Galactic archaeology, through the {\it Subaru} Strategic Survey Program (SSSP). The GCS field is not included in the PFS SSSP. However, the NIR spectrograph of PFS is especially well-suited for spectroscopic follow up for the \jasmine{} GCS field stars to obtain radial velocity and chemical abundances. We plan to apply for a {\it Subaru} Intensive Program (5 nights per semester) to follow up the \jasmine{} target stars. 

\subsubsection{ULTIMATE-{\it Subaru}}

ULTIMATE-{\it Subaru} \citep{Minowa+Koyama+Yanagisawa+20} is another next generation NIR instrument of {\it Subaru}, which is planned to be installed around 2028. ULTIMATE-{\it Subaru} is a wide-field (14'$\times$14') NIR imager and multi-objects spectrograph with Ground-Layer Adaptive Optics (GLAO), which enables a spatial resolution of FWHM$\sim$0.2" in the $K$-band. There is a planned Galactic Center Survey of $\sim$6~deg$^2$ region of the Galactic center, which covers the whole \jasmine{} Galactic center field, $J$, $H$, $K$ and the narrow-band $K_{\rm NB}$ filters with a high cadence of 4 days (1 months) for the high (low) cadence field. \jasmine{} can provide the astrometric reference stars in NIR, which would improve the astrometric accuracy of ULTIMATE-{\it Subaru} Galactic center survey data. The ULTIMATE-{\it Subaru} Galactic center survey can observe numerous stars fainter than the \jasmine{} magnitude limit. The combined data of \jasmine{} and ULTIMATE-{\it Subaru} will provide the accurate astrometric information of these faint stars. They would help to identify the star clusters in the Galactic center region from the proper motion of stars, and increase the event rate of the astrometric microlensing, which will enable the measurement of the masses of lensed objects with high precision, and help to identify BHs and exoplanets. 

\subsubsection{{\it VLT}/MOONS}

The Multi Object Optical and Near-infrared Spectrograph  \citep[MOONS:][]{Cirasuolo+Afonso+Bender+11} is a next generation instrument of the {\it Very Large Telescope (VLT)} UT1 at the European Southern Observatory (ESO) on Cerro Paranal in the Atacama Desert of Chile, and the planned first light is in 2023. MOONS is a multi-object (about 1000 fibers) NIR spectrograph with a field of view of 25~arcmin diameter. There are three channels of spectrograph, covering $RI$, $YJ$ and $H$-bands, with both low and high resolution modes. The low resolution mode covers the wavelength range of 0.65--1.8~$\mu$m with $R_{RI}>4100$, $R_{YJ}>4300$ and $R_H>6600$. High resolution modes cover 3 disconnected wavelength ranges $\lambda_{RI}=0.76$--0.89~$\mathrm{\mu}$m, $\lambda_{YJ}=0.93$--1.35~$\mathrm{\mu}$m and $\lambda_H=1.52$--1.64~$\mathrm{\mu}$m with the spectral resolution of $R_{RI}>9200$, $R_{YJ}>4300$ (fixed with the low resolution mode) and $R_H>18300$, respectively. 
MOONS science targets cover Galactic archaeology, the growth of galaxies and the first galaxies. Galactic archaeology studies by MOONS plan to take spectra for the several million stars observed by {\it Gaia} and the {\it VISTA} telescope, providing the crucial complementary information of the accurate radial velocity and detailed chemical abundances. The NIR coverage of MOONS is capable of recording the spectra of stars in the heavily obscured Galactic center region. Even with the high-resolution mode in the NIR $H$-band, a signal-to-noise ratio of more than 60 can be obtained with one hour of exposure for objects brighter than $H=15$~mag. MOONS is likely to be the most powerful instrument in 2020s, capable of taking high-resolution spectra for all the target stars in the \jasmine{} GCS field. 

\subsubsection{{\it VISTA}/4MOST}

ESO's 4-meter Multi-Object Spectroscopic Telescope \citep[4MOST:][]{deJong+Agertz+Berbel+19} will be installed on the {\it VISTA} telescope in Chile in 2024. 4MOST has 2436 fibers and about a 4.2~deg$^{2}$ field of view. There are two low-resolution and one high-resolution spectrographs. The low-resolution spectrograph covers the wavelength range of $\lambda=0.37$--0.95~$\mathrm{\mu}$m with $R\sim6500$, while the high-resolution spectrograph covers three wavelength passbands of $\lambda=0.3926$--0.4355~$\mathrm{\mu}$m, 0.5160--0.5730~$\mathrm{\mu}$m and 0.6100--0.6760~$\mathrm{\mu}$m.
The 4MOST consortium will run 10 surveys using 70\% of the available time in five years, and planned to start in 2023, taking more than 20 million low-resolution spectra and more than 3 million high-resolution spectra. 4MOST Milky Way Disc And BuLgE Low \citep[4MIDABLE-LR;][]{Chiappini+Minchev+Starkenburg+19} and High-Resolution \citep[4MIDABLE-HR;][]{Bensby+Bergmann+Rybizki+19} surveys will take the spectra of about 15 million and 2 million stars in the Milky Way, respectively. Their target includes the inner disk and bar/bulge region. Their survey focuses on the stars for which {\it Gaia} provides precise astrometry, but are too faint for {\it Gaia's} radial velocity spectrograph to provide a radial velocity. Their optical survey will not cover many stars in the \jasmine{} GCS region. However, the combination of 4MIDABLE data and Gaia data will be a powerful resource to unveil the global nature of the bar and spiral arms, and therefore highly complementary to the \jasmine{} GCS. 

\subsubsection{{\it Subaru}/IRD}

The InfraRed Doppler (IRD) spectrograph ($R\approx 70000$, $\lambda=0.950$--$1.73\,\mathrm{\mu m}$) on the {\it Subaru} telescope \citep{2018SPIE10702E..11K} is one of the most powerful instruments in the world to follow up exoplanets around M dwarfs and young stars. Besides the ``validation" of transiting-planet candidates around mid-M dwarfs identified by \jasmine{}, high-dispersion spectroscopy by IRD can play a key role in further characterization of those planets; precision radial velocity measurements by IRD would enable us to constrain precise planet masses as well as orbital eccentricities. Moreover, NIR transit spectroscopy by IRD would also allow us to constrain the stellar obliquity (spin-orbit angle) and atmospheric composition (e.g., He I and molecular species), which are supposed to reflect the dynamical and chemical evolution of exoplanetary systems \citep[e.g.,][]{2020ApJ...890L..27H}. 
Since activity-induced radial-velocity variations of host stars are suppressed in the NIR, IRD is also an ideal tool to confirm and characterize planets around young active stars. 

\subsection{Space Missions}

\subsubsection{{\it Gaia}}

ESA's {\it Gaia} \citep{Gaia+Prusti16} was launched in December 2013. The nominal mission lifetime was 5 years but the mission has been extended to the second quarter of 2025. The {\it Gaia} mission is an all-sky survey to provide the precise astrometry for more than one billion stars brighter than $G\sim21$~mag. {\it Gaia} uses a broad passband, the $G$-band that covers the wavelength range of $\lambda\sim0.33$--1.05~$\mu$m \citep{Evans+Riello+DeAngeli+18}. In the final data release after the end of the mission, the astrometric accuracy for the bright stars with $G\lesssim 13$~mag is expected to reach about 7~$\mu$as.  The astrometric accuracy of about $149$~$\mu$as is expected to be achieved for stars brighter than $G=19$~mag.\footnote{\url{cosmos.esa.int/web/gaia/science-performance}} {\it Gaia} has three instruments, the Astrometric instrument (ASTRO), the Spectrophotometer (BP/RP) and Radial Velocity Spectrograph (RVS). The ASTRO provides the five astrometric parameters, stellar position, proper motion and parallax. The Spectrophotometer consists of the BP and RP spectrophotometers. BP and RP respectively provides low-resolution ($R\sim5$--25) spectra of the wavelength ranges of $\lambda\sim0.33$--0.68~$\mathrm{\mu}$m and $\sim0.64$--1.05~$\mathrm{\mu}$m, which are used for chromaticity calibration for astrometric measurement, and estimates of the stellar parameters and dust extinction. RVS is an integral-field spectrograph with $R\sim11500$, covering $\lambda=0.845$--0.872~$\mathrm{\mu}$m \citep{Cropper+Katz+Sartoretti+18}. The main aim of the RVS is to provide the radial velocity for about 150 million stars brighter than $G_{\rm RVS}=16$~mag, depending on the spectral type of stars, where $G_{\rm RVS}$ is magnitude in the RVS passband. {\it Gaia}'s fourth data release is expected to be in 2025, and all the catalog and data will be released, including all epoch data for all sources based on 66 months of data of the nominal mission. The final fifth data release based on about 11 years of the extended mission is expected to be in 2030. Hence, late 2020s and early 2030s will be the truly golden age of the Galactic archaeology. \jasmine{} is expected to be launched in this golden age, and will provide the complementary data to the {\it Gaia} data, especially for the Galactic center stars, which the optical astrometry mission, {\it Gaia}, cannot observe.

\subsubsection{\it Nancy Grace Roman Space Telescope}
\label{sec:rst}

The {\it Nancy Grace Roman Space Telescope} ({\it Roman Space Telescope}) is a NASA observatory to study dark energy, exoplanets and infrared astrophysics \citep{Spergel+Gehrels+Baltay+15}. The telescope has a primary mirror of 2.4~m diameter. The nominal mission lifetime is six years. {The Roman Space Telescope} has the Wide Field Instrument (WFI) and the Coronagraph Instrument (CGI). WFI is a large area, 300~megapixel, NIR camera for imagaing and slitless spectroscopy with a grism ($R=435$--865) and prism ($R=70$--170). The imaging mode utilises several filters covering the wavelength range of $\lambda=0.48$--2.0~$\mathrm{\mu}$m. The most relevant survey of the {\it Roman Space Telescope} to \jasmine{}'s GCS is their Microlensing Survey, which will repeatedly observe about 2.81~deg$^2$ (with 10 fields) around $-0.5^{\circ}<l<1.8^{\circ}$ and $-2.2^{\circ}<b<-1^{\circ}$. There will be six 72 day campaigns over six years with cadence of every 15 minutes with a wide filter and 12 hours with a blue filter. The {\it Roman Space Telescope} will detect billions of bulge stars and the study to obtain the precise astrometry is ongoing \citep{WFIRST_Astrometry+19}.
The current planned survey region of Microlensing Survey of {\it Roman Space Telescope} does not cover the Galactic center where \jasmine{}'s GCS targets, because the Microlensing Survey requires it to maximize the event rate of microlensing and therefore target the region of less dust extinction to obtain a higher number density of background stars. However, the {\it Roman Space Telescope} is currently gathering the community input for the survey strategies to maximize the science output. With strong community inputs, there could be a possibility for the {\it Roman Space Telescope} to observe the \jasmine{} GCS field. Then, \jasmine{} astrometry results would be valuable to calibrate their astrometry for fainter stars observed by the {\it Roman Space Telescope}.

\subsubsection{\it James Webb Space Telescope}

The {\it James Webb Space Telescope (JWST)} is a NASA's flagship space observatory (6.5 m aperture), launched at the end of 2021 \citep{Gardner+Mather+Clampin+06}. {\it JWST} has two spectrographs in the NIR band, the NIR Spectrograph (NIRSpec) and the NIR Imager and Slitless Spectrograph (NIRISS). NIRSpec is a medium resolution spectrograph ($R=100$--2700) with a wavelength coverage of 0.6--5 $\mu$m. The saturation limit is $H \sim 10$~mag. NIRISS is a slitless spectrograph, whose spectral resolution is about $R=700$. The saturation limit is $J \sim 8.5$~mag. The mission lifetime of {\it JWST} is five year as designed, but 10 years as an optimistic goal. If the lifetime of {\it JWST} overlaps the observation period of \jasmine{}, these instruments will be the most powerful instruments to follow up exoplanets found by \jasmine{} Exoplanet Survey, to characterize the exoplanets atmosphere with detailed and precise transmission spectroscopy.  

{\it JWST} also could have a superb capability to study the structure, populations, and dynamics of the stars in the Galactic center. There is a large community proposal to observe the Galactic center field of about $1.25^\circ \times 0.25^\circ$ area around \SgrA\ with multiple-bands of NIRCam at multiple epochs
\citep{Schoedel+Longmore+Henshaw+23}. This can provide the proper motions of about 10~million faint stars in the Galactic center with the expected accuracy of $\sim0.3$~mas~yr$^{-1}$, if 5~years time baseline is achieved. The \jasmine{} GCS covers a larger area than the \textit{JWST} proposal field. \jasmine{} will provide complementary astrometry data for the brighter stars which are saturated in the \textit{JWST} observation.

\subsubsection{\it CHEOPS}
The {\it CHaraterising ExOPlanet Satellite (CHEOPS)}, launched in December 2019, is an ESA space mission dedicated for precision photometry to determine the radius of transiting exoplanets \citep{2014CoSka..43..498B}. The mission lifetime is assumed to be 3.5 years (nominal). {\it CHEOPS} is a PIT type of transiting exoplanet exploration, similar to \jasmine{}. The telescope diameter (32 cm) is similar to that of \jasmine{}. The passband of {\it CHEOPS} is visible (see figure \ref{fig:comp}) while that of \jasmine{} is NIR.  In this sense, \jasmine{} is complementary to {\it CHEOPS}. However, considering the difference of the launch date, \jasmine{} should be regarded as a successor of {\it CHEOPS} in terms of the space facility for photometric follow-up of transiting planets found by the ground-based surveys.

\subsubsection{\it TESS}

The {\it Transiting Exoplanet Survey Satellite (TESS)} is an MIT/NASA-led all-sky survey mission to find planets transiting bright nearby stars \citep{2014SPIE.9143E..20R}. {\it TESS} has four cameras, each with 10.5~cm diameter and $24\times24$~$\mathrm{deg}^2$ field of view ($\sim 2000~\mathrm{deg}^2$ in total) and each camera has four 2k$\times$2k CCDs with a pixel scale of $21''$. The detectors are sensitive from 0.6 to 1.0~$\mu$m. During the 2~year prime mission since the launch in 2018, {\it TESS} is monitoring the almost entire sky in 26 overlapping segments, and observes each segment for 27.4~days with 2~min cadence for the pre-selected 15000 stars and with 30 min cadence for the full image. Extension of the mission has been approved and {\it TESS} will keep tiling the whole sky at least till 2024.
{\it TESS} has a capability of finding Earth-sized transiting planets near the habitable zone of early- to mid-M dwarfs, and such an example has indeed been reported \citep[TOI-700d and e;][]{2020AJ....160..116G, 2023ApJ...944L..35G}. The larger telescope aperture and redder passband of \jasmine{} will make it sensitive to similar planets around later-type M dwarfs. 
Follow-up observations by \jasmine{} for planets detected by {\it TESS} may lead to finding longer-period/smaller planets that were missed by {\it TESS}, as well as to characterizing them even better through a finer sampling of the transit light curve. 

\subsubsection{\it PLATO}

{\it PLAnetary Transits and Oscillations of stars (PLATO)} is the third M-class (M3) mission under development by ESA for a planned launch in 2026 \citep{Rauer+Catala+Aerts+14}. The primary science goal is the detection and characterization of planets transiting bright solar-type stars, in particular terrestrial planets in the HZ. This will be achieved by high-precision, continuous photometric monitoring of a large number of bright stars  using a collection of small and wide-field optical telescopes. According to the {\it PLATO} definition study report,\footnote{\url{https://sci.esa.int/web/plato/-/59252-plato-definition-study-report-red-book}} the payload is planned to consist of $>20$ cameras each with 12~cm diameter and covering the wavelength range of 0.5--1.0~$\mu$m, which result in a total field of view of $\sim2000\,\mathrm{deg}^2$. Although the specific observing strategy is yet to be determined,
{\it PLATO} is likely to cover a significant fraction of the entire sky, as well as to monitor certain regions for a duration long enough \citep[$\sim$2~years;][]{Nascimbeni+Piotto+Boerner+22} to find planets in the HZ of Sun-like stars.
The duration of the nominal science operations is 4 years and may well overlap with the operation period of \jasmine{}. The main targets of {\it PLATO} are bright ($V\lesssim13$~mag) Sun-like stars, while \jasmine{} targets late-type stars fainter in the optical passband, taking advantage of the NIR photometry. Therefore, the two missions are complementary to each other. Similarly to {\it TESS}, {\it PLATO} observations might also provide transiting planet target candidates around M dwarfs that can be further characterized with NIR observations by \jasmine{}. {\it PLATO} also aims to characterise the properties, including the precise age estimates in the 10\% precision level, of the host stars from the time-series photometry using asteroseismology. The age information of a large number of stars which {\it PLATO} will observe will be a precious information for studies of Galactic archaeology \citep{Miglio+Chiappini+Mosser+17}. Hence, it will be also provide complementary data to the \jasmine{} GCS and mid-plane survey.

\subsubsection{\it ARIEL}

{\it Atmospheric Remote-sensing Infrared Exoplanet Large-survey} \citep[{\it ARIEL},][]{Tinetti+ARIEL18} is the first space telescope dedicated to the study of exoplanet atmospheres, adopted as an ESA's M4 mission, whose planned launch is in 2029. The effective size of the primary mirror of {\it ARIEL} will be $\sim$1~m, which is much smaller than {\it JWST} (6.5~m). However, {\it  ARIEL} will be able to collect fluxes in the wavelength range of 0.5--7.8 $\mu$m at one time, using five dichroic mirrors, three NIR spectrometers and three optical photometric detectors. This allows one to obtain an atmospheric spectrum with a very wide wavelength coverage from a single planetary transit or eclipse observation. 

{\it  ARIEL} will observe a thousand exoplanets with a wide range of mass and temperature, from hot Jupiters to warm/temperate Earths, in order to understand the statistical properties of exoplanetary atmospheres and planetary formation histories. While the current target list for {\it  ARIEL} already includes a large number of Jovian planets, it still lacks Neptune- and smaller-sized planets that are suitable for atmospheric study, i.e., hosted by nearby M dwarfs (Edwards et al. 2019). Although {\it TESS} has been increasing the number of such targets, it may not be enough due to its limited telescope aperture size (10~cm) and wavelength sensitivity (because of covering only the optical). Small transiting planets around nearby M dwarfs that will be discovered by {\it {JASMINE}} can thus be good targets for atmospheric characterization by {\it  ARIEL}. Given that \jasmine{} is planned to be launched ahead of {\it  ARIEL}, \jasmine{} can provide prime targets for {\it  ARIEL} in a timely manner.

\section{Summary and Conclusions}
\label{sec:sum}

We summarize that the unique capability of the \jasmine{} mission will fill the gap left by other planned and ongoing projects of the Galactic stellar surveys for Galactic archaeology and the habitable exoplanet searches in late 2020s. \jasmine{} will be the first mission to provide 10~$\mu$as-level astrometry in the NIR band with time-series photometry. \jasmine{} will offer precise astrometric information where the dust extinction is too strong for the optical astrometry mission, {\it Gaia}, to detect any stars, such as the Galactic center field and the Galactic mid-plane. The astrometric data of stars hidden behind the dust in the Galactic center and Galactic mid-plane will shed light on the formation epoch of the Galactic bar, the nature of the spiral arms and the mechanism underlying radial migration in the inner Galactic disk, which are likely to be remaining questions after {\it Gaia}. The combination of time-series photometry and precise astrometry will provide a vast opportunities of serendipitous discovery, including the possibilities of detecting IMBH, astrometric microlensing of inner disk BHs and studying the nature of star forming regions and X-ray sources. \jasmine{} will be also the only space observatory in the late 2020s which can follow up exoplanet transits detected by ground-based telescope to find the planets in the outer and habitable orbits around late-type stars, just as {\it Spitzer} space observatory contributed to revolutionising the field.

Finally, we note that \jasmine{} will be a crucial science demonstration mission for what a future NIR astrometry mission can offer. \jasmine{} will be a key mission to bridge between the successful {\it Gaia} mission and the proposed {\it Gaia}'s successor mission, {\it GaiaNIR} \citep[e.g.,][]{Hobbs+Brown+Hog+21} to be launched in the 2040s. {\it GaiaNIR} will provide the all-sky global astrometry in NIR band, including the Galactic disk, bar and bulge regions. Unprecedentedly high proper motion for the stars also observed with {\it Gaia} will be obtained, taking advantages of $\sim20$~years of baseline between the {\it Gaia} and {\it GaiaNIR} missions. Also, {\it GaiaNIR} will help maintaining and improving on the absolute astrometric quality of the celestial reference frame, which otherwise degrades with time. \jasmine{} will be an pioneering mission to open up the future $\mu$as-level NIR astrometry, and become an important milestone to demonstrate the power of NIR astrometry. About 20~years of time difference between \jasmine{} and GaiaNIR will provide superb proper motion measurements for the stars observed by \jasmine{}, including the NSD stars. With careful correction of systematic errors, the combination of \jasmine{}, {\it GaiaNIR}, and the other complementary data will offer endeavour to measure the acceleration of the stars and map the gravitational field in the Galactic center \citep[e.g.,][]{Chakrabarti+Wright+Chang+20,Chakrabarti+Stevens+Wright+22}.

\begin{ack}

We thank anonymous referees for their thorough review and helpful suggestions that have improved the manuscript.
We thank Megan Johnson and Stephen Williams for their contribution to the early draft of this manuscript.

This work presents results from the European Space Agency (ESA) space mission Gaia. Gaia data are being processed by the Gaia Data Processing and Analysis Consortium (DPAC). Funding for the DPAC is provided by national institutions, in particular the institutions participating in the Gaia MultiLateral Agreement (MLA). The Gaia mission website is https://www.cosmos.esa.int/gaia. The Gaia archive website is https://archives.esac.esa.int/gaia. This work is also based on data products from observations made with ESO Telescopes at the La Silla or Paranal Observatories under ESO programme ID 179.B-2002. Funding for the Sloan Digital Sky 
Survey IV has been provided by the 
Alfred P. Sloan Foundation, the U.S. 
Department of Energy Office of 
Science, and the Participating 
Institutions. 

SDSS-IV acknowledges support and 
resources from the Center for High 
Performance Computing  at the 
University of Utah. The SDSS 
website is www.sdss4.org. SDSS-IV is managed by the 
Astrophysical Research Consortium 
for the Participating Institutions 
of the SDSS Collaboration including 
the Brazilian Participation Group, 
the Carnegie Institution for Science, 
Carnegie Mellon University, Center for 
Astrophysics | Harvard \& 
Smithsonian, the Chilean Participation 
Group, the French Participation Group, 
Instituto de Astrof\'isica de 
Canarias, The Johns Hopkins 
University, Kavli Institute for the 
Physics and Mathematics of the 
Universe (IPMU) / University of 
Tokyo, the Korean Participation Group, 
Lawrence Berkeley National Laboratory, 
Leibniz Institut f\"ur Astrophysik 
Potsdam (AIP),  Max-Planck-Institut 
f\"ur Astronomie (MPIA Heidelberg), 
Max-Planck-Institut f\"ur 
Astrophysik (MPA Garching), 
Max-Planck-Institut f\"ur 
Extraterrestrische Physik (MPE), 
National Astronomical Observatories of 
China, New Mexico State University, 
New York University, University of 
Notre Dame, Observat\'ario 
Nacional / MCTI, The Ohio State 
University, Pennsylvania State 
University, Shanghai 
Astronomical Observatory, United 
Kingdom Participation Group, 
Universidad Nacional Aut\'onoma 
de M\'exico, University of Arizona, 
University of Colorado Boulder, 
University of Oxford, University of 
Portsmouth, University of Utah, 
University of Virginia, University 
of Washington, University of 
Wisconsin, Vanderbilt University, 
and Yale University.

This work is a part of MWGaiaDN, a Horizon Europe Marie Sk\l{}odowska-Curie Actions Doctoral Network funded under grant agreement no. 101072454 and also funded by UK Research and Innovation (EP/X031756/1).
This work was partly supported by the UK's Science \& Technology Facilities Council (STFC grant ST/S000216/1, ST/W001136/1), JSPS KAKENHI (23H00133, 21J00106), JSPS Postdoctoral Research Fellowship Program, the Spanish MICIN/AEI/10.13039/501100011033, ``ERDF A way of making Europe" by the “European Union” through grants RTI2018-095076-B-C21 and PID2021-122842OB-C21, the Institute of Cosmos Sciences University of Barcelona (ICCUB, Unidad de Excelencia ’Mar\'{\i}a de Maeztu’) through grant CEX2019-000918-M, NASA ADAP award program Number (80NSSC21K063), the Swedish National Space Agency (SNSA Dnr 74/14 and SNSA Dnr 64/17), the Royal Society (URF\textbackslash R1\textbackslash191555) and the ERC Consolidator Grant funding scheme (project ASTEROCHRONOMETRY \url{https://www.asterochronometry.eu/}, G.A. n. 772293).

\end{ack}

\bibliographystyle{apj}
\bibliography{./jwp}

\end{document}